\documentclass[resetfootnote ]{aastex702}

\usepackage{amssymb,amsfonts}
\usepackage{amsthm}
\usepackage{booktabs}
\usepackage{graphicx}
\usepackage{subcaption}
\usepackage{longtable}
\usepackage{multirow}
\usepackage{xspace}
\usepackage[font=small,labelfont=bf]{caption}
\usepackage{textcomp}
\usepackage[misc]{ifsym}
\usepackage{CJK}
\usepackage[whole]{bxcjkjatype}
\usepackage{CJKutf8}
\usepackage[utf8]{inputenc}

\usepackage{lmodern}
\usepackage{amsmath}
\usepackage{hyperref}
\usepackage{caption}
\usepackage{upgreek}
\usepackage{ulem}
\usepackage{fontawesome5}

\graphicspath{{figures/}}

\providecommand{\apjl}{Astrophys. J. Lett.} 
\providecommand{\apjs}{Astrophys. J. Suppl. Ser.} 
\providecommand{\ao}{Appl. Opt.} 
\providecommand{\aap}{Astron. Astrophys.} 

\begin{document}
\title{The 91T/99aa-like Type Ia Supernova 2019vrq, Part I: \\
Photometry, Spectroscopy, and Spectropolarimetry of a Nearly Standard Candle}

\author[orcid=0000-0002-6535-8500]{Yi Yang$^{*}$\begin{CJK*}{UTF8}{gbsn}
(杨轶)\end{CJK*}}
\affiliation{Department of Physics, Tsinghua University, Qinghua Yuan, Beijing 100084, China}
\email[show]{$^{*}$yi\_yang@mail.tsinghua.edu.cn} 

\author[0000-0002-4338-6586]{Peter Hoeflich}
\affiliation{Department of Physics, Florida State University, Tallahassee, FL 32306-4350, USA}
\email[show]{phoeflich77@gmail.com} 

\author[0000-0003-1349-6538]{J. Craig Wheeler}
\affiliation{Department of Astronomy, University of Texas, 2515 Speedway, Stop C1400, Austin, TX 78712-1205, USA}
\email[unshow]{wheel@astro.as.utexas.edu}

\author[0000-0003-1637-9679]{Dietrich Baade}
\affiliation{European Organisation for Astronomical Research in the Southern Hemisphere (ESO), Karl-Schwarzschild-Str.\ 2, 85748 Garching b.\ M{\"u}nchen, Germany}
\email[unshow]{dbaade@eso.org}

\author[0000-0001-7092-9374]{Lifan Wang}
\affiliation{Department of Physics and Astronomy, Texas A\&M University, 4242 TAMU, College Station, TX 77843, USA}
\affiliation{George P. and Cynthia Woods Mitchell Institute for Fundamental Physics \& Astronomy, Texas A\&M University, 4242 TAMU, College Station, TX 77843, USA}
\email[unshow]{lifan@tamu.edu}

\author[0000-0001-7101-9831]{Aleksandar Cikota}
\affiliation{European Organisation for Astronomical Research in the Southern Hemisphere (ESO), Alonso de Cordova 3107, Vitacura, Casilla 19001, Santiago de Chile, Chile}
\email[unshow]{aleksandar.cikota@noirlab.edu}

\author[0000-0003-4253-656X]{D. Andrew Howell}
\affiliation{Las Cumbres Observatory, 6740 Cortona Drive, Suite 102, Goleta, CA 93117-5575, USA}
\affiliation{Department of Physics, University of California, Santa Barbara, CA 93106-9530, USA}
\email[unshow]{ahowell@lco.global}

\author[0000-0001-5807-7893]{Curtis McCully}
\affiliation{Las Cumbres Observatory, 6740 Cortona Drive, Suite 102, Goleta, CA 93117-5575, USA}
\email[unshow]{cmccully@lco.global}

\author[0000-0003-3460-0103]{Alexei~V.~Filippenko}
\affiliation{Department of Astronomy, University of California, Berkeley, CA 94720-3411, USA}
\affiliation{Hagler Institute for Advanced Study, Texas A\&M University, 3572 TAMU, College Station, TX 77843, USA}
\email[unshow]{afilippenko@berkeley.edu}

\author[0000-0002-0537-3573]{Ferdinando Patat}
\affiliation{European Organisation for Astronomical Research in the Southern Hemisphere (ESO), Karl-Schwarzschild-Str.\ 2, 85748 Garching b.\ M{\"u}nchen, Germany}
\email[unshow]{fpatat@eso.org}

\author[0000-0002-5221-7557]{Chris Ashall}
\affiliation{Institute for Astronomy, University of Hawai'i at Manoa, 2680 Woodlawn Dr., Hawai'i, HI 96822, USA}
\email[unshow]{cashall@hawaii.edu}

\author[0000-0002-8255-5127]{Mattia Bulla}
\affiliation{Department of Physics and Earth Science, University of Ferrara, via Saragat 1, I-44122 Ferrara, Italy}
\affiliation{INFN, Sezione di Ferrara, via Saragat 1, I-44122 Ferrara, Italy}
\affiliation{INAF, Osservatorio Astronomico d'Abruzzo, via Mentore Maggini snc, 64100 Teramo, Italy}
\email[unshow]{mattia.bulla@unife.it}

\author[0000-0002-3653-5598]{Avishay Gal-Yam}
\affiliation{Department of Particle Physics and Astrophysics, Weizmann Institute of Science, Rehovot, Israel}
\email[unshow]{avishay.gal-yam@weizmann.ac.il}

\author[0000-0002-9154-3136]{Melissa.~L. Graham}
\affiliation{Department of Astronomy, University of Washington, Box 351580, U.W., Seattle, WA 98195, USA}
\affiliation{Institute for Data-intensive Research in Astrophysics and Cosmology, University of Washington, 3910 15th Avenue NE, Seattle, WA 98195, USA}
\email[unshow]{mlg3k@uw.edu}

\author[0000-0002-1125-9187]{Daichi Hiramatsu}
\affiliation{Department of Astronomy, University of Florida, Bryant Space Science Center, Gainesville, FL 32611-2055, USA}
\email[unshow]{dhiramatsu@ufl.edu}


\author[0000-0001-5965-0997]{Divya Mishra}
\affiliation{Department of Physics and Astronomy, Texas A\&M University, 4242 TAMU, College Station, TX 77843, USA}
\affiliation{George P. and Cynthia Woods Mitchell Institute for Fundamental Physics \& Astronomy, Texas A\&M University, 4242 TAMU, College Station, TX 77843, USA}
\email[unshow]{divya2410mishra@gmail.com}

\author[0000-0003-0209-9246]{Estefania Padilla Gonzalez}
\affiliation{Department of Physics and Astronomy, Johns Hopkins University, Baltimore, MD 21218, USA}
\email[unshow]{epadill7@jh.edu}

\author[0000-0002-7472-1279]{Craig~M.~Pellegrino}
\affiliation{NASA Goddard Space Flight Center, 8800 Greenbelt Road, Greenbelt, MD 20771, USA}
\email[unshow]{craig.m.pellegrino@nasa.gov}

\author[0000-0001-6797-1889]{Steve Schulze}
\affiliation{Department of Particle Physics and Astrophysics, Weizmann Institute of Science, Rehovot, Israel}
\email[unshow]{steve.schulze@weizmann.ac.il}

\begin{abstract}
Among the various major subtypes of Type Ia supernovae (SNe), the luminous 1991T-like~\citep{1992ApJ...384L..15F, 1992AJ....103.1632P} and 1999aa-like~\citep{2000ApJ...539..658K, 2004AJ....128..387G} events stand out as particularly important for constraining the explosion mechanism. Their white dwarf (WD) progenitors are thought to have masses close to or exceeding the Chandrasekhar limit. 
We present spectrophotometric and polarimetric time-series observations of the overluminous Type Ia SN\,2019vrq. Between $-$9 and $+$12 days relative to the light-curve peak, its continuum polarization is consistent with zero, as seen in most normal-bright SNe\,Ia and despite a possible rise up to 0.2\% toward longer wavelengths as seen in subluminous SNe\,Ia. 
Polarization across major spectral features is marginal ($\lesssim$0.3\%) with the exception of the high-velocity Ca\,{\sc ii} near-infrared triplet (NIR3), which reaches $\sim$0.5\%. This low polarization indicates a substantially spherically symmetric electron-scattering photosphere and minimal large-scale ejecta asymmetry. 
Early-time spectra reveal hot, doubly ionized ejecta dominated by Si\,{\sc iii} and blended iron-group features. Intermediate-mass-element lines, such as Ca\,{\sc ii} and Si\,{\sc ii}, are more clearly seen than in 91T-like events, though these lines remain weaker than in normal SNe\,Ia. 
The intrinsically higher peak luminosity of overluminous 91T/99aa-like SNe\,Ia~\citep{2022ApJ...938...83Y, 2024ApJS..273...16P}, their comparatively short formation timescale for a massive WD progenitor -- a channel that becomes more common toward high redshift, and the moderate viewing-angle dependence of their spectrophotometric properties as inferred from their persistently low polarization, together suggest that this subclass may serve as a promising standard candle at high redshift. 
\end{abstract}


\section{Introduction}~\label{sec:main}
Despite being a critical step on the cosmic distance ladder that led to the discovery of the accelerating expansion of the universe~\citep{1998AJ....116.1009R, 1999ApJ...517..565P, 2000AIPC..540..227F, 2016ApJ...826...56R, 2022ApJ...934L...7R}, the spectrophotometric signatures of Type Ia supernovae (SNe\,Ia) appear to be diverse, hinting at a rather inhomogeneous population of their progenitor white dwarfs (WDs; see reviews by \citealp{2011NatCo...2..350H, 2013FrPhy...8..116H, 2014ARA&A..52..107M, 2017suex.book.....B, 2017hsn..book.1151H, 2017hsn..book..317T, 2023RAA....23h2001L}). 
The tight correlation between the maximum luminosity of most SNe\,Ia and their light-curve (LC) shape, particularly the 15-day post-peak magnitude decline in the $B$ band ($\Delta m_{15}(B)$), enables a precise and systematic determination of their intrinsic brightness~\citep{1993ApJ...413L.105P}. This LC width-luminosity relationship facilitates SNe\,Ia as cosmic distance indicators up to a redshift of $z\approx1$--2~\citep{2000ApJ...544L.111C, 2001ApJ...560...49R, 2004ApJ...607..665R, 2007ApJ...659...98R}
Exploring the boundaries and identifying the outliers is critical for determining the nature of the explosions and measuring the cosmological parameters. 
The discovery of SN\,1991T (hereafter ``91T'';~\citealp{1992ApJ...384L..15F, 1992AJ....103.1632P, 1992ApJ...387L..33R}) and 1999aa (hereafter ``99aa'';~\citealp{2001ApJ...546..734L}) established an important subtype whose spectroscopic signatures during the rising phase present significantly shallower absorption lines from intermediate-mass elements (IMEs; that is, from Si to Ca) compared to those observed from most SNe\,Ia. Instead, strong blueshifted features of iron-group elements (IGEs; e.g., Fe\,{\sc iii}\,$\lambda\lambda$4397, 4421, 4432, and Fe\,{\sc iii}\,$\lambda$5129) can be identified from $\sim$4 days before  maximum light~\citep{2024ApJS..273...16P}.

Modeling efforts under the assumption of local thermodynamic equilibrium (LTE) suggest higher temperatures and an IGE-rich outermost ejecta layer of 91T/99aa-like SNe compared to that inferred for normal events~\citep{1992ApJ...387L..33R, 1992ApJ...397..304J, 1995A&A...297..509M, 2014MNRAS.445..711S}. 
The overluminous 91T/99aa-like events also exhibit $\gtrsim$0.4--0.5 brighter peak magnitude than normal-bright SNe\,Ia~\citep{2014MNRAS.445..711S}, indicating that they are powered by the radioactive decay of $\gtrsim$0.8\,M$_{\odot}$ of $^{56}$Ni synthesized in the explosion of their massive WD progenitors~\citep{2014MNRAS.440.1498S}. 
Such high production of IGEs suggests progenitor masses close to or larger than the Chandrasekhar limit ($M_{\rm Ch}\approx 1.375$\,M$_{\odot}$). The relatively short time required by the formation of high-mass WDs makes such overluminous SNe likely to be the most common subtype in the early universe. Their relatively high peak luminosity allows them to be observed more easily as standard candles at large distances~\citep{2022ApJ...938...83Y}.

The $M_{\rm Ch}$ large-amplitude radially pulsating delayed-detonation (PDD) scenario was previously invoked to explain the relatively slow rise and decline LCs of SN\,1990N~\citep{1992A&A...253L...9K, 1992A&A...259..549H}, whose pre-peak spectral evolution displayed a prominent Si\,{\sc ii}\,$\lambda$6355 line and thus more resembles spectroscopically normal SNe\,Ia~\citep{1991ApJ...371L..23L, 1997ARA&A..35..309F}. 
Such a mechanism was discussed in more detail by \citet{1994ApJ...427..315A}, and a comprehensive modeling of explosions and the LCs under the PDD framework has been carried out by \citet{1996ApJ...457..500H}.
The pulsation affects the early burning phase by puffing up the WD, and the detonation phase is initiated by fallback of the outer layers and the hydrodynamical instabilities at the interface between fallback and expanding layers. The unburned layers depend on the amplitude of the pulsation. 
However, large-amplitude pulsations produce unburned C/O layers of $\geq$0.2\,M$_{\odot}$, inconsistent with the prompt emergence of weak lines owing to explosive nuclear burning at day\,$-9$, compared to the roughly day\,$-4$ emergence expected for normal-bright SNe\,Ia~\citep{2007ApJ...666.1083Q}.


The abundance structure of the ejecta contains fingerprints left behind from the propagation of the burning front through the WD progenitor. Any deviation from spherical symmetry may impose a notable viewing-angle dependence, thus contributing to the persistent intrinsic scatter among their photometric and spectroscopic properties~\citep{1991A&A...246..481H, 2003ApJ...591.1110W, 2003ApJ...593..788K}. 
A detailed probe of the ejecta properties is offered by time-series spectroscopy combined with polarimetry. Spectroscopy delivers not only information about the chemical composition but also line profiles that reflect the physical conditions of the line-forming regions, while polarization maps out the geometry orthogonal to the line of sight. 
The expanding ejecta within 1--2 weeks after explosion can be approximated by an electron-scattering atmosphere as long as the photosphere is formed in IME layers~\citep{2023MNRAS.520..560H}. 
Continuum polarization may arise when the photosphere deviates from spherical symmetry. An incomplete cancellation of electric ($E$) vectors is then seen when integrating across the photosphere. 
Polarization measured across a given spectral feature traces the geometrical distribution of the corresponding line-forming region within a photosphere dominated by a quasicontinuum with weak lines and electron scattering~\citep{2008ARA&A..46..433W, 2017hsn..book.1017P, 2023MNRAS.520..560H, 2024MNRAS.528.3875M}. 
The photosphere also recedes into deeper layers of the ejecta over time as the latter expands and becomes progressively optically thin. 


Based on the analysis of \citet{2024ApJ...969...80C}, 91T-like SNe\,Ia are expected to constitute an increasing fraction of the SN\,Ia population with increasing redshift. 
The birthplaces of members of this subclass are correlated with galaxies undergoing active star formation, suggesting a young population with a short delay time between star formation and explosion~\citep{2000AJ....120.1479H, 2017hsn..book..317T, 2022ApJ...938...47P}.
Specifically, the relative abundance of 91T-like events rises from $\sim$5\% in the local universe to $\sim$30\% at higher redshifts~\citep{2024ApJ...969...80C}, which could introduce systematic biases in cosmological distance measurements. 
Owing to their intrinsically higher peak luminosities (0.2--0.5\,mag brighter than normal SNe\,Ia; \citealt{2022ApJ...938...83Y, 2024ApJS..273...16P}, overluminous 91T/99aa-like SNe\,Ia remain observable to greater cosmological distances, making them valuable standard candles for probing the expansion history of the universe at early epochs. 

However, direct observational tests of this predicted rise remain limited. Both observational efforts \citep{2001ApJ...546..734L} and Monte~Carlo simulations of high-redshift search strategies suggest that 91T-like events might comprise $\sim$20\% of high-$z$ discoveries. 
But confirmed 91T-like objects remained scarce among the first $\sim$100 high-$z$ SNe\,Ia studied, with the observed fraction later estimated at $\lesssim$5\%~\citep{2009A&A...507...85B}. This tension may stem from sparse temporal sampling near maximum light, together with the narrow pre-maximum window over which reliable spectroscopic classification -- based on the diagnostic Si\,{\sc ii}/Ca\,{\sc ii} weakness and Fe\,{\sc iii} dominance -- remains possible \citep{2004AJ....128..387G}. 
One potential criterion for discriminating 91T/99aa-like from normal SNe\,Ia is ultraviolet (UV) color, given their pronounced blueness before maximum light~\citep{2013ApJ...779...23M}. The hot, highly ionized ejecta of 91T-like events suppress line blanketing by IGEs, an effect most pronounced at UV wavelengths. This diagnostic weakens considerably in $B-V$ near maximum light (see, e.g., \citealp{2022ApJ...938...83Y, 2024ApJS..273...16P}). 
A detectability simulation performed specifically for 91T/99aa-like SNe\,Ia, accounting for these cadence and color-sampling constraints, would help determine whether the observed and predicted high-$z$ rates can be reconciled, and would establish the true completeness of overluminous SNe\,Ia as standard candles for early-universe cosmology. 
A high detectability at high redshift may offer opportunities to refine models of SN\,Ia progenitor systems and explosion mechanisms across cosmic time. 


Here we report spectropolarimetric observations of the Type Ia SN\,2019vrq, which was discovered by the All-Sky Automated Survey for Supernovae (ASAS-SN~\citealp{2014ApJ...788...48S}) at $g\approx18$\,mag on 2019-11-28 05:46 UTC (MJD 58815.24; \citealp{2019TNSTR2470....1S}). 
We also analyze the spectrophotometric evolution of the SN, incorporating some data previously introduced by \citet{2022ApJ...938...83Y}, 
which established the 91T/99aa-like nature of SN\,2019vrq based on its photometric and spectroscopic properties. That comprehensive analysis of 16 91T/99aa-like events also demonstrated that SNe\,Ia belonging to this subclass are excellent distance indicators and can be used to place galaxies in the Hubble flow to within 12\% in distance. 

This paper is organized as follows. Observations and data reduction are detailed in Section~\ref{sec:obs}. Section~\ref{sec:analysis} describes the photometric, spectroscopic, and bolometric behavior of the SN. We investigate its spectropolarimetric properties in Section~\ref{sec:pol}. A discussion and brief summary are given in Section~\ref{sec:summary}. 
We defer the more detailed three-dimensional (3D) radiation-hydrodynamical simulations of the 91T/99aa-like SN\,2019vrq to Paper~II (Hoeflich et al. 2026), where we show that a low-amplitude PDD of a near-$M_{\rm Ch}$ WD can reproduce most of the observed properties and quantify the direction-dependent luminosity, with applications to cosmology. 

\section{Observations}~\label{sec:obs}
\subsection{Optical Photometry}~\label{sec:obs_phot}
The host of SN\,2019vrq has  $z = 0.01308$ \citep{2009MNRAS.399..683J} and was classified as a spiral galaxy (morphological type S; \citealp{2022MNRAS.509.6028C}). Classification spectra obtained a day later match well with 91T/99aa-like events before the LC peak~\citep{2019TNSCR2483....1I, 2019TNSCR2898....1Z}. 
Motivated by its proximity, our observing campaign was started at about $-9$ days relative to the $B$-band maximum light (see \citealp{2022ApJ...938...83Y}).  
Photometric observations of SN\,2019vrq conducted using the Sinistro cameras on the Las Cumbres Observatory (LCO) Global Network of 1\,m telescopes have been thoroughly discussed by \citet{2022ApJ...938...83Y}. 

We carried out independent photometry on the images preprocessed by the BANZAI automatic pipeline~\citep{2018SPIE10707E..0KM}. 
For each frame, the point-spread function (PSF) was determined from the ten brightest, isolated field stars relatively close to the SN, using the full width at half-maximum intensity (FWHM) as the fitting radius. Rather than subtracting cross-band template images to remove the contamination of the host-galaxy light, the background was estimated by fitting a median pixel value within an annulus centered at the SN with inner and outer radii of 4\farcs0 and 5\farcs5, respectively. 
These annulus sizes were justified by the fact that the residuals measured from the PSF-subtracted field stars were consistent with the noise beyond $\sim1.5\times$FWHM, which was measured to be better than 2\farcs5 in all LCO images. Moreover, because the SN remained more than ten times brighter than the ambient host nebulosity ($m(B)\lesssim18$\,mag) throughout the phases of interest for this study ($\lesssim$+100\,d), the iterative procedure between the PSF fitting and background estimation provides a reasonable estimate of the underlying background of the SN. 

\begin{figure}[ht]
\centering
  \includegraphics[width=0.7\textwidth]{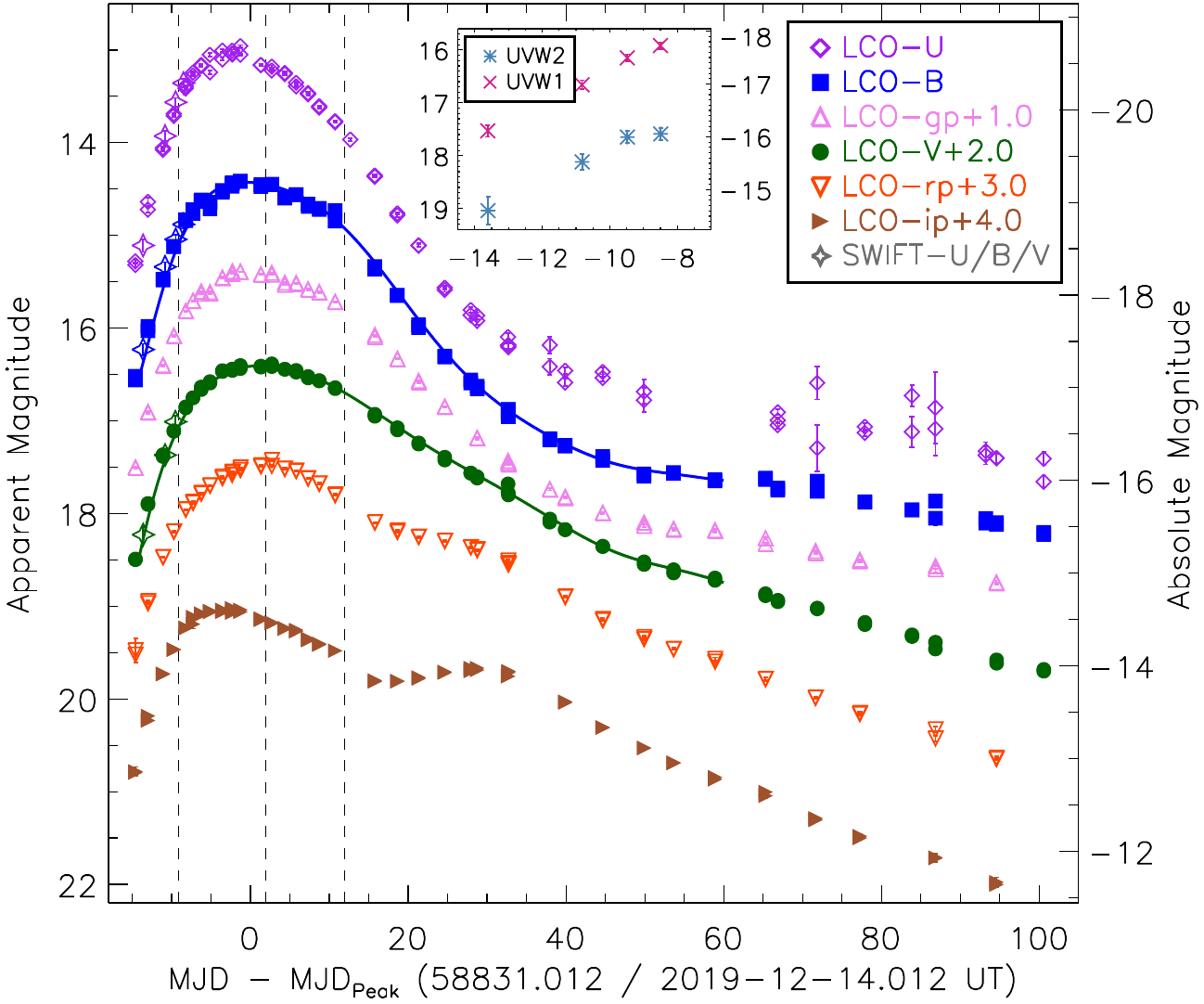}
  \caption{UV-optical LCs of SN\,2019vrq. Vertical dashed lines mark the epochs of VLT spectropolarimetry. Solid blue and green curves represent polynomial fits to the $B$- and $V$-band LCs, respectively. The top insets present {\it Swift} $uvw2$ and $uvw1$ photometry obtained at early phases. }~\label{fig:lc} 
\end{figure}

By querying the magnitudes of ten local comparison stars from the AAVSO Photometric All-Sky Survey (APASS) DR9 Catalogue~\citep{2015AAS...22533616H}, we calibrated the instrumental $UBV$ and $g'r'i'$ magnitudes of SN\,2019vrq in the Johnson system~\citep{1966CoLPL...4...99J} (Vega magnitudes) and in the SDSS photometric system~\citep{1996AJ....111.1748F} (AB magnitudes; \citealp{1983ApJ...266..713O}), respectively. The final $UBg'Vri'$-band calibrations were derived from the median of the differences between the catalog and instrumental magnitudes.
Fig.~\ref{fig:lc} shows the LCO $UBg'Vr'i'$-band LCs of SN\,2019vrq from roughly days $-14$ to $+$101. Early photometry acquired by the Ultra-violet/Optical Telescope (UVOT;~\citealp{2005SSRv..120...95R}) on the Neil Gehrels {\it Swift} Observatory~\citep{2004ApJ...611.1005G} is also displayed. All LCs presented here have not been corrected for Galactic or host-galaxy extinction.

\subsection{Optical Spectroscopy}~\label{sec:obs_spec}
The spectral time series of SN\,2019vrq consists of 19 low-resolution optical spectra spanning $-$10 to $+$78 days relative to $B$ maximum light, including three total-flux spectra derived from the VLT spectropolarimetric observations. The remaining data presented here of LCO optical spectra obtained with the FLOYDS spectrographs mounted on the 2\,m Faulkes Telescopes North and South at Haleakala, USA (FTN) and Siding Spring, Australia (FTS), through the Global Supernova Project~\citep{2013PASP..125.1031B}. For all LCO observations, a 2$\arcsec$-wide slit was adopted at the parallactic angle~\citep{1982PASP...94..715F}. 
The spectra were extracted, reduced, and calibrated following standard procedures using the FLOYDS pipeline\footnote{\url{https://github.com/svalenti/FLOYDS_pipeline}} \citep{2014MNRAS.438L.101V}.

\begin{figure}[t!]
    \centering
    \includegraphics[trim={0.0cm 0.0cm 0.0cm 0.0cm},clip,width=1.0\textwidth]{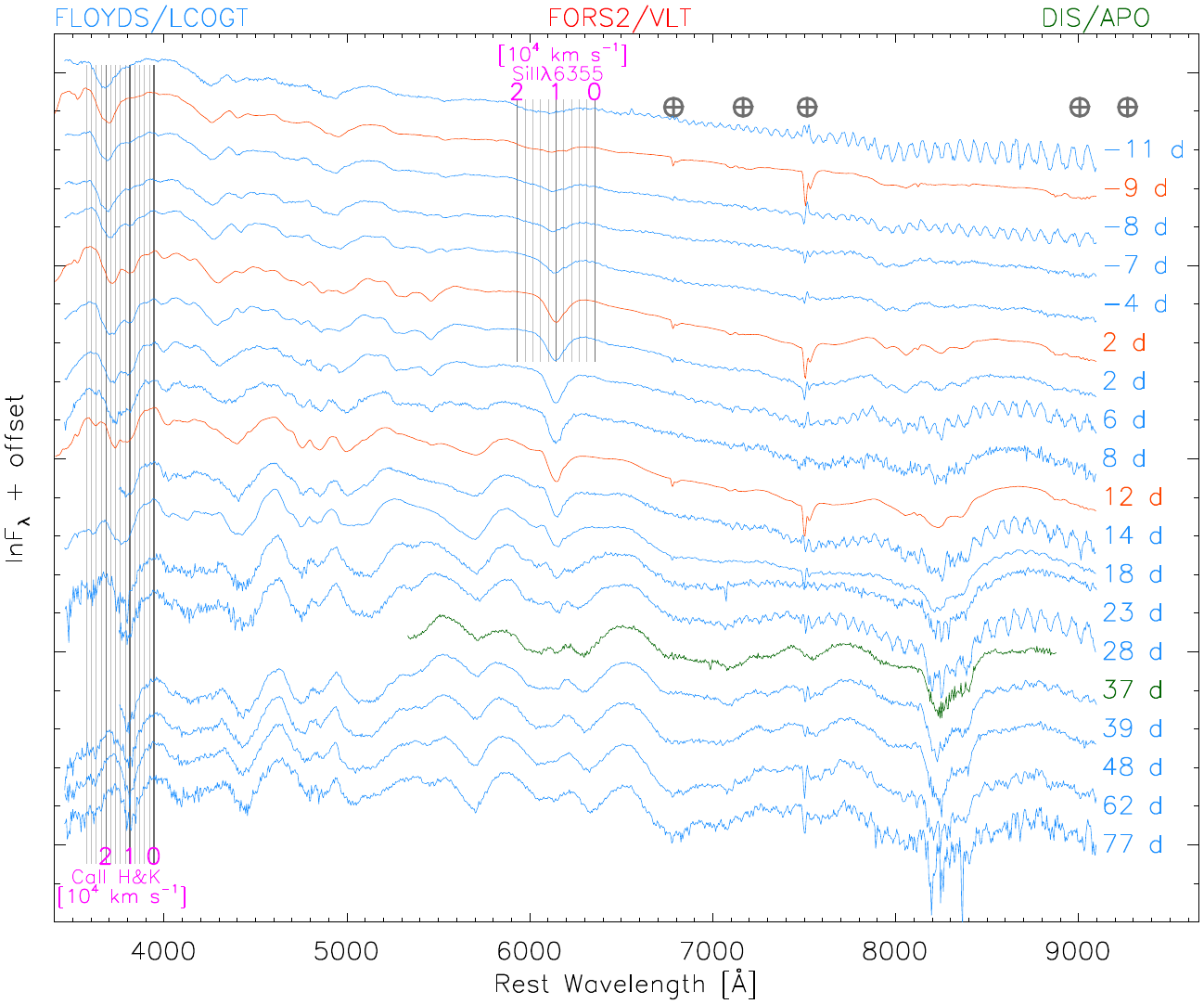}
    \vspace{-0.0 cm}
    \caption{Spectral time series of SN\,2019vrq. Phases and instruments are labeled on the right and indicated on the top. 
    Ca\,{\sc ii}\,H\&K and Si\,{\sc ii}\,$\lambda$6355 features are indicated by the text near the thick vertical gray lines, which measure the rest-frame velocity over the associated profile. The distance between adjacent gray lines denotes a 2000\,km\,s$^{-1}$ interval. Several strong telluric lines are marked by crossed circles. All spectra were corrected for the redshift of the host galaxy and rebinned to 5\,\AA. 
    }~\label{Fig_spec} 
\end{figure}

One low-resolution spectrum of SN\,2019vrq was also obtained using the Dual Imaging Spectrograph (DIS\footnote{\url{https://www.apo.nmsu.edu/arc35m/Instruments/DIS/}}), mounted on the 3.5\,m Astrophysics Research Consortium (ARC) telescope at Apache Point Observatory. The observations employed the B400 and R300 gratings, with central wavelengths of 4500 and 7500\,\AA, respectively. The instrument was aligned to the parallactic angle. The data were processed using conventional reduction techniques and calibrated against a standard star observed on the same night.

Throughout the paper, the wavelength scales of all SN spectra were corrected to the rest frame. In Fig.~\ref{Fig_spec} we present the spectral time sequence of SN\,2019vrq. A log of the spectroscopic observations of SN\,2019vrq is given in Table~\ref{Table_log_specpol}.

\begin{table}[ht]
\normalsize
{\centering
\captionsetup{name= Table.}
\caption{Log of Spectroscopy and Spectropolarimetry of SN\,2019vrq.}\label{Table_log_specpol}
\begin{tabular}{c|ccccccc}
\hline
\# & MJD   & Phase$^a$ & Spectral Range & $R$ (blue/red)             & Exp. Time$^b$ & Airmass &  Instrument/Telescope \\
      & [day] & [day]     &  [\AA]         & $\lambda/\Delta \lambda$ &  [s]             &         &   \\
\hline
 1  &  58820.4418  &  $-$10.6 &  3300$-$10180  &  435/449  &  2700                  &  1.10  &  FLOYDS/LCO 2.0 m FTS \\
 2  &  58822.0755  &  $-$8.9  &  3413$-$9333   &  440      &  540$\times$4 &  1.04  &  VLT/FORS2 \\
 3  &  58823.4400  &  $-$7.6  &  3300$-$10180  &  478/477  &  2700                  &  1.09  &  FLOYDS/LCO 2.0 m FTS \\
 4  &  58824.3183  &  $-$6.7  &  3300$-$10180  &  555/509  &  1200                  &  1.27  &  FLOYDS/LCO 2.0 m FTN \\
 5  &  58827.3043  &  $-$3.7  &  3146$-$10885  &  629/545  &  1200                  &  1.28  &  FLOYDS/LCO 2.0 m FTN \\
 6  &  58833.0440  &  $+$2.0  &  3413$-$9333   &  440      &  360$\times$4 &  1.04  &  VLT/FORS2 \\
 7  &  58833.2348  &  $+$2.2  &  3300$-$10180  &  558/151  &  1200                  &  1.48  &  FLOYDS/LCO 2.0 m FTN \\
 8  &  58836.5196  &  $+$5.5  &  3300$-$10180  &  474/457  &  900                   &  1.09  &  FLOYDS/LCO 2.0 m FTS \\
 9  &  58839.3495  &  $+$8.3  &  3146$-$10885  &  623/538  &  900                   &  1.30  &  FLOYDS/LCO 2.0 m FTN \\
10  &  58843.0527  &  $+$12.0 &  3413$-$9333   &  440      &  540$\times$4 &  1.01  &  VLT/FORS2 \\
11  &  58844.5648  &  $+$13.6 &  3014$-$10626  &  495/458  &  900                   &  1.36  &  FLOYDS/LCO 2.0 m FTS \\
12  &  58849.3364  &  $+$18.3 &  3300$-$10180  &  553/527  &  1200                  &  1.35  &  FLOYDS/LCO 2.0 m FTN \\
13  &  58854.2035  &  $+$23.2 &  3146$-$10885  &  641/546  &  1200                  &  1.35  &  FLOYDS/LCO 2.0 m FTN \\
14  &  58859.4533  &  $+$28.4 &  3014$-$10626  &  478/453  &  1200                  &  1.09  &  FLOYDS/LCO 2.0 m FTS \\
15  &  58868.0527  &  $+$37.0 &  5300$-$8900   &  2$"$.31\,pixel$^{-1}$  &  900             &  1.58  &  DIS/ARC 3.5 m \\
16  &  58870.4507  &  $+$39.4 &  3014$-$10626  &  526/488  &  1800                  &  1.17  &  FLOYDS/LCO 2.0 m FTS \\
17  &  58879.2141  &  $+$48.2 &  3300$-$10180  &  554/514  &  1800                  &  1.26  &  FLOYDS/LCO 2.0 m FTN \\
18  &  58893.4453  &  $+$62.4 &  3300$-$10180  &  462/455  &  2700                  &  1.58  &  FLOYDS/LCO 2.0 m FTS \\
19  &  58908.4014  &  $+$77.4 &  3300$-$10180  &  462/443  &  2700                  &  1.55  &  FLOYDS/LCO 2.0 m FTS \\
\hline
\end{tabular}\\
}
{$^a$}{Relative to the $B$-band maximum at MJD 58831.012.}\\
{$^b$}{For VLT spectropolarimetry, the values indicate the integration time at each retarder plate angle $\times$ 4 angles.}
\end{table}

\subsection{VLT Spectropolarimetry}~\label{sec:vlt}
Spectropolarimetry of SN\,2019vrq was performed using the FOcal Reducer and low-dispersion Spectrograph 2 (FORS2; \citealp{1998Msngr..94....1A}) on Unit Telescope 1 (UT1, Antu) of the ESO Very Large Telescope (VLT). 
Exposures were conducted in the Polarimetric Multi-Object Spectroscopy (PMOS) mode, following a series of $N=4$ integrations at retarder-plate angles of $\theta_{i} = i\times$22\fdg5, in which $i$ runs from 0 to 3. In our observations, $\theta_{i}$ yields 0, 22.5, 45, and 67.5 degrees. In each frame, the incident light was divided into ordinary ($o$-) and extraordinary ($e$-) beams -- that is, $f^{o}(\theta_{i})$ and $f^{e}(\theta_{i})$. 
All observations were carried out using the 300V grism and a 1\farcs0-wide slit. This provides a spectral-resolving power of $R \approx 440$, corresponding to an intrinsic size of $\sim$14\,\AA\ or $\sim$680\,km\,s$^{-1}$ for each resolution element near Si\,{\sc ii}\,$\lambda$6355~\citep{Anderson_etal_2018}. 
Because second-order contamination in spectropolarimetry has been characterized as negligible unless the source is very blue~\citep{2010A&A...510A.108P}, we did not use the GG435 order-sorting filter, 
thereby also extending the coverage into the blue. The slit was always aligned with the north celestial meridian, in which case the instrument position angle, $\chi$, is always zero; however, in most cases the airmass was quite low, so there shouldn't be much differential light lost from atmospheric dispersion. A log of the VLT spectropolarimetry of SN\,2019vrq is also included in Table~\ref{Table_log_specpol}.

After preprocessing the raw spectropolarimetry data for the spectra of each set of ordinary and extraordinary beams following the standard routine, which includes bias subtraction and flat-field correction, we extract the flux spectra 
with IRAF\footnote{{IRAF} is distributed by the National Optical Astronomy Observatories, which are operated by the Association of Universities for Research in Astronomy, Inc., under cooperative agreement with the U.S. National Science Foundation.}~\citep{1986SPIE..627..733T, 1993ASPC...52..173T}. 
Cosmic rays were rejected by implementing a Laplacian edge-detection algorithm (LACosmic~\citealp{2001PASP..113.1420V}). The calibration of the wavelength of each spectral trace was carried out, with a typical root-mean-square (RMS) accuracy of $\sim 0.25$\,\AA. 
The complete workflow for the VLT spectropolarimetry data analysis, including calculation of the Stokes $Q$ and $U$ parameters and the polarization spectra~\citep{2006PASP..118..146P, 2007MNRAS.381..201M}, correction for the instrumental polarization~\citep{2007ASPC..364..503F, 2014A&A...561A..82S, 2017MNRAS.464.4146C}, and the bias correction required to derive the true polarization level~\citep{1985A&A...142..100S, 1997ApJ...476L..27W} is described by \citet{2017MNRAS.464.4146C}, \citet{2020ApJ...902...46Y}, and  \citet{2023MNRAS.519.1618Y}. 

\begin{figure}
    \centering
    \includegraphics[width=0.9\textwidth]{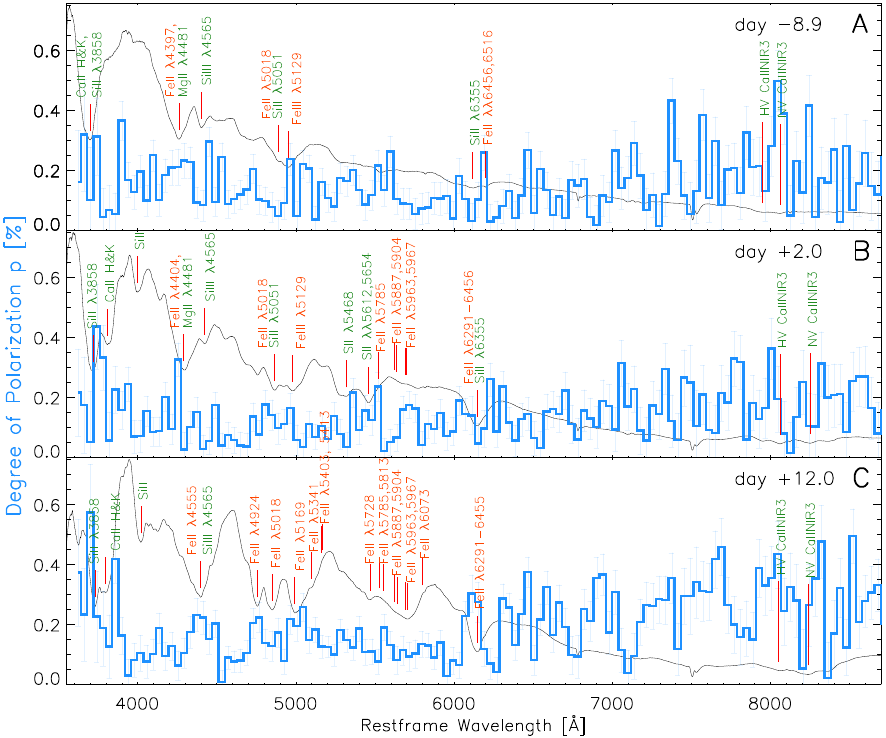}
    \vspace{-0.0 cm}
    \caption{\textbf{Polarization (blue histograms) and scaled total-flux spectra (black curves) of SN\,2019vrq.} Polarization spectra were computed after subtracting the ISP. Major spectral features are labeled with blueshifted velocities of $\sim10,000$\,km\,s$^{-1}$. 
    Note that a high-velocity (HV) component of the Ca\,{\sc ii}\,NIR3 exhibits a discernible peak in the polarization spectra, particularly at day\,$-$8.9, which is rather weak in the corresponding total-flux spectrum. 
    }\label{fig:pol_spec}
\end{figure}

The continuum polarization of SN\,2019vrq at all three epochs of VLT observations was estimated as the error-weighted mean Stokes $Q$ and $U$ parameters within certain wavelength ranges (Section~\ref{sec:contpol}), where strong blanketing by numerous bound-bound transitions of IGEs depolarizes the emission~\citep{1996ApJ...457..500H, 2001ApJ...556..302H, 2013MNRAS.433L..20M}. After subtracting the interstellar polarization (ISP; Section~\ref{sec:isp}), the continuum polarization is consistent with zero and displays no significant temporal evolution from days $-$9 to $+$12 (see Section~\ref{sec:contpol}). 
The peak polarizations of the characteristic Si\,{\sc ii}\,$\lambda$6355 and Ca\,{\sc ii} near-infrared triplet (NIR3) features are $\lesssim$0.3\% (Fig.~\ref{fig:pol_spec}). 

\section{Data Analysis}~\label{sec:analysis}
\subsection{Photometric Properties}~\label{sec:lc}
A simultaneous fit to the $BVgri$-band LCs of SN\,2019vrq using the SuperNovae in object-oriented Python (SNooPy) package~\citep{2011AJ....141...19B} suggests a $B$-band LC-peak time of MJD $t_{B\rm max}= 58830.773 \pm 0.053 \pm 0.340$, consistent with the result from a high-order polynomial ﬁt to the $B$-band LC between days\,$-$15 and $+$60, which indicates $t_{\rm max}(B) = 58831.012\pm0.695$ and $\Delta m_{15}(B)=0.794\pm0.072$\,mag, with the K-correction applied. 
For each parameter derived from SNooPy, the first and second uncertainties reflect the statistical and systematic errors, respectively. 
These values are consistent with the analysis carried out in \citet{2022ApJ...938...83Y}. In Fig.~\ref{fig:lc}, vertical dashed lines indicate the epochs of VLT spectropolarimetric observations. 

\begin{figure}[ht]
    \centering
    \captionsetup{name= Fig.}
    \includegraphics[trim={0.0cm 0.0cm 0.0cm 0.0cm},clip,width=1.0\textwidth]{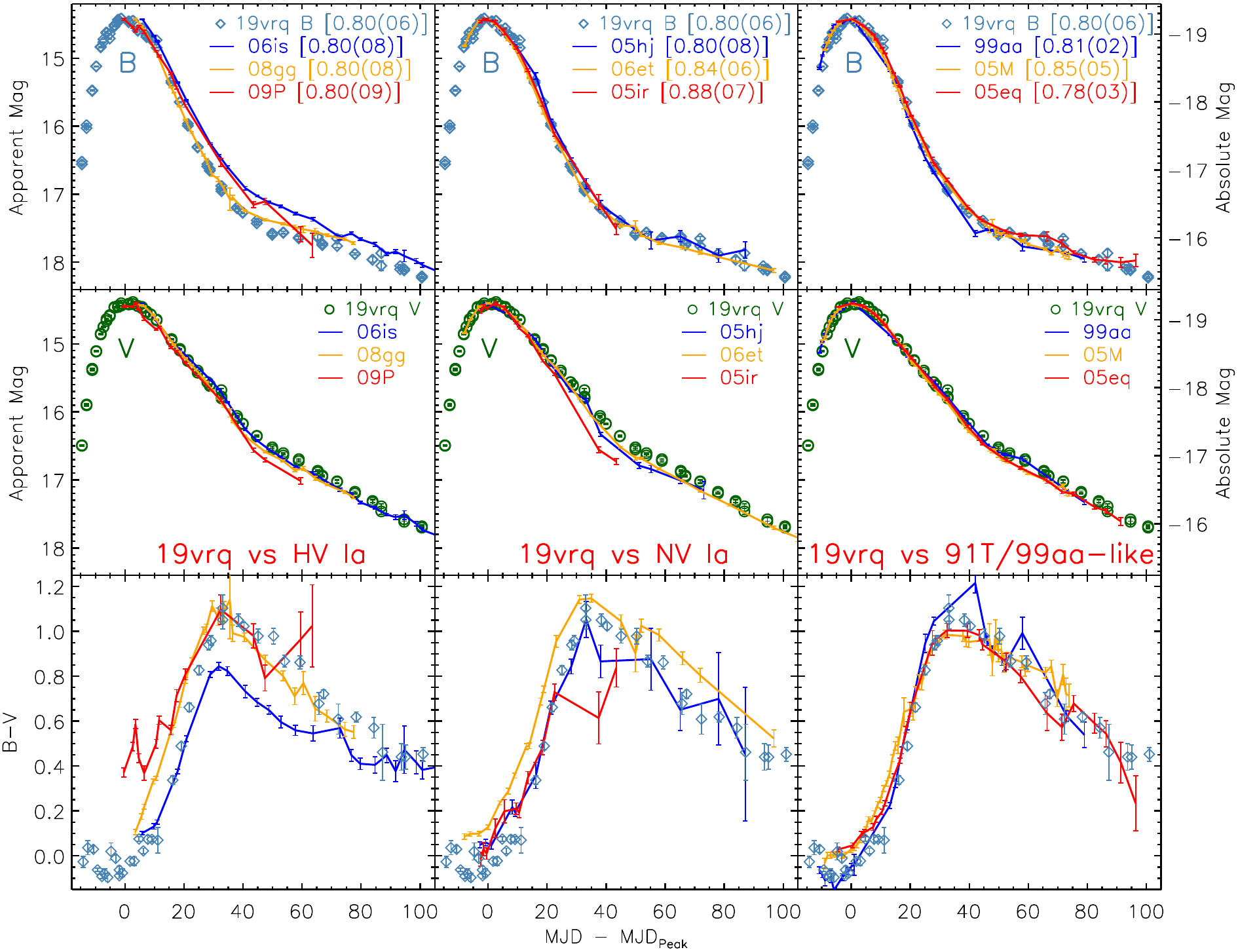}
    \vspace{-0.2 cm}
    \caption{The $B$ (top) and $V$ (middle) LCs and $B-V$ color curve (bottom row) of SN\,2019vrq compared to those of high-velocity (HV, left column), normal-velocity (NV, middle column), and overluminous 91T/99aa-like SNe\,Ia (right column). The LCs of the comparison SNe have been shifted to align with the $B$- and $V$-band peaks of SN\,2019vrq. 
    All photometry and color measurements are corrected for the host galaxy and Galactic reddening. For each SN, the legend indicates its $B$-band LC decline rate $\Delta m_{15}(B)$.
    The LCs and color evolution of SN\,2019vrq are very similar to those of 91T/99aa-like SNe. 
    The $B-V$ color evolution of SN\,2019vrq appears to be bluer compared to those of HV and NV SNe\,Ia before $B$ maximum light.}~\label{Fig_lc_compare} 
\end{figure}

Fig.~\ref{Fig_lc_compare} compares the $B$- and $V$-band LCs (upper row) and the $B-V$ color curve of SN\,2019vrq with that of selected SNe\,Ia of different subtypes that show similar $\Delta m_{15}(B)$. 
The left, middle, and right columns present the HV SNe\,2006is, 2008gg, 2009P, normal-velocity (NV) SNe\,2005hj, 2006et, 2005ir, and overluminous 91T/99a-like SNe\,2009aa, 2005M, 2005eq, respectively. 
All comparison LCs were selected from the Carnegie Supernova Project (CSP) Data Releases 2~\citep{2011AJ....142..156S} and 3~\citep{2017AJ....154..211K}. All photometry has been corrected for reddening in the Milky Way, and host-galaxy reddening corrections have also been applied where applicable. 

The shapes of the $B$ and $V$ LCs of SN\,2019vrq are not distinguishable from the HV and NV SNe\,Ia that are spectroscopically normal and exhibit similar $\Delta m_{15}(B)$. 
However, the early color evolution of SN\,2019vrq as indicated by the $B-V$ color curves is bluer compared to those of HV and NV SNe\,Ia before and around the $B$-band peak, consistent with other 91T/99aa-like SNe. These events occur at the luminous end of thermonuclear explosions and produce higher ejecta temperatures that delay the recombination of IGEs in the ejecta (e.g., Fe\,{\sc iii} to Fe\,{\sc ii}). 
In contrast, recombination is faster in cooler ejecta, leading to a rapid development of numerous Fe\,{\sc ii} and Co\,{\sc ii} lines, thus shifting the spectral energy distribution (SED) to redder wavelengths~\citep{2007ApJ...656..661K}.

\subsection{Pseudobolometric Luminosity}~\label{sec:bolo_construct}
To compare the bolometric properties of SN\,2019vrq with those of the overluminous 91T/99aa subclass, we computed its pseudobolometric LC following a methodology similar to that detailed in Appendix C of \citet{2020ApJ...902...46Y}. We outline the key steps and highlight several important considerations below. 

First, we queried the NASA/IPAC NED Galactic Extinction Calculator and obtained a Galactic reddening value of $E(B-V)^{\rm MW}_{\rm 19vrq} = 0.0396\pm0.006$ mag toward the SN\,2019vrq line of sight, based on the extinction map from \citet{2011ApJ...737..103S}. We also adopt a host-galaxy reddening of $E(B-V)^{\rm Host}_{\rm 19vrq} = 0.02\pm0.02$ mag~\citep{2022ApJ...938...83Y}. Extinction corrections at the pivot wavelengths of the {\it Swift} $uvw2$, $uvw1$, and $UBVg'r'i'$-band photometry were estimated assuming the standard $R_{V} = 3.1$ extinction law~\citep{1989ApJ...345..245C}. 

Second, we registered the spectral template time series for SNe\,Ia~\citep{2007ApJ...663.1187H} to the nearest photometric phase of SN\,2019vrq. By warping each template spectrum to match the LCO $UBVg'r'i'$ photometry obtained at different phases, which we denote as the ``SED-warp'' method, we derived a temporal sequence of the corresponding SED. 
The SED-warp approach effectively captures key spectral features, offering a more accurate representation than the ``SED-dot'' method, which connects mean photon flux densities in photometric bands with line segments~\citep{2016AJ....152..102B, 2020ApJ...902...46Y}. 
These two SED construction techniques provide overall consistent estimates of the bolometric luminosity during the near-peak evolution of SNe\,Ia~\citep{2020ApJ...902...46Y}.

\begin{figure}[ht]
    \centering
    \captionsetup{name= Fig.}
    \includegraphics[trim={0.0cm 0.0cm 0.0cm 0.0cm},clip,width=0.8\textwidth]{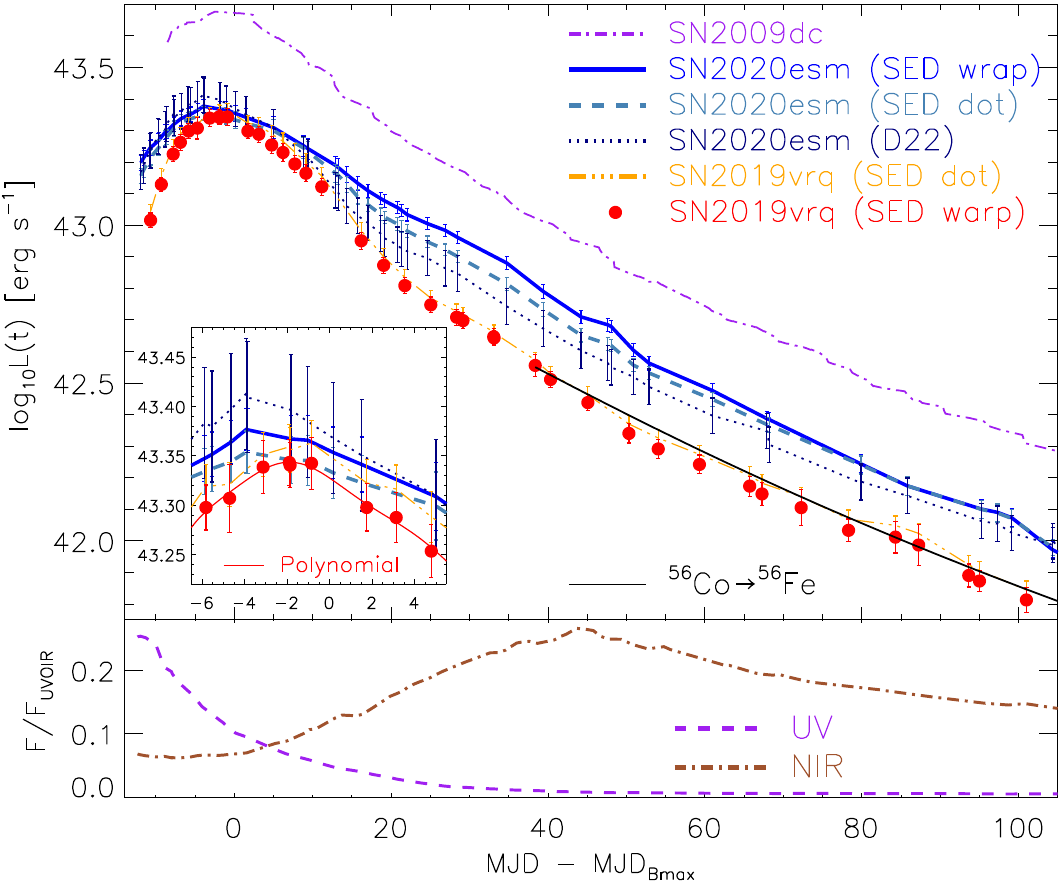}
    \vspace{-0.0 cm}
    \caption{Pseudobolometric LC of SN\,2019vrq calculated by integrating the UVOIR SED between 1600 and 24,000\,\AA, compared to that of the super-Chandrasekhar SN\,2009dc and 2020esm. 
Our results are overall consistent with those reported by \citet{2022ApJ...927...78D} (hereafter D22), although the log\,$L$ derived in this work is systematically higher by $\sim$0.1\,dex than that of D22 after day\,15. 
The lower-left inset zooms in on phases between days\,$-6.5$ and $+5.5$, showing a polynomial fit to the near-peak LC derived using the SED warp method. 
SNe\,2019vrq and 2020esm exhibit similar peak luminosities and similar trends in their bolometric luminosity evolution, except that the latter shows a slower rise and decline, and has been estimated to have a higher $^{56}$Ni mass. The lower panel displays the ratio of the UV (1600--3000\,\AA, purple dashed line) and NIR (10,000--24,000\,\AA, brown dotted-dashed line) fluxes to the total UVOI flux estimated in UVOIR (1600--24,000\,\AA). 
}~\label{Fig_bolo} 
\end{figure}

SN\,2019vrq was observed by {\it Swift} UVOT in the UV between days $-$14 and $-$8. Owing to the limited phase coverage in the UV, we scaled the $uvw2$ and $uvw1$ LCs of SN\,2020esm, a super-Chandrasekhar event with a $^{56}$Ni mass of 1.23$_{-0.14}^{+0.14}$\,M$_{\odot}$~\citep{2022ApJ...927...78D}, to match the four epochs of UV observations of SN\,2019vrq obtained during its rise. 
In the absence of near-infrared (NIR) data for SN\,2019vrq, we estimated the fraction of the NIR flux compared to the total bolometric luminosity ($F_{\rm NIR} / F_{\rm UVOIR}$) of SN\,2020esm, and assume that SN\,2019vrq has a similar temporal evolution in the fraction of the flux in the NIR. 
We computed the pseudobolometric LCs $L(t)$ of SNe\,2020esm and 2019vrq by integrating their SEDs constructed by both the SED-dot and the SED-warp approaches over the UV-optical-NIR (UVOIR) wavelength range (1600--24,000\,\AA). The results are given in Table~\ref{tab:bolo_lc}. 
As presented in the inset of Fig.~\ref{Fig_bolo}, the near-peak bolometric LCs derived from the two methods are in good agreement within the uncertainties. 
As a validation, the results for SN\,2020esm were also compared to those reported by \citet{2022ApJ...927...78D}, revealing overall good consistency ($\lesssim1\sigma$ discrepancy) except for systematically higher luminosities roughly after day 15. 
Fig.~\ref{Fig_bolo} also displays the SED-warp pseudobolometric LC for the super-Chandrasekhar Type Ia SN\,2009dc based on photometry reported by \citet{2011MNRAS.410..585S}, \citet{ 2011MNRAS.412.2735T}, \citet{2014Ap&SS.354...89B}, and \citet{2015ApJS..220....9F}.

\subsection{Bolometric Properties}~\label{sec:bolo_properties}
As the ejecta expand, they gradually become more transparent to $\gamma$-rays, whereas the energy carried by positrons can be treated as fully trapped prior to $\lesssim$120 days after the SN explosion~\citep{2012ApJ...757...12S, 2014MNRAS.440.1498S}. 
Adopting the transparency time $t_{0}$ to be when the $\gamma$-ray optical depth of the ejecta drops to unity~\citep{2014MNRAS.440.1498S}, we fit the bolometric LC of SN\,2019vrq by accounting for the decay energy carried by charged leptons and X-rays. The luminosity contribution from a single decay chain is expressed as 
\begin{equation}\label{Eqn_bolo}
L_A (t_{e}) = 2.221 \frac{1}{A} \frac{\lambda_A}{\mathrm{day^{-1}}} \frac{M(A)}{M_{\odot}}
\frac{q^{\gamma}_{A} f^{\gamma}_{A}(t_{e}) + q^l_{A} + q^X_{A}}{\mathrm{keV}} \mathrm{exp} (-\lambda_A t_e) \times 10^{43} \mathrm{erg^{-1}}, 
\end{equation}
where $A$ denotes the mass number of the decaying isotope. We adopt $A=56$ considering a single $^{56}$Co$\rightarrow ^{56}$Fe decay chain that primarily powers the LCs of SNe\,Ia within the first few months after explosion. Other energy sources that contribute to the luminosity include $q^{\gamma}_{A}$, $q^{l}_{A}$, and $q^{X}_{A}$, which denote the average energies per decay carried by $\gamma$-rays, charged leptons, and X-rays, respectively. The time since the explosion is represented by $t_{e}$. 
The fraction of the $\gamma$-rays contributing to the luminosity is given by $f^{\gamma}_{A}(t_{e}) = 1 - {\rm exp}[-(\frac{t_{0}}{t_{e}})^2]$, where $\lambda_{A}$ is the inverse mean lifetime, namely $\lambda_{A} = \tau_A^{-1} = \mathrm{ln(2)}/t_{1/2, A}$. The term $M(A)$ represents the total mass of a specific decaying element. In the case of our fitting to the bolometric LC of SN\,2019vrq, it denotes the mass of radioactive $^{56}$Co initially synthesized by the thermonuclear explosion of the WD.

We adopt the values of $\lambda_{A}$, $q^{l}_{A}$, and $q^{X}_{A}$ from Table~1 of \citet{2009MNRAS.400..531S} and Table~2 of \citet{2014ApJ...792...10S}. By fitting the pseudobolometric LC $\gtrsim$50 days after the explosion with the model powered by a single $^{56}$Co$\rightarrow ^{56}$Fe decay chain, we estimate $M(^{56}{\rm Ni}) = 0.77\pm0.16$\,M$_{\odot}$ and $t_{0}=53.0\pm7.9$\,d for SN\,2019vrq. 
Our estimated parameters also place SN\,2019vrq in the parameter space defined by the mass of $^{56}$Ni and the $\gamma$-ray transparency timescale of the ejecta close to other 91T/99aa-like events \citep{2014ApJ...792...10S}.

By fitting the bolometric luminosity of SN\,2019vrq before $t=+15$\,days relative to  $B$-band maximum light with a high-order polynomial function, we find that the bolometric LC peaked around $t_{\rm rise}^{\rm bol} = t_{\rm rise} (B) -1.9\pm2.4$\,d at log\,$L$ = $43.342\pm0.024$ erg s$^{-1}$. 
The uncertainty was estimated by adding, in quadrature, the statistical error in the fitting procedure and a $\sim3.5\%$ systematic difference in $L(t)$ between the SED-warp and SED-dot procedures.

The difference between the dates of bolometric and $B$-band maximum light of SN\,2019vrq is quite consistent with the distribution of 19 well-studied SNe from the Nearby Supernova Factory~\citep{2014ApJ...792...10S}. 
Adopting the relationship between the rise time measured in the $B$ band and that of the bolometric LCs~\citep{2011MNRAS.416.2607G, 2014MNRAS.440.1498S}, 
\begin{align}
    t_{\rm rise}(B) = 17.5 - 5[\Delta m_{15}(B) - 1.1]\,{\rm d},
\end{align}~\label{Eqn_trise}
and taking $t_{\rm rise}(B)=19.03\pm0.36$\,days for SN\,2019vrq, we infer the rise time of its bolometric LC to be $t_{\rm rise}^{\rm bol} = 17.13\pm2.43$\,days.

Radiation from the transformation of radioactive $^{56}$Co to stable $^{56}$Fe is produced via electron capture (81\%) or positron decay (19\%). This fact enables an independent estimate of the total mass of radioactive $^{56}$Ni initially synthesized in the SN ejecta. 
Following the semi-analytic description of \citet{1994ApJS...92..527N}, the peak bolometric luminosity powered by radioactive $^{56}$Ni can be written as 
\begin{equation}
L_{\rm max} = \bigg{[}6.45 \times {\rm exp}(-{t_{\rm rise}^{\rm bol}/{8.8 \rm \ day})} + 1.45 \times {\rm exp(}-{t_{\rm rise}^{\rm bol}{/111.3 \rm \ day}}) \bigg{]} \times 
\bigg{(} \frac{M_{\rm Ni}}{M_{\odot}} \bigg{)} \times 10^{43} {\rm erg \ s^{-1}}. 
\label{Eqn_ni_mass}
\end{equation}
We estimate the mass of $^{56}$Ni initially synthesized in the SN ejecta to be $M(^{56}{\rm Ni})=1.02_{-0.14}^{+0.14}$\,M$_{\odot}$, consistent within 1$\sigma$ with the value of 0.77$\pm$0.16\,M$_{\odot}$ derived from the late-time bolometric LC. 
Detailed modeling of the monochromatic and bolometric LCs of SN\,2019vrq is carried out in Paper~II, where a PDD scenario is shown to provide a plausible fit to its slower rise and decline relative to normal SNe\,Ia.

\subsection{Spectroscopic Properties}~\label{sec:spec}
\begin{figure}[ht]
    \centering
    \captionsetup{name= Fig.}
    \includegraphics[trim={0.0cm 0.0cm 0.0cm 0.0cm},clip,width=1.0\textwidth]{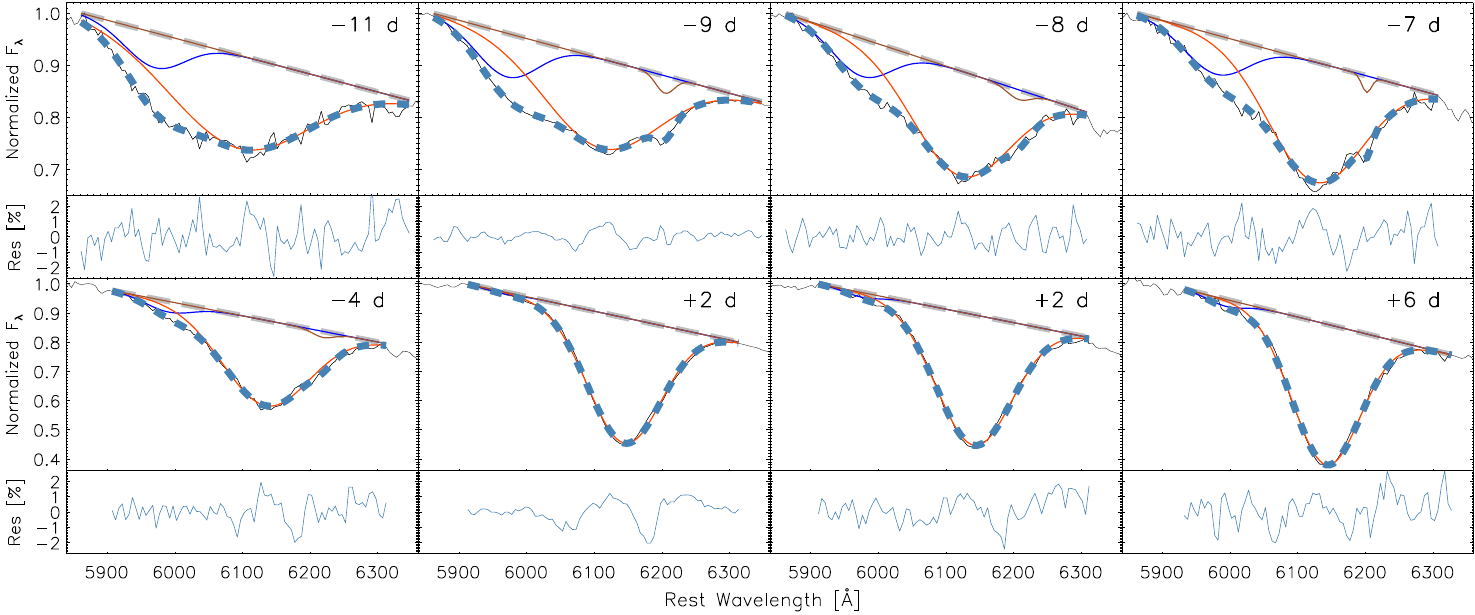}
    \vspace{-0.2 cm}
    \caption{Investigating the 5900--6400\,\AA\ absorption profile of Si\,{\sc ii}\,$\lambda$6355 in SN\,2019vrq through a multicomponent Gaussian fitting approach. The analysis was carried out over the rest-frame wavelength range $\sim$5950--6350\,\AA. In each panel, the pseudocontinuum is marked by the long-dashed line segment. The blue and orange curves show the HV and NV components, respectively. 
    The purple dotted lines in the upper and the lower subpanels display the fit result and the residual, the latter multiplied by a factor of 100. A weak absorption feature at a velocity similar to that of the NV component of Si\,{\sc ii}\,$\lambda$6355 is also tentatively detected from days $-$9 to $-$3, which is most likely due to blending with the Fe\,{\sc ii}\,$\lambda\lambda$6078,\,6293,\,6358,\,6456,\,6516 transitions (see Paper~II). 
    The fits shown in the top row suggest that the HV component, if present, may persist until the SN reaches its $B$-band maximum. 
    This time evolution can be reproduced by a layered ionization structure, as supported by both simulations and by the spectral time series (see Paper~II). The HV component vanishes by maximum light as Si\,{\sc iii} recombines. The presence of at least three broad features can be attributed to blending, predominantly from IGE lines. 
In both the outer and photospheric layers, Si is mostly in the {\sc ii} ionization stage, with a {\sc iii} ionization stage present in between. 
The early profiles are blended with Fe\,{\sc iii} (blue) and Si\,{\sc iii} (magenta, see table) at early times, and with Fe/Co\,{\sc ii} at later epochs; note the possible implications for the interpretation of the $Q-U$ diagram (see Fig.~\ref{Fig_quall}). Note also the temporary blueshift, which may be understood as the dominant line transitioning from Si to Fe/Co between days\,$-$7 and $+$5 (see Paper~II).
}~\label{Fig_fitgauss} 
\end{figure}

Model-independent first-order interpretations are given using the general assumption that the absorption component of the overlapping P~Cygni profiles can be characterized by Gaussians, with the minimum tracing the expansion velocity of the ejecta component associated with that line opacity. 
Such an approach based on multiple Gaussian profiles is commonly adopted to disentangle blended components from the same ion, which we also carried out to investigate the Si\,{\sc ii}/Fe{\sc ii} profile of SN\,2019vrq (Fig.~\ref{Fig_fitgauss}). 
However, the shape of an absorption feature is set by the spatial distribution of the associated line opacity, which, in principle, is intrinsically non-Gaussian~\citep{2018MNRAS.476.1299M, 2019MNRAS.484.4785M}. Caution is therefore advised when inferring line properties from multi-Gaussian fits to such profiles. 

Moreover, we note that the Doppler shifts of absorption minima are not reliable tracers of the true expansion velocity at early times, since most features forming close to the photosphere reflect only the projected velocity along the line of sight. 
For instance, Doppler-shift velocities can underestimate the true expansion velocity by up to a factor of 2, particularly when the photospheric radius is comparable to the extent of the line-forming region (see, e.g., \citealp{2023MNRAS.520..560H}, and references therein). 
The similarity of the Doppler shifts among different elements ($\sun$10,000\,km\,s$^{-1}$ at day\,$-$10) results from the extended photosphere and overlapping line-forming regions. Consequently, a low measured Doppler velocity does not necessarily correspond to a low bulk expansion speed (see Section~\ref{sec:spec}). 

\begin{figure}[ht]
    \centering
    \captionsetup{name= Fig.}
    \includegraphics[trim={0.0cm 0.0cm 0.0cm 0.0cm},clip,width=1.0\textwidth]{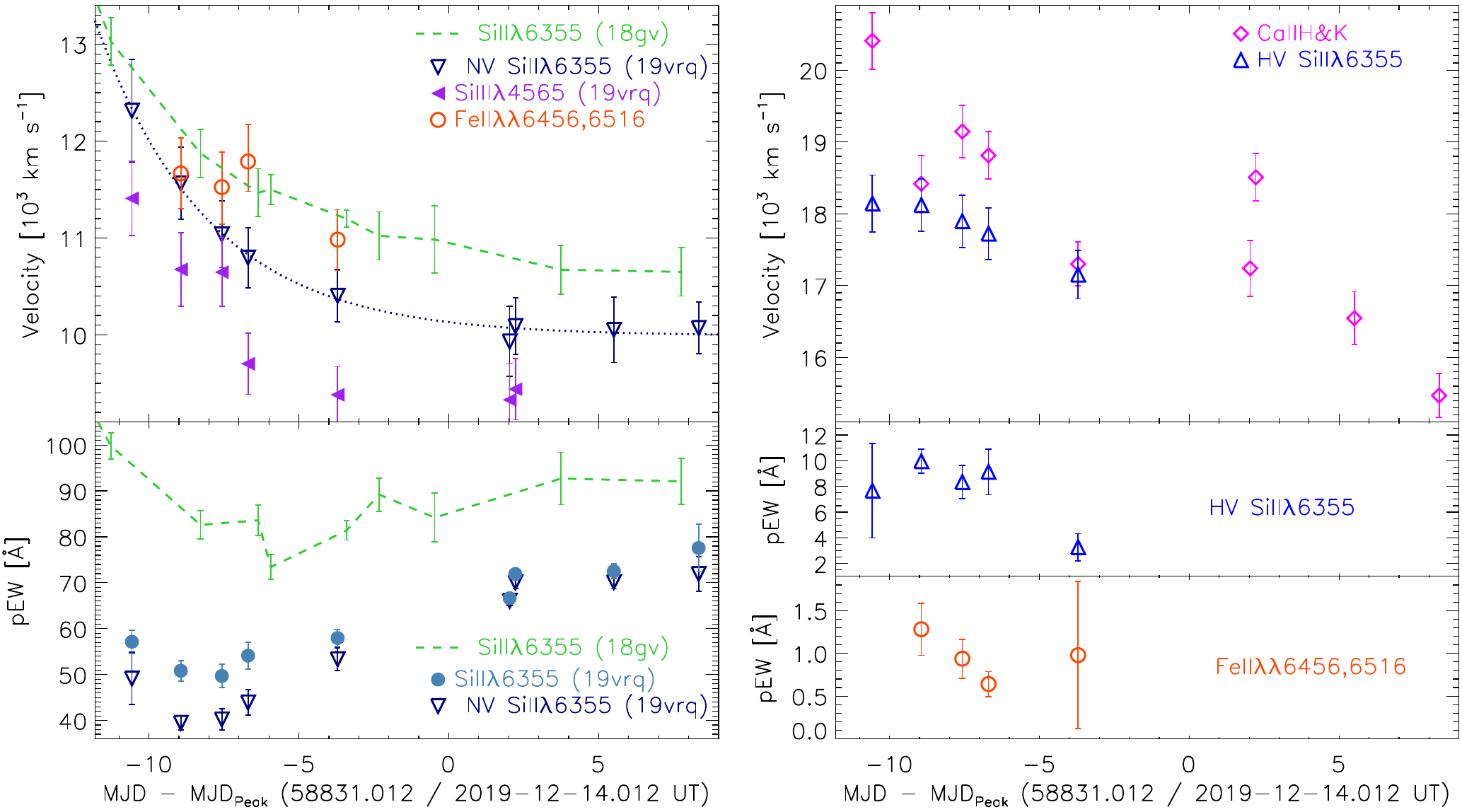}
    \vspace{-0.0 cm}
    \caption{Time evolution of the Doppler velocities relative to the rest wavelengths of lines of SN\,2019vrq as traced by various spectral features. 
    The upper-left panel shows the velocity measured from the NV component of the Si\,{\sc ii}\,$\lambda$6355 and the Si\,{\sc ii}\,$\lambda$4565 line of SN\,2019vrq. We also present the values inferred for the Fe\,{\sc ii}\,$\lambda\lambda$6456,\,6516 line complex through a multi-Gaussian fitting procedure. 
The navy dashed curve shows the exponential fit to the photospheric velocity of SN\,2019vrq represented by the absorption minimum of Si\,{\sc ii}\,$\lambda$6355. The time evolution of the corresponding pEWs of the entire Si\,{\sc ii}\,$\lambda$6355 profile and its NV component is shown in the lower-left panel. The dashed-green lines in the left subpanels present the measurements of the normal Type Ia SN\,2018gv~\citep{2020ApJ...902...46Y}. 
The right panel displays the velocities of the Ca\,{\sc ii}\,H\&K and the HV component of Si\,{\sc ii}\,$\lambda$6355. Fitted pseudo-equivalent-width (pEW) values of the HV Si\,{\sc ii}\,$\lambda$6355 and Fe\,{\sc ii}\,$\lambda\lambda$6456,6516 are also shown. 
}~\label{Fig_velo} 
\end{figure}

The upper panels of Fig.~\ref{Fig_velo} present the photospheric velocities, as indicated by the absorption minima of several blueshifted lines including Si\,{\sc ii}\,$\lambda$6355, Si\,{\sc\,}iii\,$\lambda$4565, and Ca\,{\sc ii}\,H\&K. 
In fact, the absorption feature near Si\,{\sc ii}\,$\lambda$6355 suffers from blending with numerous Fe{\sc ii} lines with rest wavelengths of 6291--6456\,\AA\ (see, e.g., Fig.~\ref{fig:pol_spec} and Paper~II for more details). The strength of these Fe{\sc ii} features increases over time, and they become the dominant contributor by day\,$+$2. 
Therefore, the multiple Gaussian fitting process only provides a rough estimate of the location of the absorption minima. Nevertheless, it facilitates comparison with the same observables reported in the literature. One should be aware that this treatment affects how these observables map onto the parameter space describing the formation of the polarization signal and the geometrical configuration of the ejecta.

Unlike the Si\,{\sc ii}\,$\lambda$6355 velocity inferred from the absorption minimum, which decreased moderately from $\sim$12,400\,km\,s$^{-1}$ at day\,$-$10 to 10,100\,km\,s$^{-1}$ at day\,$-$9 (see Fig.~\ref{Fig_velo}), the Ca\,{\sc ii}\,H\&K line exhibits an overall significantly higher velocity, spanning a broad range from $\sim$21,300\,km\,s$^{-1}$ to $\sim$15,500\,km\,s$^{-1}$. 
Note that the Doppler shift is a lower limit on the expansion velocity at a given time and is measured relative to a fixed rest wavelength. These curves only provide an approximate tracer of the radial distribution of the associated line opacities and their relative locations within the SN ejecta. 
Additionally, the dominant axis fitted across the Si\,{\sc ii}\,$\lambda$6355 line would suffer from contamination of Fe\,{\sc ii} lines, an effect that cannot be captured by a multi-Gaussian fitting approach.

\begin{figure}[ht]
    \centering
    \includegraphics[trim={0.0cm 0.0cm 0.0cm 0.0cm},clip,width=1.0\textwidth]{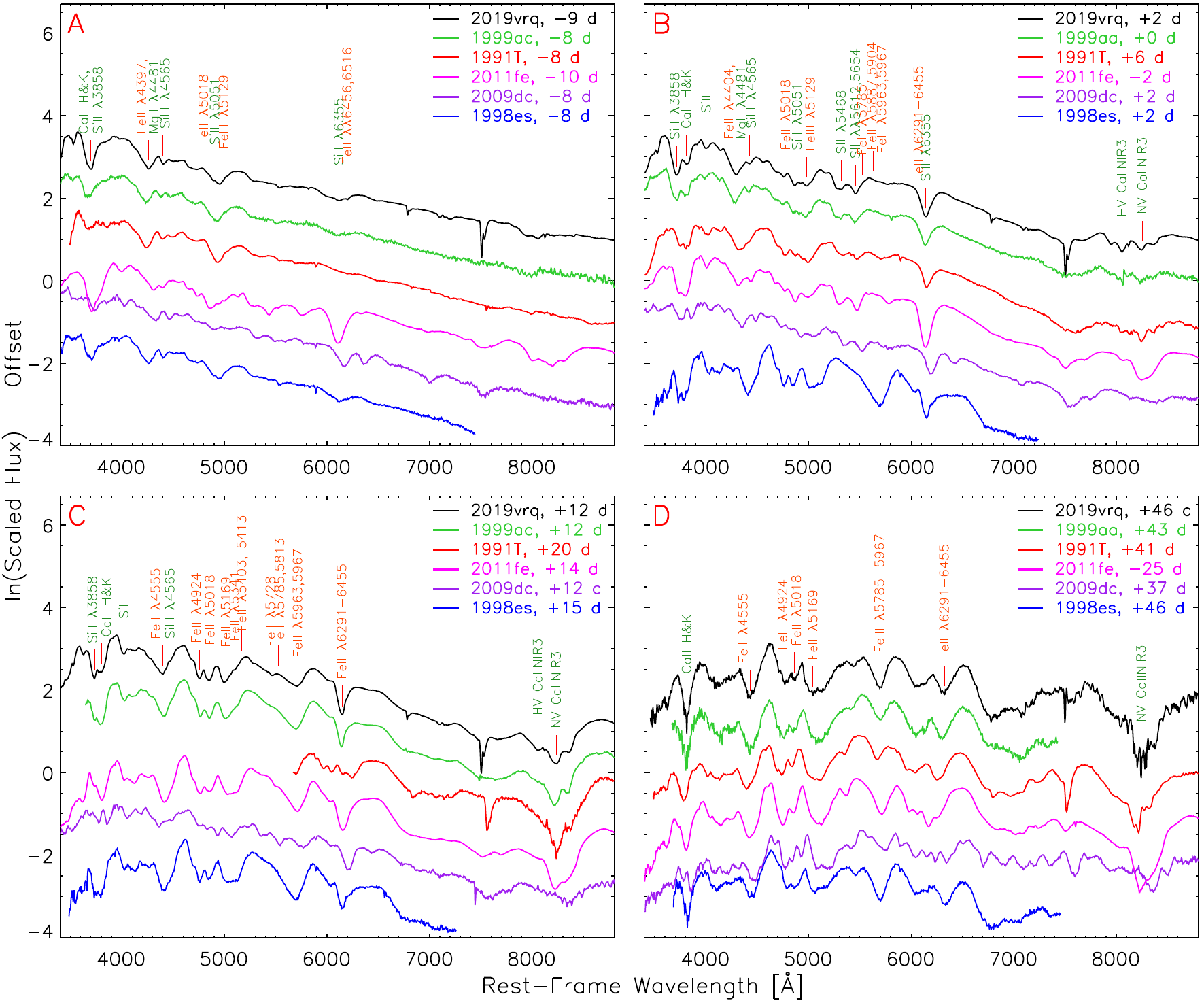}
   \caption{Spectra of SN\,2019vrq compared to other SNe\,Ia at similar phases. Panels A--D present the comparisons for days\,$-$9 (upper left), $+$2 (upper right), $+$12 (lower left), and $+$46 (lower right) spectra of SN\,2019vrq to those of SNe\,1999aa, 1991T, 2011fe, 2009dc, and 1998es at similar phases, respectively.}~\label{Fig_spec_compare} 
\end{figure}

Figs.~\ref{Fig_spec_compare}A--D compare the spectra of SN\,2019vrq at days $-$9, $+$0, $+$12, and $+46$ (respectively) with those of SN\,1999aa~\citep{2013MNRAS.430.1030S, 2008AJ....135.1598M}, the 99aa-like SN\,1998es~\citep{2008AJ....135.1598M, 2012MNRAS.425.1789S}, SN\,1991T~\citep{1992ApJ...384L..15F}, the 91T-like SN\,2009dc~\citep{2011MNRAS.412.2735T}, and the spectroscopically normal SN\,2011fe~\citep{2013A&A...554A..27P} at similar phases. Major spectral features are also identified, as guided by the detailed simulations (see Paper II). 

In Fig.~\ref{Fig_spec_compare}A we show that the $-$9\,d spectrum of SN\,2019vrq is dominated by a blue continuum with superimposed absorption features. 
These can be attributed to a blend of Ca\,{\sc ii}\,H\&K (rest-frame wavelengths $\lambda_{0}=$3969, 3934\,\AA) and Si\,{\sc ii}\,$\lambda$3858 lines, the Si\,{\sc ii}\,$\lambda$6355 feature, and the Si\,{\sc\, iii}\,$\lambda \lambda \lambda\,$4553,\,4568,\,4575 multiplet (hereafter Si\,{\,iii}\,$\lambda$4565), together with several prominent lines from Fe\,{\sc ii}/Fe\,{\sc iii}, Mg\,{\sc ii}, and other IMEs (see Paper~II). 
In particular, the pre-peak spectra of SN\,2019vrq and the comparison 91T/99aa-like SNe display significantly shallower IME features than those of SN\,2011fe, suggesting a substantial deficit of IMEs in the outer ejecta layers of 91T/99aa-like SNe\,Ia. 
These pre-peak spectra are dominated by doubly ionized species, particularly Si\,{\sc iii}, which may indicate high temperatures in their outer ejecta. 
A corresponding high ionization is expected, which is consistent with the nondetection of neutral or singly ionized C and O at day\,$-$9, in contrast to the prominent C/O lines seen in the cooler SN\,2020esm at a similar phase~\citep{2022ApJ...927...78D}. 

The IGE lines of SN\,2019vrq also show little evolution between days\,$-$11 and $-$7 (Fig.~\ref{Fig_spec}). In particular, the profile at $\sim$4500\,\AA, which is formed primarily by the blanketing of iron lines, remains nearly unchanged over this interval. As we discuss in more detail in Paper~II, the persistence of these slowly evolving IGE lines implies chemically homogeneous outer layers of $\sim$0.1\, M$_{\odot}$ with a primordial, twice-solar-abundance pattern. This spectroscopic signature is compatible with the spectral evolution expected in the 
PDD model, whereas a ``classical'' DD scenario is not favored. 

Fig.~\ref{Fig_spec_compare}B shows that as SN\,2019vrq reaches peak brightness, the IME features also become more prominent. 
However, they remain weaker than in other spectroscopically normal events, such as SN\,2018gv~\citep{2020ApJ...902...46Y}. 
We also investigated the shape and temporal evolution of the Si\,{\sc ii}\,$\lambda$6355 line by fitting the continuum-subtracted profile with multiple Gaussian functions. As shown in Fig.~\ref{Fig_fitgauss}, a shallow HV component can be tentatively detected in the pre-peak spectra of SN\,2019vrq.

In the following, we discuss the spectra in more detail. First, as presented in the upper-left panel of Fig.~\ref{Fig_velo}, the velocity evolutions of Si\,{\sc iii}\,$\lambda$4565 and Si\,{\sc ii}\,$\lambda$6355 display marked differences.  
The former shows an overall lower velocity and a faster decline than the latter. Given their different excitation potentials, namely 19\,eV for Si\,{\sc iii}\,$\lambda$4565 versus 8\,eV for Si\,{\sc ii}\,$\lambda$6355, such a discrepancy is consistent with non-local-thermodynamic-equilibrium (NLTE) effects, similar to the implications of the Si\,{\sc iii}\,$\lambda$4565 velocity evolution in other 91T/99aa-like events~\citep{2024ApJS..273...16P}. 
Second, the upper-right panel of Fig.~\ref{Fig_velo} shows that the HV Si\,{\sc ii}\,$\lambda$6355 component exhibits a similar but slightly lower velocity than the Ca\,{\sc ii}\,H\&K. 
As the SN evolves past the LC peak, the HV component of Si\,{\sc ii}\,$\lambda$6355 becomes indiscernible (Fig.~\ref{Fig_fitgauss}). Starting a few weeks after $B$-band maximum light, as illustrated in Fig.~\ref{Fig_spec_compare}C--D, the spectral evolution of SN\,2019vrq and other 91T/99aa-like SNe becomes rather similar to that of SN\,2011fe.

\subsection{SN~2019vrq as an Overluminous SN~Ia}~\label{sec:obs_overluminous}
Table~\ref{Table_sn} summarizes the basic properties of SN\,2019vrq. Throughout the paper, all phases are given relative to $B$ maximum light at MJD $58831.012\pm1.075$ (Section~\ref{sec:lc}). 
After correcting for Galactic and host extinction, we obtain $\Delta m_{15}(B) = 0.794\pm0.072$\,mag and a $B$-band peak absolute magnitude of $M_{\rm max}(B)=-19.44\pm 0.08$\,mag. 
Together with a clear secondary peak in the $i$-band LC (Fig.~\ref{fig:lc}) and a nickel mass of $M(^{56}$Ni) $\approx 0.8$--1.0\,$M_{\odot}$ estimated from the UVOIR bolometric LC (Sec.~\ref{sec:bolo_construct} and~\ref{sec:bolo_properties}), these photometric properties place SN\,2019vrq in the regime of overluminous SNe\,Ia. Its spectroscopic evolution also shows a strong resemblance to that of 99aa-like events (Sec.~\ref{sec:spec}, \citealp{2017hsn..book..317T}).

The early-time spectra of SN\,2019vrq are characterized by a blue continuum with broad absorption lines from IMEs and IGEs. Key features include lines from Ca\,{\sc ii} and Si\,{\sc ii}, while the blue part of the spectrum is dominated by blanketing from numerous IGE lines. As shown in Fig.~\ref{fig:pol_spec}A, the wiggles around 4500\,\AA\ are most likely produced by several Co\,{\sc ii} lines. The presence of these IGE features at this early phase of day\,$-$8.9 (hereafter day\,$-9$) indicates prompt outward mixing of radioactive $^{56}$Ni synthesized in the explosion. 
Compared to 91T-like SNe, SN\,2019vrq also shows clearer IME lines, such as Ca\,{\sc ii} and Si\,{\sc ii}, making it resemble the overluminous 99aa-like subclass more closely, though still with a deficit of IMEs in the outer ejecta relative to normal events. 

The early spectra of SN\,2019vrq also show Fe\,{\sc iii} and Si\,{\sc iii}\,$\lambda$4565 (see line identifications in Fig.~\ref{fig:pol_spec}, indicating high temperatures in the ejecta. This high energy suppresses the formation of singly ionized species that are common in normal SNe\,Ia~\citep{1995A&A...297..509M, 2024ApJS..273...16P}. We identify no signs of neutral or singly ionized carbon and oxygen. This characteristic, together with the overluminous nature of SN\,2019vrq, is compatible with a substantial amount of $^{56}$Ni synthesized in the ejecta. 
The near- and post-peak spectral evolution of SN\,2019vrq becomes similar to that of normal SNe\,Ia. A detailed analysis of the spectroscopic properties of SN\,2019vrq with interpretations based on simulations will be presented in Paper~II. 

\begin{table}[ht]
\begin{center}
\normalsize	
\caption{Basic Properties of SN\,2019vrq independent of explosion models.}~\label{Table_sn}
\begin{tabular}{c|c}
\hline
\hline
Parameter           &  Value \\
\hline
  $B_{\rm max}$     &  14.46$\pm$0.06\,mag \\
  $V_{\rm max}$     &  14.40$\pm$0.03\,mag \\
$z_{\rm host}$      &  $z=$0.013079 \\ 
$E(B-V)$            &  0.060$\pm$0.02\,mag \\
$A_{V}$             &  0.19$\pm$0.08\,mag \\
$\mu_{V}=m-M-A_{V}$ & $+$33.65$\pm$0.09\,mag \\
$\Delta m_{15}(B)$  &  0.794$\pm$0.072\,mag \\
$M_{\rm peak}(B)$   &  $-19.44 \pm 0.08$\,mag \\
$t_{B\rm max}$      &  MJD\,58831.012$\pm$1.075 \\
$t_{\rm rise}(B)$   &  19.03$\pm$0.36\,day \\
$t_{\rm rise}^{\rm bol}$  &  17.13$\pm$2.43\,day \\
log$L_{\rm peak}^{\rm bol}$    &  43.342$\pm$0.024\,erg\,s$^{-1}$ \\
$M(^{56}{\rm Ni})$ (from peak/tail)  &  $1.02_{-0.14}^{+0.14}$ / 0.77$_{-0.16}^{+0.16}$\,M$_{\odot}$ \\
 \hline
\end{tabular}~\label{tab:parameters}
\end{center}
\end{table}

\section{Polarization Properties}~\label{sec:pol}
\subsection{Interstellar Polarization}~\label{sec:isp}
Photons traversing interstellar matter (ISM) along the SN-Earth line of sight become partially polarized through dichroic extinction by nonspherical, paramagnetic dust grains aligned by interstellar magnetic fields. This introduces a baseline interstellar polarization (ISP) that is superimposed on the intrinsic polarization, which can be present in both the MW and the SN host galaxy.  
Estimating the ISP is challenging, since it requires an unpolarized source that samples the same SN-Earth line of sight. Given the extragalactic nature of the SN, in most cases the only available beacon is the SN itself. Various methods have been proposed to identify the wavelength ranges and evolution stages of different types of SNe that can be treated as intrinsically unpolarized (see, e.g., ~\citealp{2017MNRAS.469.1897S, 2020MNRAS.494..885S}, and \citealp{2023MNRAS.519.1618Y}).

\begin{figure}[ht]
    \centering
    \captionsetup{name= Fig.}
    \includegraphics[trim={0.0cm 0.0cm 0.0cm 0.0cm},clip,width=0.6\textwidth]{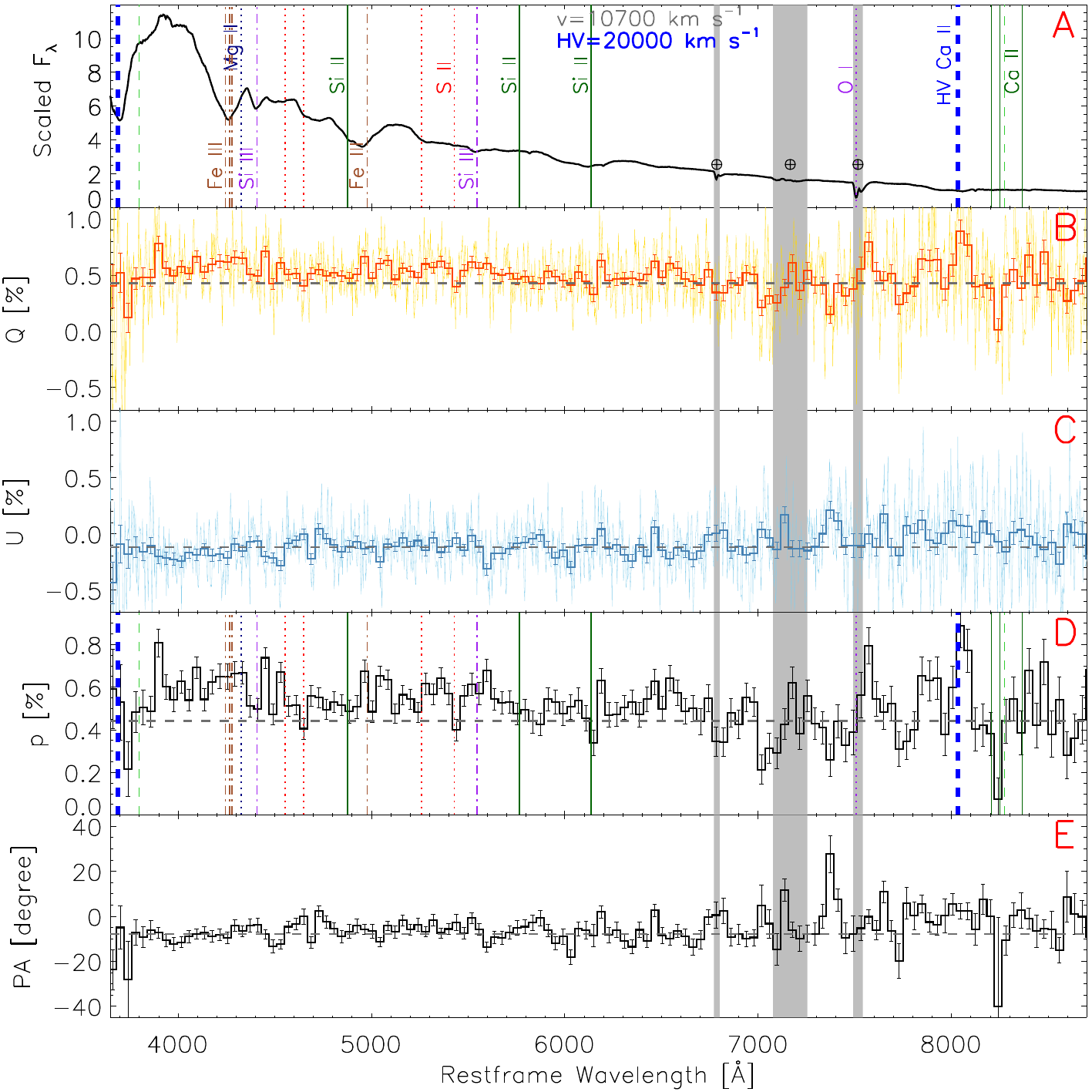}
    \vspace{-0.0 cm}
    \caption{Spectropolarimetry of SN\,2019vrq on day $-$8.9 (Epoch 1). From top to bottom, the panels display (A) the arbitrarily scaled total-flux spectrum; (B, C) the intensity-normalized Stokes $Q$ and $U$ parameters, respectively; (D) the polarization degree ($p$); and (E) the polarization position angle (PA). No ISP subtraction has been applied. All polarization data have been rebinned to 40\,\AA\ for display purposes.}~\label{Fig_iqu_ep1} 
\end{figure}
\begin{figure}[ht]
   \begin{minipage}[t]{0.49\textwidth}
     \centering
     \includegraphics[width=1.0\linewidth]{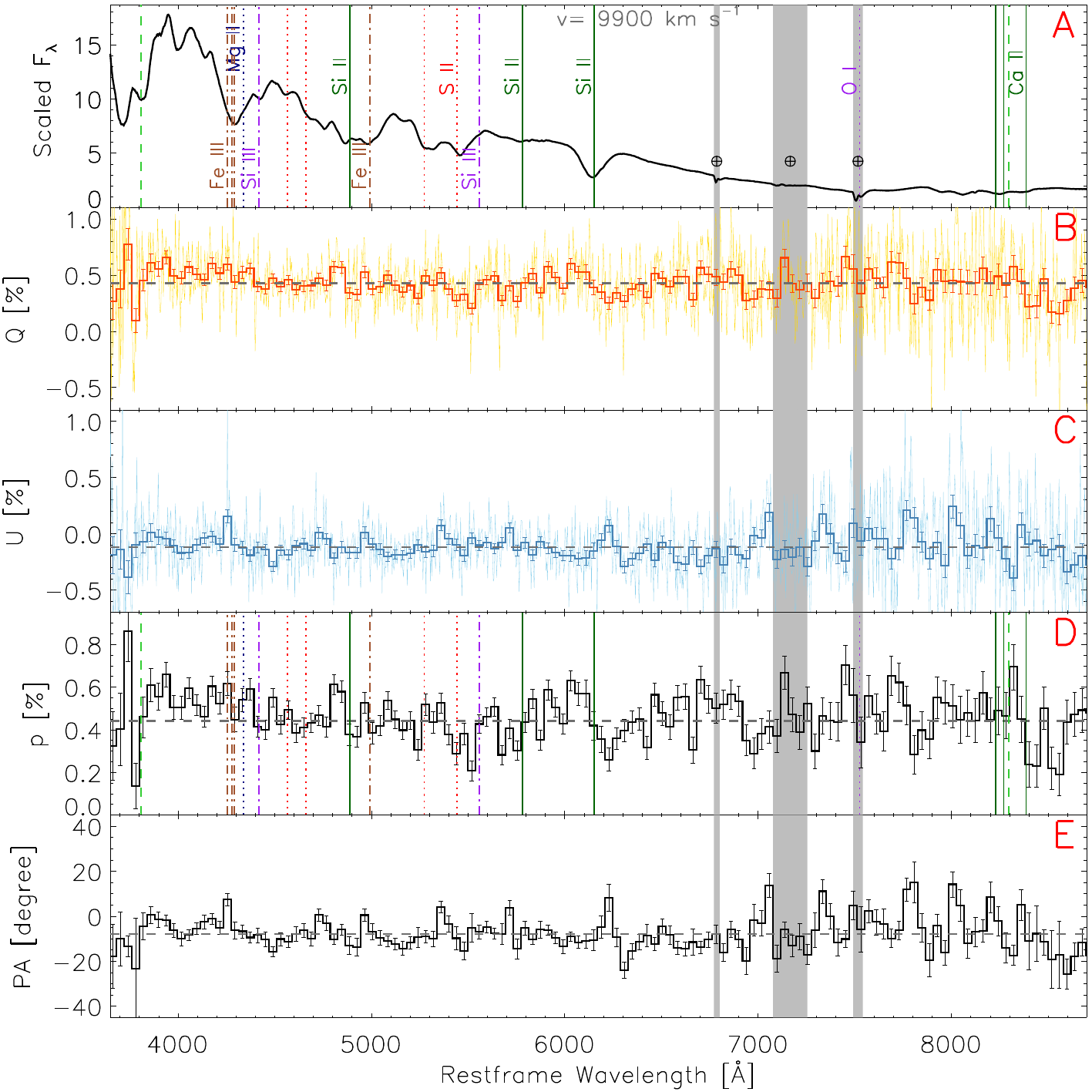}
     \vspace{-0.0 cm}
     \captionsetup{name= Fig.}
\caption{Same as Fig.~\ref{Fig_iqu_ep1} but for day $+$2.0 (Epoch 2).}~\label{Fig_iqu_ep2}
   \end{minipage}\hfill
   \begin{minipage}[t]{0.49\textwidth}
     \centering
     \includegraphics[width=1.0\linewidth]{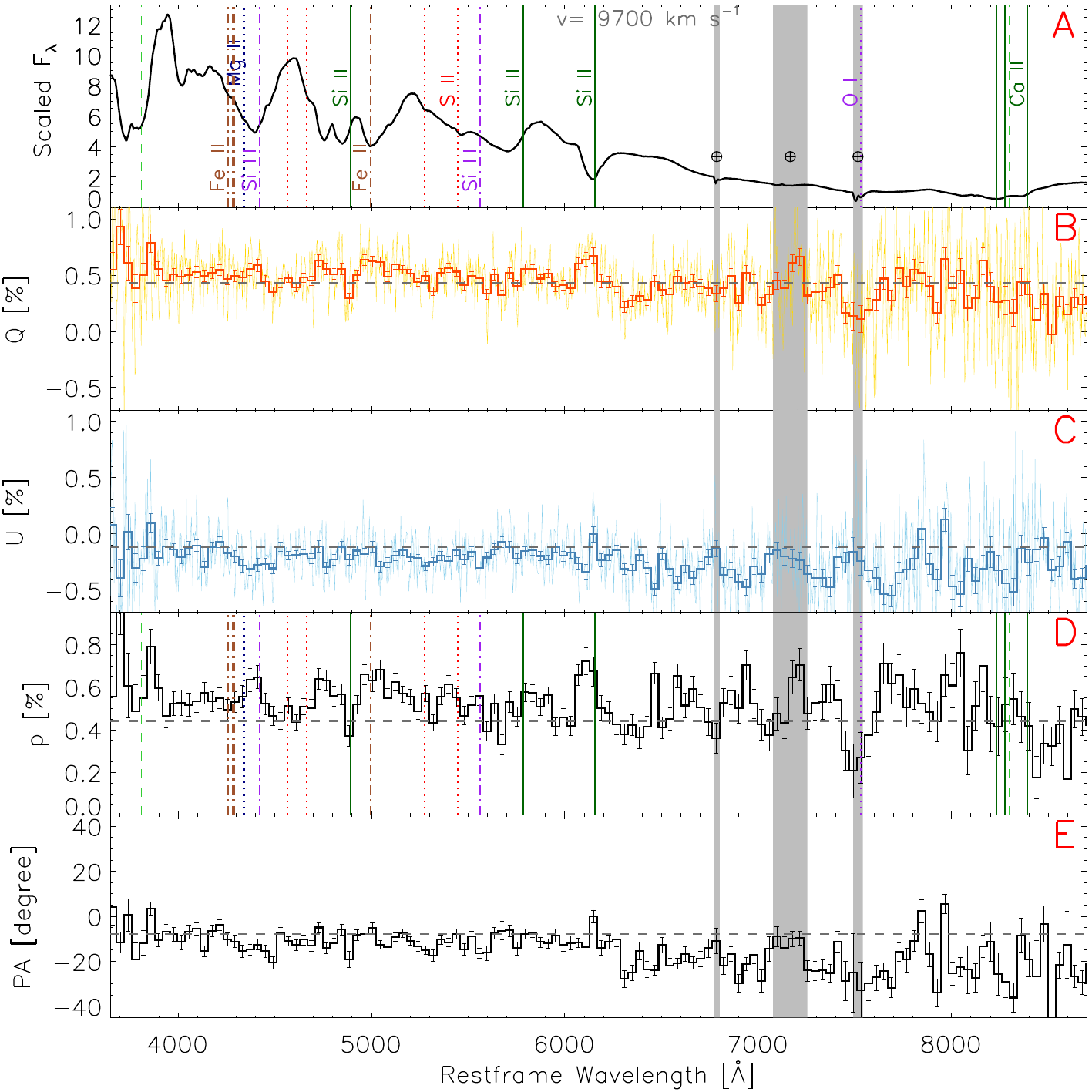}
     \vspace{-0.0 cm}
     \captionsetup{name= Fig.}
     \caption{Same as Fig.~\ref{Fig_iqu_ep1} but for day $+$12.0 (Epoch 3).}~\label{Fig_iqu_ep3}
   \end{minipage}
\end{figure}

For SN\,2019vrq, we estimate the ISP by calculating the error-weighted mean polarization within the wavelength range 4800--5600\,\AA\ at $+2$\,d. It has been shown that near the peak luminosity of a SN\,Ia, this part of the spectrum is substantially depolarized owing to blanketing by numerous IGE absorption features~\citep{2001ApJ...556..302H, 2006PASP..118..722C, 2009A&A...508..229P, 2010ApJ...722.1162M, 2013MNRAS.433L..20M}. 
This yields $Q_{\rm ISP}=0.426\pm0.099$\%, $U_{\rm ISP}=-0.119\pm0.095$\%, and therefore $p_{\rm ISP}=0.45\pm0.10$\%, consistent with the ISP estimated based on the imaging polarimetry of SN\,2019vrq obtained at day\,$+$70.0 nebular phases~\citep{2022MNRAS.509.6028C}, when  electron scattering becomes negligible in the substantially diluted ejecta (e.g., \citealp{2001ApJ...556..302H, 2001ApJ...550.1030W}). 
We also compare our estimated ISP with the upper limit implied by partially aligned dust grains in the Galaxy and in the host of SN\,2019vrq as constrained by the interstellar extinction. Assuming that both components follow an $R_{V}=3.1$ extinction law~\citep{1989ApJ...345..245C}, we adopt $E(B-V)^{\rm MW}_{\rm 19vrq} = 0.0396\pm0.006$ mag and $E(B-V)^{\rm Host}_{\rm 19vrq} = 0.02\pm0.02$\,mag (see \citealp{2022ApJ...938...83Y}, and also Section~\ref{sec:bolo_construct}). 
The upper limit on the ISP~\citep{1975ApJ...196..261S} is then $p_{\rm ISP}\leq9\times E(B-V)^{\rm total}_{\rm 19vrq} = 0.54\pm0.19$\%, consistent with our estimated value. 
Spectropolarimetry of SN\,2019vrq obtained at days\,$-$9, $+$2, and $+$12, together with the associated total-flux spectra, is shown in Figs.~\ref{Fig_iqu_ep1}--\ref{Fig_iqu_ep3}, respectively, with the level of the ISP indicated by horizontal dashed lines in panels B--D.

\subsection{Continuum Polarization}~\label{sec:contpol}
After subtracting the ISP, we estimate the continuum polarization intrinsic to SN\,2019vrq over three wavelength regions: A, 4300--4800\,\AA; B, 5300--5800\,\AA; and C, 6600--7100\,\AA. 
The continuum in Regions A and B is dominated by a plethora of bound-bound transitions from numerous IGEs, which substantially depolarize the emission~\citep{1996ApJ...457..500H}. All three regions are free of spectral features found to be significantly polarized~\citep{2009A&A...508..229P}.

\begin{figure}[ht]
    \centering
    \captionsetup{name= Fig.}
    \includegraphics[trim={0.0cm 0.0cm 0.0cm 0.0cm},clip,width=0.9\textwidth]{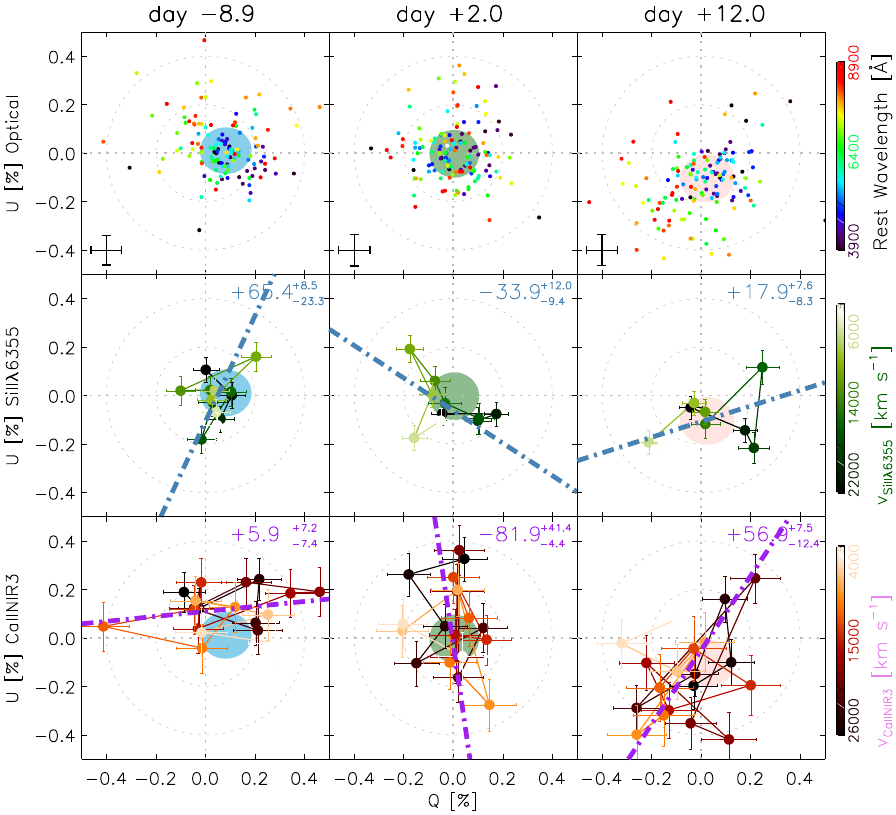}
    \vspace{-0.0 cm}
    \caption{\textbf{Polarization of SN\,2019vrq shown on the Stokes $Q-U$ plane.} 
The top row displays the data over the wavelength range 3900--8900\,\AA. Filled ellipses indicate the error-weighted mean polarization and its 1$\sigma$ standard deviation calculated over this same wavelength range. 
The middle and bottom rows present the polarization measured for the Si\,{\sc ii}\,$\lambda$6355 and the Ca\,{\sc ii}\,NIR3 features over the velocity ranges of 24,000--1000\,km\,s$^{-1}$ and 27,000--1000\,km\,s$^{-1}$, respectively. For each row, the wavelength of each 40\,\AA\ bin is indicated by the color bar to the right. 
The navy and purple dot-dashed lines in the middle and the bottom rows represent linear fits to the green- and red-color-coded data points, respectively, which cover the Si\,{\sc ii}\,$\lambda$6355 and the Ca\,{\sc ii}\,NIR3 features. 
The degree of the slope and uncertainties are indicated by the text labeled in the upper-right corner of each subpanel. 
Note that Si\,{\sc ii}\,$\lambda$6355 is blended at all epochs; thus, the Doppler velocity shown may not reflect the true geometrical position within the ejecta. For example, significant blends of Si\,{\sc ii}/Co\,{\sc ii}/Fe\,{\sc ii} can be seen at days\,$+2.0$ and $+$12.0. We also note that a dominant axis can be tentatively identified in both Si\,{\sc ii} at day\,$-$9 and in Ca\,{\sc ii}\,NIR3 at all three epochs, which may be evidence for a large-scale symmetry. 
}~\label{Fig_quall}
\end{figure}

We estimate the continuum polarization of SN\,2019vrq by calculating the error-weighted mean Stokes parameters across the three regions，A, B, and C and at three epochs, separately. As shown in the upper row of Fig.~\ref{Fig_quall}, which presents the Stokes parameters derived within each 40\,\AA\ wavelength bin, the scatter in the distribution of data points provides the main source of uncertainty in the continuum polarization. 
We therefore computed the error-weighted standard deviation of the polarization measured within each wavelength range and assigned it as the uncertainty on the continuum polarization. A similar method is also applied to estimate the continuum polarization level when calculated over the entire wavelength range of interest,  3900--8900\,\AA. 
The results of this calculation are indicated by the filled cyan, green, and pink circles in the upper row of Fig.~\ref{Fig_quall}, respectively, and are also listed in Table~\ref{Table_contpol}, together with the continuum polarization measured across Regions A--C.

\begin{table}[ht]
\normalsize	
\captionsetup{name= Table.}
\caption{Continuum Polarization of SN\,2019vrq Over Different Wavelength Ranges~\label{Table_contpol}}
\begin{tabular}{c|c|c|c|c|c}
\hline
      Phase\,[d] & $Q/U$\,[\%] & 4300--4800\,\AA & 5300--5800\,\AA &  6600--7100\,\AA & 3900--8900\,\AA \\
\hline
$-$8.9 & $Q^{\rm Cont}$ & 0.103$\pm$0.087 &  0.108$\pm$0.064 & $-$0.019$\pm$0.101  & 0.082$\pm$0.104  \\
       & $U^{\rm Cont}$ & 0.006$\pm$0.101 & $-$0.001$\pm$0.078 &  0.017$\pm$0.091 & 0.010$\pm$0.096 \\
 \hline
$+$2.0 & $Q^{\rm Cont}$  & 0.007$\pm$0.057 & $-$0.051$\pm$0.096 & $-$0.002$\pm$0.109 & 0.004$\pm$0.100 \\
      & $U^{\rm Cont}$ & $-$0.015$\pm$0.077 & 0.043$\pm$0.076 & $-$0.037$\pm$0.127 & $-$0.003$\pm$0.098 \\
\hline
$+$12.0 & $Q^{\rm Cont}$ & 0.046$\pm$0.078 & 0.036$\pm$0.085 &  $-$0.032$\pm$0.082 & 0.025$\pm$0.110  \\
        & $U^{\rm Cont}$ & $-$0.099$\pm$0.070 &  $-$0.073$\pm$0.061 &   $-$0.185$\pm$0.103 & $-$0.103$\pm$0.098 \\
 \hline
\end{tabular}
\end{table}

\subsection{Line Polarization}~\label{sec:linepol}
Polarized spectral features indicate a deviation from spherical symmetry in the line-forming regions. In the context of SNe\,Ia, line polarization is generally associated with IMEs and is often interpreted as arising from specific line opacities unevenly covering the underlying, Thomson-scattering-dominated photosphere, although most features are a blend of many transitions. 
The spatial distribution of these line opacities and the degree of their departure from spherical symmetry provides a signature of the propagation of the burning front. 
Note that lines and continua are formed in the same region, with the latter arising as a quasicontinuum of many weak line blends where electron-scattering opacity does not dominate (see, e.g.,~\citealp{1995ApJ...443...89H, 1998ApJ...496..454B}). 
For the case of SN\,2019vrq, the observed scatter in the $Q-U$ plane (see Fig.~\ref{Fig_quall}), both in time and wavelength, suggests the absence of a single dominant, global axis 
and/or the presence of blends from different elements. Narrow features in the polarization spectra with widths comparable to the spectral resolution may simply reflect the limited signal-to-noise ratio (S/N) of the data. 
Alternatively, if such patterns are real, these fluctuations may be induced by small-scale, optically thick structures, either individual lines or blends of line transitions. However intriguing, without unambiguous identification from high-S/N polarimetry, one should be cautious in interpreting such small-scale modulations as physically significant. 

As presented in the upper row of Fig.~\ref{Fig_quall} that displays the polarization data over the wavelength range 3900--8900\,\AA, the absence of a clearly defined dominant axis indicates the lack of a universal symmetry axis across all wavelengths of interest, throughout the continuum. 
The slope of such a dominant axis, when present, measures the polarization position angle. The distribution of the data cloud shown in the upper row of Fig.~\ref{Fig_quall} is also randomly scattered around the origin, suggesting the photosphere intersecting with a spherically symmetric electron-scattering atmosphere from days\,$-$8.9 to $+$12.0. 
The middle and bottom rows of Fig.~\ref{Fig_quall} display the polarization measured across the Si\,{\sc ii}\,$\lambda$6355 and the Ca\,{\sc ii}\,NIR3 features, respectively, over the velocity ranges 24,000--1000\,km\,s$^{-1}$ and 27,000--1000\,km\,s$^{-1}$.

We attempt to perform error-weighted linear least-squares ﬁts to the data points across these spectral features. The Si\,{\,II}\,$\lambda$6355 polarization at day $-$8.9 shows no conspicuous deviation from the clustering in the continuum, which is centered near the origin, indicating a spherically symmetric line opacity distribution in the outermost $\sim5 \times 10^{-2}$\,M$_{\odot}$ mass of the ejecta (see the detailed modeling in Paper~II). 
As illustrated by the green-color-coded data points in the middle row of Fig.~\ref{Fig_quall}, the polarization across the Si\,{\sc ii}/Fe\,{\sc ii} profile tentatively traces a counterclockwise loop from higher to lower velocities at all three epochs. The feature is already dominated by Fe\,{\sc ii} by day\,$+2.0$, and becomes almost entirely dominated by Fe\,{\sc ii} by day\,$+12.0$. 
On both days $-$8.9 and $+$2.0, the overall $p\lesssim0.3$\% can be attributed to the quasicontinuum not being scattering-dominated in these IGE-rich regions~\cite{2023MNRAS.520..560H}, or to the presence of structures that are optically thin in the corresponding bound-bound line opacities (see Paper~II). 
The distribution of the data points can barely be described by a dominant axis, as reflected by the large uncertainties of the PA noted in the text labeled in the upper-right corner of each subpanel. 

{Note that these tentatively identified dominant axes are governed by narrow Si\,{\sc ii} features at 6000--12,000\,km\,s$^{-1}$. A well-defined symmetric axis for the global Si\,{\sc ii} opacity distribution (i.e., extending down to 6000\,km\,s$^{-1}$) is unlikely. 
The Ca\,{\sc ii}\,NIR3 feature displays an overall higher polarization than the Si\,{\sc ii}/Fe\,{\sc ii} profile, though never exceeding 0.4\% across the three epochs. Given the broad uncertainties in the slopes of the dominant axes labeled in the corresponding subpanels of Fig.~\ref{Fig_quall}, and the large uncertainty of each data point owing to limited S/N, we suggest that any line-opacity distribution in the ejecta of SN\,2019vrq cannot be described by a well-defined axisymmetric configuration.}

\subsection{Comparison with Other Overluminous SNe~Ia}~\label{sec:polcompare}
\begin{figure}[ht]
    \centering
    \captionsetup{name= Fig.}
    \includegraphics[trim={0.0cm 0.0cm 0.0cm 0.0cm},clip,width=0.75\textwidth]{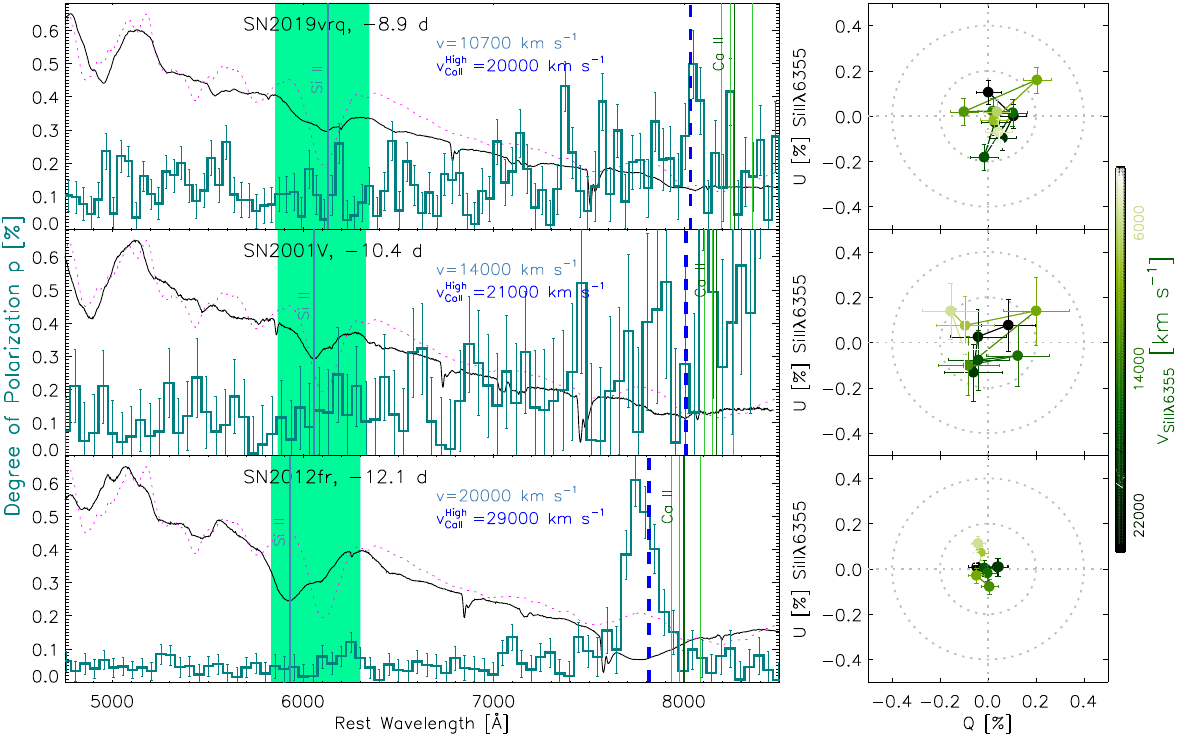}
    \vspace{-0.0 cm}
    \caption{Intrinsic polarization of SN\,2019vrq on day $-$8.9 (Epoch 1, top row) compared to that of SN\,2001V at day $-$10.4 (middle row) and SN\,2012fr at day $-$12.1 (bottom row). 
Polarization spectra of SN\,2019vrq and the two comparison SNe are binned to 40 and 50\,\AA, respectively. In each subpanel of the left column, the black line shows the arbitrarily scaled total-flux spectra. The vertical solid blue lines mark Si\,{\sc ii}\,$\lambda$6355 at the velocity $v_{\rm Siii}$ labeled in each subpanel. The vertical long-dashed blue and green lines mark the mean wavelength of the Ca\,{\sc ii}\,NIR3 feature at HV and at the photospheric velocity traced by $v_{\rm Siii}$, respectively. The magenta dotted line overplots the scaled spectrum of SN\,2011fe at day $-$10 for comparison. 
The right column shows the polarization of the Si\,{\sc ii}\,$\lambda$6355 line over the velocity range 24,000--1000\,km\,s$^{-1}$ in the $Q-U$ plane. The velocities are indicated by the color bar on the right. Some weak patterns are common to all three objects, although the S/N in SN\,2001V is too low to resolve any detail. The shared feature across all objects is a lack of strong polarization in Si\,{\sc ii}. 
}~\label{Fig_polcompare_ep1} 
\end{figure}

\begin{figure}[ht]
    \centering
    \captionsetup{name= Fig.}
    \includegraphics[trim={0.0cm 0.0cm 0.0cm 0.0cm},clip,width=0.75\textwidth]{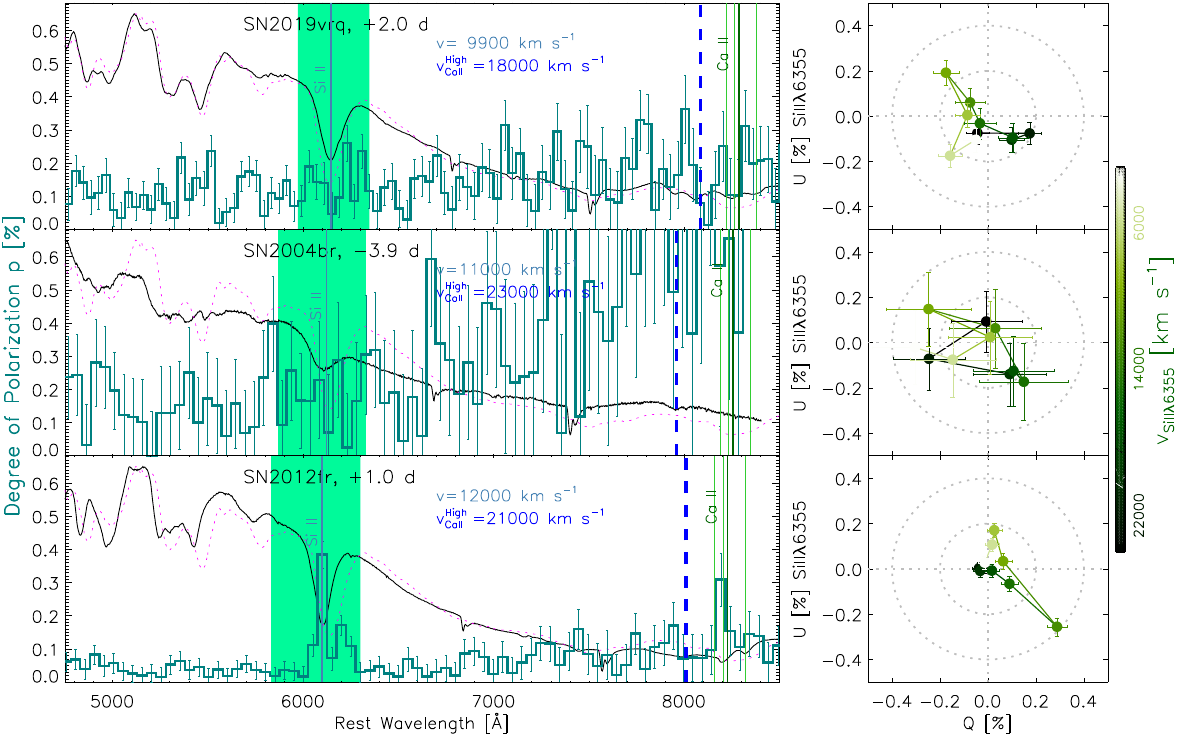}
    \vspace{-0.0 cm}
    \caption{Intrinsic polarization of SN\,2019vrq on day $+$2.0 (Epoch 2, top row) compared to that of SN\,2004br on day $-$3.0 (middle row) and SN\,2012fr on day $+$1.0 (bottom row). The layout of the figure is the same as  Fig.~\ref{Fig_polcompare_ep1}.}~\label{Fig_polcompare_ep2} 
\end{figure}

\begin{figure}[ht]
    \centering
    \captionsetup{name= Fig.}
    \includegraphics[trim={0.0cm 0.0cm 0.0cm 0.0cm},clip,width=0.75\textwidth]{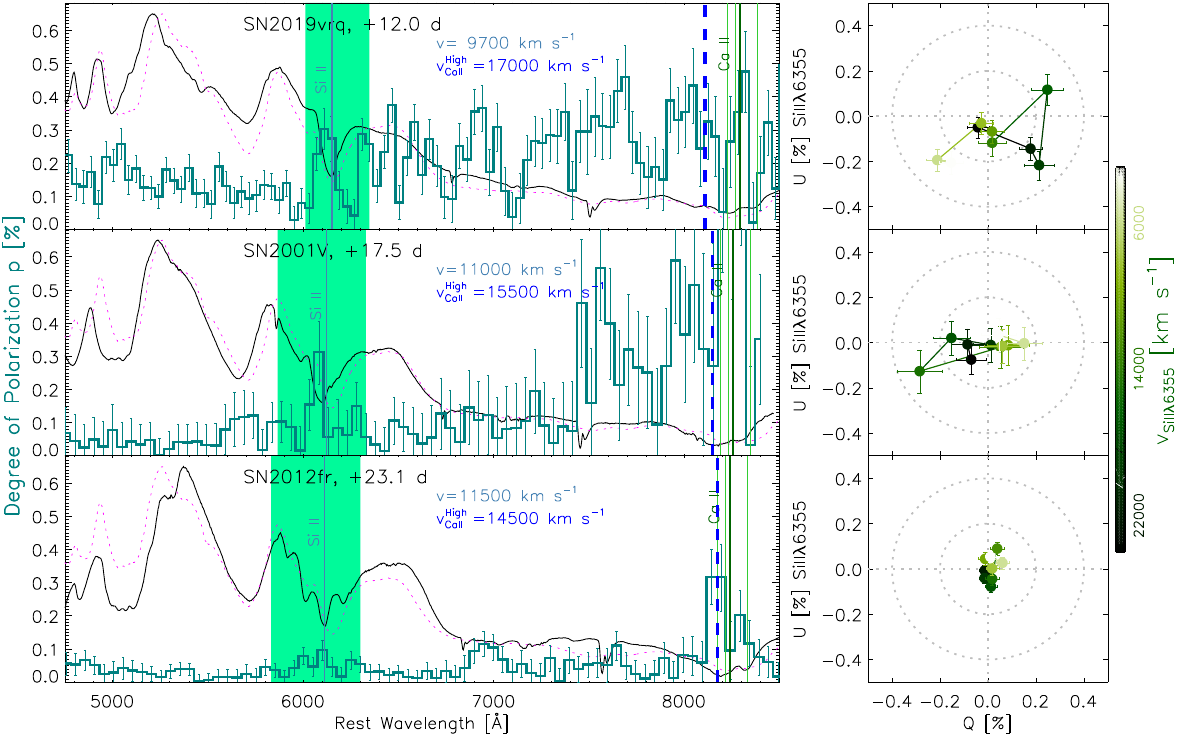}
    \vspace{-0.0 cm}
    \caption{Intrinsic polarization of SN\,2019vrq on day $+$12.0 (Epoch 3, top row) compared to that of SN\,2001V on day $+$17.5 (middle row) and SN\,2012fr at day $+$23.1 (bottom row). The layout of the figure is the same as  Fig.~\ref{Fig_polcompare_ep1}.}~\label{Fig_polcompare_ep3} 
\end{figure}

We also compare the polarization of SN\,2019vrq with that reported for other SNe\,Ia having $\Delta m_{15}(B) \lesssim 0.8$\,mag measured at similar phases. Our comparison sample includes SNe\,2001V ($\Delta m_{15}(B)=0.73\pm0.03$; \citealp{2013MNRAS.433.2240G, 2007Sci...315..212W, 2019MNRAS.490..578C}), 2004br ($\Delta m_{15}(B)=0.68\pm0.15$; \citealp{2010ApJS..190..418G, 2016A&A...594A..13P, 2019MNRAS.490..578C}), and 2012fr ($\Delta m_{15}(B)=0.82\pm0.03$; \citealp{2018ApJ...859...24C, 2019MNRAS.490..578C}). 
The results are presented in Figs.~\ref{Fig_polcompare_ep1}--\ref{Fig_polcompare_ep3}. All polarization spectra were adopted from \citet{2019MNRAS.490..578C} and corrected for the ISP. 
In general, the absence of significantly polarized Si\,{\sc ii}\,$\lambda$6355 is also seen in the premaximum spectropolarimetry of SNe\,2001V and 2004br. An exception is given by SN\,2012fr, which shows significantly higher polarization of Ca\,{\sc ii}\,NIR3, but still a low polarization across Si\,{\sc ii}\,$\lambda$6355 (Section~\ref{sec:linepol}, Figs.~\ref{Fig_polcompare_ep1}--\ref{Fig_polcompare_ep3}). 
The spectroscopic evolution of SN\,2012fr is more similar to the average behavior of normal-bright events. 
In Paper~II, we demonstrate that the low polarization of the blended Fe\,{\sc ii}/Si\,{\sc ii} absorption feature in 91T/99aa-like events is a natural consequence of the significantly lower Si\,{\sc ii} optical depth compared to normal SNe\,Ia. 

The upper, middle, and lower panels of Fig.~\ref{Fig_polcompare_ep1} present the polarization of SNe\,2019vrq at day\,$-$8.9, 2001V at day\,$-$10.4, and 2012fr at day\,$-$12.1, respectively. The left column shows the total-flux and the polarization spectra. The former is also scaled and overlaid with the day\,$-$10 flux spectrum of SN\,2011fe. 
In Table~\ref{Table_polcompare} we summarize the spectropolarimetric comparison of SN\,2019vrq with SNe\,2001V, 2004br, and 2012fr across three epochs (Figs.~\ref{Fig_polcompare_ep1}--\ref{Fig_polcompare_ep3}). At all three epochs, the flux spectra of the comparison objects are also scaled and overlaid with the corresponding-phase spectrum of SN\,2011fe, providing a normal-SN\,Ia reference for both line depth and polarization behavior. 


\begin{table}[ht]
\centering
\caption{Comparison of SN\,2019vrq spectropolarimetry with SN\,2001V, SN\,2004br, and SN\,2012fr at similar phases.}
\label{Table_polcompare}
\begin{tabular}{lllll}
\hline\hline
Epoch & SN & Phase & Key Si\,{\sc ii}\,$\lambda$6355 / Ca\,{\sc ii}\,NIR3 properties & Peak line polarization \\
\hline
\multirow{1}{*}{1 (Fig.~\ref{Fig_polcompare_ep1})}
 & 2019vrq & day\,$-$8.9  & Shallowest Si\,{\sc ii}\,$\lambda$6355 profile of the three & $p\lesssim0.2$\% \\
 & 2001V   & day\,$-$10.4 & Significantly deeper Si\,{\sc ii}\,$\lambda$6355 than SN\,2011fe & $p\lesssim0.2$\% \\
 & 2012fr  & day\,$-$12.1 & Distinct HV component in Si\,{\sc ii}\,$\lambda$6355 and Ca\,{\sc ii}\,NIR3 & $\sim$0.6\% \\
\hline
\multirow{1}{*}{2 (Fig.~\ref{Fig_polcompare_ep2})}
 & 2019vrq & near peak    & Flux similar to SN\,2011fe, but shallower Si\,{\sc ii}\,$\lambda$6355 & $\lesssim$0.2\% \\
 & 2004br  & day\,$-$3.9  & Low S/N; consistent with \citet{2019MNRAS.490..578C} & $\lesssim$0.3\% \\
 & 2012fr  & day\,$+$1.0  & HV component indiscernible by this phase & $\sim$0.4\% \\
\hline
\multirow{1}{*}{3 (Fig.~\ref{Fig_polcompare_ep3})}
 & SN\,2019vrq & post-peak    & Photosphere in NSE-dominated region & $\lesssim$0.3\% \\
 & SN\,2001V   & day\,$+$17.5 & Same as SN\,2019vrq & $\lesssim$0.3\% \\
 & SN\,2012fr  & day\,$+$23.1 & Same as SN\,2019vrq & N/A (nebular transition) \\
\hline
\end{tabular}
\end{table}

Before maximum light (Fig.~\ref{Fig_polcompare_ep1}), SN\,2012fr stands apart from SNe\,2019vrq and 2001V through its distinct HV component and markedly higher peak polarization. This distinct HV component, which also manifests more structured IME features in SN\,2012fr, against the persistently shallow, weakly polarized profiles of SNe\,2019vrq and 2001V, supports the classification of SN\,2012fr as a normal SN\,Ia near the luminous end of the 
\citet{1993ApJ...413L.105P} relation, rather than a genuine 91T/99aa-like event \citep{2013MNRAS.433L..20M}. 

SN\,2019vrq is also consistent with the $\Delta m_{15}(B)-p_{\rm Si\,{\,II}}$ relationship~\citep{2007Sci...315..212W, 2010ApJ...725L.167M, 2019MNRAS.490..578C}. In contrast to the significantly polarized Si\,{\sc ii}\,$\lambda$6355 line that peaks between days $-$10 and 0 observed in normal SNe\,Ia, the persistently low line polarization places SN\,2019vrq at the brightest and least polarized end of the $\Delta m_{15}(B)-p_{\rm Si\,{\,II}}$ relation. The unpolarized continuum of SN\,2019vrq is consistent with that of slowly evolving events ($\Delta m_{15}(B) \lesssim 0.8$\,mag), a regime for which polarimetry exists only for SNe\,2001V, 2004br, and 2012fr~\citep{2019MNRAS.490..578C, 2013MNRAS.433L..20M}.

Near maximum light (Fig.~\ref{Fig_polcompare_ep2}), the flux spectrum of SN\,2019vrq closely resembles that of SN\,2011fe aside from a shallower Si\,{\sc ii}\,$\lambda$6355 line, while the HV component of SN\,2012fr has already faded and its peak polarization coincides in wavelength with the flux absorption minimum. Both SNe\,2019vrq and 2004br exhibit minimal chemical asymmetry, as indicated by their persistently low line polarization, whereas SN\,2012fr develops significantly more structured IME features. 

By $\sim2$--3 weeks after the maximum light, polarization is low across all three SNe as compared in Fig.~\ref{Fig_polcompare_ep3}. As the photosphere recedes into the 
nuclear-statistical-equilibrium (NSE)-dominated inner ejecta and the Si\,{\sc ii}, the Fe\,{\sc ii} becomes dominant in the Si\,{\sc ii}\,$\lambda$6355 profile displayed earlier, which also have vanished entirely in SN\,2012fr. At this phase, line polarization measurements lose their meaning as strong IGE lines emerge and the SN enters the nebular phase. 


To summarize the properties of SN\,2019vrq and the other overluminous SNe discussed above, their polarization shows no significant elongation in any direction in the $Q-U$ plane. 
Additionally, the polarization data are scattered around the origin (Fig.~\ref{Fig_quall}). This scatter is likely to be induced by inhomogeneous, small-scale structures that produce only a moderate departure from spherical symmetry. 
We therefore infer that neither the global geometry of the electron-scattering atmosphere nor that of the line opacity exhibits an apparent global axial symmetry. 
These characteristics place SN\,2019vrq in spectropolarimetric Class N1 according to the classification scheme proposed by \citet{2008ARA&A..46..433W}. 

\section{Discussion and Summary}~\label{sec:summary}
In this paper, we present a comprehensive photometric, spectroscopic, and spectropolarimetric study of SN\,2019vrq, an overluminous SN\,Ia that displays strong resemblance to the 91T/99aa-like subclass. The pseudobolometric properties of the SN exhibit good agreement with the super-Chandrasekhar SNe\,2009dc and 2020esm. A nickel mass of $M(^{56}$Ni) $\approx0.8$--1.0\,$M_{\odot}$ synthesized in the ejcta of SN\,2019vrq can also be inferred, which appears to be smaller than that estimated for super-Chandrasekhar events. 
Spectroscopically, as revealed by the high-S/N VLT data, the early-time spectra of SN\,2019vrq (from day\,$-$8.9 onward) are dominated by a blue continuum with various broad absorption features from IMEs and IGEs. The detection of prominent IGE blanketing at such early phases (e.g., the blended Co{\sc ii} features near 4500\,\AA) indicates prompt outward mixing of radioactive $^{56}$Ni synthesized during the explosion (see Paper~II for details). 

Relative to 91T-like events, SN\,2019vrq exhibits comparatively clearer Ca\,{\sc ii} and Si\,{\sc ii} lines, positioning it closer to the 99aa-like subclass, though a deficit of IME features persists in the outer ejecta relative to spectroscopically normal SNe\,Ia such as SN\,2011fe. 
The presence of doubly ionized species, particularly Si\,{\sc iii} in the pre-peak spectra, indicates elevated temperatures in the outer ejecta. This is consistent with the suppression of neutral and singly ionized C and O features expected under such conditions. It is also in contrast to cooler super-Chandrasekhar events like SN\,2020esm, which show prominent C/O lines. 
By tracking the time evolution of photospheric velocities using the absorption minima of various IME features, we find that Si\,{\sc iii}\,$\lambda$4565 displays a systematically lower velocity and a more rapid decline than Si\,{\sc ii}\,$\lambda$6355. Such behavior may be attributed to NLTE effects given their differing excitation potentials. 
A shallow HV component of Si\,{\sc ii}\,$\lambda$6355 is tentatively identified in the pre-peak spectra via multi-Gaussian decomposition, though we caution that such fits provide only approximate line-property estimates given the intrinsically non-Gaussian nature of line-opacity distributions and the significant blending from Fe\,{\sc ii} lines redward of the Si\,{\sc ii} feature. This blending effect becomes progressively dominant by day\,$+$2 and persists through later epochs. By a few weeks after $B$-band maximum light, the spectral evolution of SN\,2019vrq and other 91T/99aa-like events converges toward that of spectroscopically normal SNe\,Ia.

After subtracting the ISP estimated from the depolarized continuum near peak brightness, our spectropolarimetric analysis reveals a low degree of intrinsic polarization throughout the observed epochs. The continuum polarization across three line-free wavelength regions shows no well-defined dominant axis, and the data cluster tightly around the origin in the Stokes $Q-U$ plane. This signature is consistent with a largely spherically symmetric electron-scattering photosphere from day $-$8.9 through $+$12.0. 
Line polarization associated with Si\,{\sc ii}\,$\lambda$6355 remains weak ($p_{\rm Siii}\lesssim0.2$--0.4\%) at all epochs. The polarization modulation across this feature also traces a tentative counterclockwise loop with increasing wavelength/decreasing velocity in the $Q-U$ plane. However, such patterns are also complicated by blending with Fe\,{\sc ii} and Co\,{\sc ii}. The Ca\,{\sc ii}\,NIR3 shows somewhat higher but still modest peak polarization ($\lesssim$0.4\%). Comparison with other slowly declining SNe\,Ia ($\Delta m_{15}(B)\lesssim0.8$\,mag) shows that SN\,2019vrq is consistent with the established $\Delta m_{15}(B)-p_{\rm Si\,{\,II}}$ relation~\citep{2007Sci...315..212W}. 
This combination of an overluminous, slowly declining LC, a high $^{56}$Ni yield, hot, doubly ionized early-time spectra reminiscent of 91T/99aa-like events, and a notably minimal polarization signature makes SN\,2019vrq an addition to the still-limited sample of overluminous SNe\,Ia with high-quality spectropolarimetric coverage.

In Paper~II, we test these observational signatures against 3D radiation-hydrodynamical simulations of a 
radially pulsating delayed-detonation (PDD, \citealp{1992A&A...253L...9K, 1995ApJ...444..831H, 1996ApJ...457..500H, 2009ApJ...695.1244B, 2009ApJ...695.1257B})
of a near-$M_{\rm Ch}$ WD. In particular, the theoretical analysis is carried out using the HYDrodynamical RAdiation code (HYDRA, e.g.,~\citealp{1993A&A...268..570H, 1995ApJ...440..821H}), which employs detailed hydrodynamics, nuclear networks, magnetic fields, transport of high-energy photons and nonthermal leptons, and full NLTE atomic treatments.

Under this PDD framework, we find the following. 
\begin{itemize}
    \item A near-M$_{\rm Ch}$ WD first ignites a subsonic deflagration near its center. This burning, however, releases too little energy to unbind the WD, in contrast to a normal delayed detonation, in which the deflagration eventually transitions to a supersonic detonation. 
    \item This initial deflagration causes the WD to expand while remaining gravitationally bound. As the WD reaches a maximum radius, the expansion stalls, and the outer envelope then settles inward. 
    \item As the infalling matter recompresses the WD's interior, the resulting deceleration drives Rayleigh-Taylor instabilities, leading to mixing of burned and unburned material. 
    \item The WD then becomes fully unbound after the triggering of the detonation, and the lower density during detonation burning favors abundance structures that are quantitatively different from the nuclear products of ~~classical'' delayed detonations for 91T/99aa-like events.
\end{itemize}
{In Paper~II we show that this model reproduces both the LCs and spectra of SN\,2019vrq, yielding a $^{56}$Ni mass 
consistent with the observationally derived estimates presented in Paper~I.}

Next-generation, wide-field, high-$z$ surveys such as the Nancy Grace Roman Space Telescope's High-Latitude Time-Domain Survey (HLTDS;~\citealp{2018ApJ...867...23H, 2021arXiv211103081R}) and the Chinese Space Station Survey Telescope (CSST; \citealp{2026SCPMA..6939501C}) may open a new window for SN cosmology through 91T/99aa-like events. 
For instance, the HLTDS is expected to discover thousands of SNe\,Ia at $z\approx 0.5$--2, including a substantially larger sample of 91T/99aa-like events than is currently available. 
However, identifying them via sparsely sampled multiband photometry alone is complicated by the redshift-dependent mapping between the rest-frame diagnostic bandpass and the observed filter set, a limitation that manifests differently for Roman and CSST owing to their differing filter sets. The principal photometric discriminant established at low redshift is the pronounced pre-maximum blueness of the rest-frame near-UV ($\sim$2600,\AA; e.g., {\it Swift} $uvw1-v$) color relative to normal SNe\,Ia, whereas the corresponding optical $B-V$ contrast near maximum light is comparatively weak~\citep{2013ApJ...779...23M, 2024ApJS..273...16P}. The Wide Field Instrument (WFI) of the Roman Space Telescope, however, offers no filter blueward of $R$ (central wavelength 6340\,\AA; \citealp{2025arXiv250510574O}). Therefore, the diagnostic rest-frame UV window redshifts into Roman's own bandpass only for $z\gtrsim1.3$--1.5, leaving the low-$z$ portion of the Roman HLTDS ($z\lesssim1$) effectively blind to this signature. The intrinsically weaker optical $B-V$ contrast offers a correspondingly weaker diagnostic in this regime.

CSST, by contrast, provides native near-UV and $u$-band coverage (2550--4300,\AA) via both its Survey Camera and Multi-Channel Imager~\citep{2023SCPMA..6629511L, 2026RAA....26e5020Z}, placing the diagnostic signature around rest-frame 2600\,\AA\ within its observed bandpass for $z\approx0.2$--0.5, thereby complementing the low-$z$ regime inaccessible to Roman HLTDS. We also note that the peak-magnitude excess characteristic of the subclass ($\sim$0.2--0.5\,mag; \citealp{2022ApJ...938...47P, 2024ApJS..273...16P}) is comparable to the intrinsic scatter in standardized SN\,Ia luminosities. Photometric classification based on brightness alone is therefore degenerate with ordinary population scatter unless an independent, unbiased distance estimate is available. This caveat may call for dedicated spectroscopic classification efforts using large ground-based telescopes.

These two missions are best regarded as complementary rather than individually sufficient for the identification problem of 91T/99aa-like SNe\,Ia. CSST's comparatively coarse cadence (10--30\,day, or a 4--14\,day cadence in its time-domain program; \citealp{2023SCPMA..6629511L}) will undersample the brief pre-maximum interval over which the UV color anomaly is most pronounced, and its shallower redshift reach of $z\lesssim1.3$ limits its utility at  high redshifts, which are most relevant to dark-energy constraints. Roman's HLTDS cadence of $\sim$5\,days~\citep{2025arXiv250510574O} and deeper reach ($z\approx2.5$) are better matched to sampling the narrow diagnostic phase window and to extending such classification into the cosmologically critical, matter-dominated regime. This also requires reliable template SED libraries for K-corrections~\citep{1996PASP..108..190K, 2002PASP..114..803N, 2007ApJ...663.1187H} that extend to represent the peculiar, high-ionization SEDs of 91T/99aa-like ejecta. 

A joint strategy, in which CSST establishes low-redshift classification priors and refined SED templates for the subclass while Roman applies matched-cadence, K-corrected and time-dilation-corrected photometric selection at $z\gtrsim1$, has the potential to mitigate the $z$-dependent limitations inherent to either survey alone. 
This approach may help resolve whether the apparent deficit of confirmed 91T/99aa-like events at high redshift, relative to local and Monte~Carlo-predicted rates~\citep{2001ApJ...546..734L}, reflects any unique population evolution or a detectability and classification bias intrinsic to single-survey photometric selection.
The combined sample will enable tests of possible redshift and host-environment evolution in the luminosities and spectroscopic properties of 91T/99aa-like SNe\,Ia. Such evolution could introduce redshift-dependent biases in the Hubble diagram if their standardized luminosities differ from those of normal events. Conversely, incorporating spectroscopic subtype into hierarchical SN\,Ia population models may allow these events to be retained as distance indicators while mitigating associated systematic uncertainties, therefore providing a critical test of the consistency of SN\,Ia standardization and its impact on future precision cosmology.

Finally, we note an additional caution regarding the potential use of 91T/99aa-like events in high-$z$ cosmology: no unambiguous, spectroscopically confirmed example of this subclass has yet been reported at $z\gtrsim1$. Consequently, neither their true relative rate nor their cosmological utility can yet be established. In the local universe, 91T/99aa-like SNe\,Ia account for only a few percent of the SN\,Ia population~\citep{2001ApJ...546..734L}. Their progenitors are thought to trace young stellar populations and short delay times, so their fraction may rise considerably toward higher redshift as the cosmic star-formation rate increases. 
Since most high-$z$ SN\,Ia samples are selected via photometric classification rather than spectroscopy, an evolving fraction of 91T/99aa-like events could enter such samples without being flagged. Any intrinsic luminosity differences associated with this subclass might then be mistaken for genuine evolution in the standardized SN\,Ia population. 

This concern is supported by evidence that the width-luminosity relation is not strictly linear across the full range of decline rates and spectroscopic subtypes. \citet{2006ApJ...641...50W} identified that the difference between two independent estimates of peak magnitude varies with decline rate in a manner consistent with a bilinear instead of a linear dependence. 
The recent work of \citet{2022ApJ...938...83Y} demonstrated that 91T/99aa-like SNe remain $\sim$0.2\,mag brighter than normal SNe\,Ia even after applying full LC-shape and color standardization. They also found that the Hubble residuals of these overluminous events correlate with the pEW of Si\,{\sc ii}$\lambda$6355 through a broken linear relation. 
If such subtype-dependent nonlinearity is not modeled explicitly, any redshift-dependent change in the 91T/99aa fraction, whether real or introduced by incomplete classification, will translate into a systematic, redshift-correlated bias in the derived distance moduli. Extending the spectroscopic characterization of 91T/99aa-like events to high redshift, and establishing whether the nonlinear standardization relation itself evolves with redshift, therefore both remain essential. 

\section*{Acknowledgments}
All VLT spectropolarimetry data presented in this study are based on observations collected at the European Organisation for Astronomical Research in the Southern Hemisphere under ESO program 0104.D-0109 (PI Y.\,Yang) and can be accessed via \url{https://archive.eso.org/cms.html}. 
This work makes use of observations from the Las Cumbres Observatory network.
{IRAF} is distributed by the National Optical Astronomy Observatories, which are operated by the Association of Universities for Research in Astronomy, Inc., under cooperative agreement with the U.S. NSF.
PyRAF, PyFITS, and STSCI$\_$PYTHON are products of the Space Telescope Science Institute (STScI), which is operated by the Association of Universities for Research in Astronomy, Inc., under NASA contract NAS5-26555. This research has made use of NASA's Astrophysics Data System Bibliographic Services, the SIMBAD database, operated at CDS, Strasbourg, France, and the NASA/IPAC Extragalactic Database (NED) which is operated by the Jet Propulsion Laboratory, California Institute of Technology, under contract with NASA. 
The authors used Claude to check for grammar and to improve the readability of some paragraphs in this manuscript. 
The authors reviewed and edited the content to ensure accuracy and take full responsibility for the final text. 

Y.Y.'s research is partially supported by the Tsinghua University Dushi Program, and 
previously through a Benoziyo Prize Postdoctoral Fellowship and the Bengier-Winslow-Robertson Fellowship. 
P.H. acknowledges the support of 
NSF awards AST-1715133 and  AST-2306395 for enabling the development of the methods and code HYDRA used for the simulations. The simulations have been performed on the Astrophysics group's computer cluster at FSU. 
J.C.W. is supported by NSF grant AST-1813825. 
The LCO team is supported by NSF grants AST-2308113 and AST-1911151. 
The work of A.C. is supported by NOIRLab, which is managed by the Association of Universities for Research in Astronomy (AURA) under a cooperative agreement with the NSF. 
The research group of A.V.F. at U.C. Berkeley received financial assistance from the Christopher R. Redlich Fund, as well as donations from Gary and Cynthia Bengier, Clark and Sharon Winslow, 
Sanford Robertson, 
and numerous other donors. 
M.B. acknowledges the Department of Physics and Earth Science of the University of Ferrara for financial support through the FIRD 2025 grant. 
A.G.-Y.’s research is supported by the ISF GW excellence centre, an IMOS space infrastructure grant and BSF/Transformative and GIF grants, as well as the Andr\'{e} Deloro Institute for Space and Optics Research, the Center for Experimental Physics, a WIS-MIT Sagol grant, the Norman E Alexander Family M Foundation ULTRASAT Data Center Fund, and Yeda-Sela; A.G.-Y. is the incumbent of the Arlyn Imberman Professorial Chair. 
D.H. is supported by STScI grants HST-GO-17770.002, JWST-GO-12468.001, and JWST-GO-09964.001.

\bibliographystyle{aasjournal}
\bibliography{sn-bibliography}

@BOOK{2017suex.book.....B,
       author = {{Branch}, David and {Wheeler}, J. Craig},
        title = "{Supernova Explosions}",
         year = 2017,
          doi = {10.1007/978-3-662-55054-0},
       adsurl = {https://ui.adsabs.harvard.edu/abs/2017suex.book.....B},
    publisher = {Springer-Verlag Berlin and Heidelberg GmbH \& Co. KG, Germany}
}

@ARTICLE{1998AJ....116.1009R,
       author = {{Riess}, Adam G. and {Filippenko}, Alexei V. and {Challis}, Peter and {Clocchiatti}, Alejandro and {Diercks}, Alan and {Garnavich}, Peter M. and {Gilliland}, Ron L. and {Hogan}, Craig J. and {Jha}, Saurabh and {Kirshner}, Robert P. and {Leibundgut}, B. and {Phillips}, M.~M. and {Reiss}, David and {Schmidt}, Brian P. and {Schommer}, Robert A. and {Smith}, R. Chris and {Spyromilio}, J. and {Stubbs}, Christopher and {Suntzeff}, Nicholas B. and {Tonry}, John},
        title = "{Observational Evidence from Supernovae for an Accelerating Universe and a Cosmological Constant}",
      journal = {\aj},
         year = 1998,
        month = sep,
       volume = {116},
       number = {3},
        pages = {1009-1038},
          doi = {10.1086/300499},
archivePrefix = {arXiv},
       eprint = {astro-ph/9805201},
 primaryClass = {astro-ph},
       adsurl = {https://ui.adsabs.harvard.edu/abs/1998AJ....116.1009R}
}

@ARTICLE{2016ApJ...826...56R,
       author = {{Riess}, Adam G. and {Macri}, Lucas M. and {Hoffmann}, Samantha L. and {Scolnic}, Dan and {Casertano}, Stefano and {Filippenko}, Alexei V. and {Tucker}, Brad E. and {Reid}, Mark J. and {Jones}, David O. and {Silverman}, Jeffrey M. and {Chornock}, Ryan and {Challis}, Peter and {Yuan}, Wenlong and {Brown}, Peter J. and {Foley}, Ryan J.},
        title = "{A 2.4\% Determination of the Local Value of the Hubble Constant}",
      journal = {\apj},
         year = 2016,
        month = jul,
       volume = {826},
       number = {1},
          eid = {56},
        pages = {56},
          doi = {10.3847/0004-637X/826/1/56},
archivePrefix = {arXiv},
       eprint = {1604.01424},
 primaryClass = {astro-ph.CO},
       adsurl = {https://ui.adsabs.harvard.edu/abs/2016ApJ...826...56R}
}

@ARTICLE{1999ApJ...517..565P,
       author = {{Perlmutter}, S. and {Aldering}, G. and {Goldhaber}, G. and {Knop}, R.~A. and {Nugent}, P. and {Castro}, P.~G. and {Deustua}, S. and {Fabbro}, S. and {Goobar}, A. and {Groom}, D.~E. and {Hook}, I.~M. and {Kim}, A.~G. and {Kim}, M.~Y. and {Lee}, J.~C. and {Nunes}, N.~J. and {Pain}, R. and {Pennypacker}, C.~R. and {Quimby}, R. and {Lidman}, C. and {Ellis}, R.~S. and {Irwin}, M. and {McMahon}, R.~G. and {Ruiz-Lapuente}, P. and {Walton}, N. and {Schaefer}, B. and {Boyle}, B.~J. and {Filippenko}, A.~V. and {Matheson}, T. and {Fruchter}, A.~S. and {Panagia}, N. and {Newberg}, H.~J.~M. and {Couch}, W.~J. and {Project}, The Supernova Cosmology},
        title = "{Measurements of {\ensuremath{\Omega}} and {\ensuremath{\Lambda}} from 42 High-Redshift Supernovae}",
      journal = {\apj},
         year = 1999,
        month = jun,
       volume = {517},
       number = {2},
        pages = {565-586},
          doi = {10.1086/307221},
archivePrefix = {arXiv},
       eprint = {astro-ph/9812133},
 primaryClass = {astro-ph},
       adsurl = {https://ui.adsabs.harvard.edu/abs/1999ApJ...517..565P}
}

@ARTICLE{1996ApJ...457..500H,
       author = {{Hoeflich}, P. and {Khokhlov}, A.},
        title = "{Explosion Models for Type IA Supernovae: A Comparison with Observed Light Curves, Distances, H 0, and Q 0}",
      journal = {\apj},
         year = 1996,
        month = feb,
       volume = {457},
        pages = {500},
          doi = {10.1086/176748},
archivePrefix = {arXiv},
       eprint = {astro-ph/9602025},
 primaryClass = {astro-ph},
       adsurl = {https://ui.adsabs.harvard.edu/abs/1996ApJ...457..500H}
}

@ARTICLE{2008ARA&A..46..433W,
       author = {{Wang}, L. and {Wheeler}, J.~C.},
        title = "{Spectropolarimetry of supernovae.}",
      journal = {\araa},
         year = 2008,
        month = sep,
       volume = {46},
        pages = {433-474},
          doi = {10.1146/annurev.astro.46.060407.145139},
archivePrefix = {arXiv},
       eprint = {0811.1054},
 primaryClass = {astro-ph},
       adsurl = {https://ui.adsabs.harvard.edu/abs/2008ARA&A..46..433W}
}

@ARTICLE{1991A&A...246..481H,
       author = {{Hoflich}, P.},
        title = "{Asphericity effects in scatterring dominated photospheres.}",
      journal = {\aap},
         year = 1991,
        month = jun,
       volume = {246},
        pages = {481},
       adsurl = {https://ui.adsabs.harvard.edu/abs/1991A&A...246..481H}
}

@ARTICLE{1998Msngr..94....1A,
       author = {{Appenzeller}, I. and {Fricke}, K. and {F{\"u}rtig}, W. and {G{\"a}ssler}, W. and {H{\"a}fner}, R. and {Harke}, R. and {Hess}, H. -J. and {Hummel}, W. and {J{\"u}rgens}, P. and {Kudritzki}, R. -P. and {Mantel}, K. -H. and {Meisl}, W. and {Muschielok}, B. and {Nicklas}, H. and {Rupprecht}, G. and {Seifert}, W. and {Stahl}, O. and {Szeifert}, T. and {Tarantik}, K.},
        title = "{Successful commissioning of FORS1 - the first optical instrument on the VLT.}",
      journal = {The Messenger},
         year = 1998,
        month = dec,
       volume = {94},
        pages = {1-6},
       adsurl = {https://ui.adsabs.harvard.edu/abs/1998Msngr..94....1A}
}

@ARTICLE{Anderson_etal_2018,
   author = {{Anderson}, J. {\it et al}},
    title = "{Very Large Telescope Paranal Science Operations FORS2 User Manual}",
  journal = {EUROPEAN SOUTHERN OBSERVATORY},
    year = 2018,
   month = { },
   volume = {Doc. No. VLT-MAN-ESO-13100-1543},
    pages = { },
     doi = { },
   adsurl = { }
}

@ARTICLE{2010A&A...510A.108P,
       author = {{Patat}, F. and {Maund}, J.~R. and {Benetti}, S. and {Botticella}, M.~T. and {Cappellaro}, E. and {Harutyunyan}, A. and {Turatto}, M.},
        title = "{VLT spectropolarimetry of the optical transient in NGC 300. Evidence of asymmetry in the circumstellar dust}",
      journal = {\aap},
         year = 2010,
        month = feb,
       volume = {510},
          eid = {A108},
        pages = {A108},
          doi = {10.1051/0004-6361/200913083},
archivePrefix = {arXiv},
       eprint = {0908.0942},
 primaryClass = {astro-ph.SR},
       adsurl = {https://ui.adsabs.harvard.edu/abs/2010A&A...510A.108P}
}

@INPROCEEDINGS{1986SPIE..627..733T,
       author = {{Tody}, Doug},
        title = "{The IRAF Data Reduction and Analysis System}",
    booktitle = {Instrumentation in astronomy VI},
         year = 1986,
       editor = {{Crawford}, David L.},
       series = {Society of Photo-Optical Instrumentation Engineers (SPIE) Conference Series},
       volume = {627},
        month = jan,
        pages = {733},
          doi = {10.1117/12.968154},
       adsurl = {https://ui.adsabs.harvard.edu/abs/1986SPIE..627..733T}
}

@INPROCEEDINGS{1993ASPC...52..173T,
       author = {{Tody}, Doug},
        title = "{IRAF in the Nineties}",
    booktitle = {Astronomical Data Analysis Software and Systems II},
         year = 1993,
       editor = {{Hanisch}, R.~J. and {Brissenden}, R.~J.~V. and {Barnes}, J.},
       series = {Astronomical Society of the Pacific Conference Series},
       volume = {52},
        month = jan,
        pages = {173},
       adsurl = {https://ui.adsabs.harvard.edu/abs/1993ASPC...52..173T}
}

@ARTICLE{1985A&A...142..100S,
       author = {{Simmons}, J.~F.~L. and {Stewart}, B.~G.},
        title = "{Point and interval estimation of the true unbiased degree of linear polarization in the presence of low signal-to-noise ratios}",
      journal = {\aap},
         year = 1985,
        month = jan,
       volume = {142},
       number = {1},
        pages = {100-106},
       adsurl = {https://ui.adsabs.harvard.edu/abs/1985A&A...142..100S}
}

@ARTICLE{1997ApJ...476L..27W,
       author = {{Wang}, Lifan and {Wheeler}, J. Craig and {H{\"o}flich}, Peter},
        title = "{Polarimetry of the Type IA Supernova SN 1996X}",
      journal = {\apjl},
         year = 1997,
        month = feb,
       volume = {476},
       number = {1},
        pages = {L27-L30},
          doi = {10.1086/310495},
archivePrefix = {arXiv},
       eprint = {astro-ph/9609178},
 primaryClass = {astro-ph},
       adsurl = {https://ui.adsabs.harvard.edu/abs/1997ApJ...476L..27W}
}

@ARTICLE{2019MNRAS.490..578C,
       author = {{Cikota}, Aleksandar and {Patat}, Ferdinando and {Wang}, Lifan and {Wheeler}, J. Craig and {Bulla}, Mattia and {Baade}, Dietrich and {H{\"o}flich}, Peter and {Cikota}, Stefan and {Clocchiatti}, Alejandro and {Maund}, Justyn R. and {Stevance}, Heloise F. and {Yang}, Yi},
        title = "{Linear spectropolarimetry of 35 Type Ia supernovae with VLT/FORS: an analysis of the Si II line polarization}",
      journal = {\mnras},
         year = 2019,
        month = nov,
       volume = {490},
       number = {1},
        pages = {578-599},
          doi = {10.1093/mnras/stz2322},
archivePrefix = {arXiv},
       eprint = {1908.07526},
 primaryClass = {astro-ph.HE},
       adsurl = {https://ui.adsabs.harvard.edu/abs/2019MNRAS.490..578C}
}

@ARTICLE{2017MNRAS.464.4146C,
       author = {{Cikota}, Aleksandar and {Patat}, Ferdinando and {Cikota}, Stefan and {Faran}, Tamar},
        title = "{Linear spectropolarimetry of polarimetric standard stars with VLT/FORS2}",
      journal = {\mnras},
         year = 2017,
        month = feb,
       volume = {464},
       number = {4},
        pages = {4146-4159},
          doi = {10.1093/mnras/stw2545},
archivePrefix = {arXiv},
       eprint = {1610.00722},
 primaryClass = {astro-ph.IM},
       adsurl = {https://ui.adsabs.harvard.edu/abs/2017MNRAS.464.4146C}
}

@ARTICLE{2006PASP..118..146P,
       author = {{Patat}, Ferdinando and {Romaniello}, Martino},
        title = "{Error Analysis for Dual-Beam Optical Linear Polarimetry}",
      journal = {\pasp},
         year = 2006,
        month = jan,
       volume = {118},
       number = {839},
        pages = {146-161},
          doi = {10.1086/497581},
archivePrefix = {arXiv},
       eprint = {astro-ph/0509153},
 primaryClass = {astro-ph},
       adsurl = {https://ui.adsabs.harvard.edu/abs/2006PASP..118..146P}
}

@ARTICLE{2007MNRAS.381..201M,
       author = {{Maund}, Justyn R. and {Wheeler}, J. Craig and {Patat}, Ferdinando and {Baade}, Dietrich and {Wang}, Lifan and {H{\"o}flich}, Peter},
        title = "{Spectropolarimetry of the Type Ib/c SN 2005bf}",
      journal = {\mnras},
         year = 2007,
        month = oct,
       volume = {381},
       number = {1},
        pages = {201-210},
          doi = {10.1111/j.1365-2966.2007.12230.x},
archivePrefix = {arXiv},
       eprint = {0707.2237},
 primaryClass = {astro-ph},
       adsurl = {https://ui.adsabs.harvard.edu/abs/2007MNRAS.381..201M}
}

@ARTICLE{2003ApJ...593..788K,
       author = {{Kasen}, Daniel and {Nugent}, Peter and {Wang}, Lifan and {Howell}, D.~A. and {Wheeler}, J. Craig and {H{\"o}flich}, Peter and {Baade}, Dietrich and {Baron}, E. and {Hauschildt}, P.~H.},
        title = "{Analysis of the Flux and Polarization Spectra of the Type Ia Supernova SN 2001el: Exploring the Geometry of the High-Velocity Ejecta}",
      journal = {\apj},
         year = 2003,
        month = aug,
       volume = {593},
       number = {2},
        pages = {788-808},
          doi = {10.1086/376601},
archivePrefix = {arXiv},
       eprint = {astro-ph/0301312},
 primaryClass = {astro-ph},
       adsurl = {https://ui.adsabs.harvard.edu/abs/2003ApJ...593..788K}
}

@ARTICLE{2011NatCo...2..350H,
       author = {{Howell}, D. Andrew},
        title = "{Type Ia supernovae as stellar endpoints and cosmological tools}",
      journal = {Nature Communications},
         year = 2011,
        month = jun,
       volume = {2},
          eid = {350},
        pages = {350},
          doi = {10.1038/ncomms1344},
       adsurl = {https://ui.adsabs.harvard.edu/abs/2011NatCo...2..350H}
}

@ARTICLE{2013FrPhy...8..116H,
       author = {{Hillebrandt}, W. and {Kromer}, M. and {R{\"o}pke}, F.~K. and {Ruiter}, A.~J.},
        title = "{Towards an understanding of Type Ia supernovae from a synthesis of theory and observations}",
      journal = {Frontiers of Physics},
         year = 2013,
        month = apr,
       volume = {8},
       number = {2},
        pages = {116-143},
          doi = {10.1007/s11467-013-0303-2},
       adsurl = {https://ui.adsabs.harvard.edu/abs/2013FrPhy...8..116H}
}

@ARTICLE{2014ARA&A..52..107M,
       author = {{Maoz}, Dan and {Mannucci}, Filippo and {Nelemans}, Gijs},
        title = "{Observational Clues to the Progenitors of Type Ia Supernovae}",
      journal = {\araa},
         year = 2014,
        month = aug,
       volume = {52},
        pages = {107-170},
          doi = {10.1146/annurev-astro-082812-141031},
       adsurl = {https://ui.adsabs.harvard.edu/abs/2014ARA&A..52..107M}
}

@INCOLLECTION{2017hsn..book.1017P,
       author = {{Patat}, Ferdinando},
        title = "{Introduction to Supernova Polarimetry}",
    booktitle = {Handbook of Supernovae},
         year = 2017,
       editor = {{Alsabti}, Athem W. and {Murdin}, Paul},
        pages = {1017},
          doi = {10.1007/978-3-319-21846-5_110},
       adsurl = {https://ui.adsabs.harvard.edu/abs/2017hsn..book.1017P},
publisher = {Springer Cham}
}

@ARTICLE{2020ApJ...902...46Y,
       author = {{Yang}, Yi and {Hoeflich}, Peter and {Baade}, Dietrich and {Maund}, Justyn R. and {Wang}, Lifan and {Brown}, Peter. J. and {Stevance}, Heloise F. and {Arcavi}, Iair and {Burke}, Jamison and {Cikota}, Aleksandar and {Clocchiatti}, Alejandro and {Gal-Yam}, Avishay and {Graham}, Melissa. L. and {Hiramatsu}, Daichi and {Hosseinzadeh}, Griffin and {Howell}, D. Andrew and {Jha}, Saurabh W. and {McCully}, Curtis and {Patat}, Ferdinando and {Sand}, David. J. and {Schulze}, Steve and {Spyromilio}, Jason and {Valenti}, Stefano and {Vink{\'o}}, J{\'o}zsef and {Wang}, Xiaofeng and {Wheeler}, J. Craig and {Yaron}, Ofer and {Zhang}, Jujia},
        title = "{The Young and Nearby Normal Type Ia Supernova 2018gv: UV-optical Observations and the Earliest Spectropolarimetry}",
      journal = {\apj},
         year = 2020,
        month = oct,
       volume = {902},
       number = {1},
          eid = {46},
        pages = {46},
          doi = {10.3847/1538-4357/aba759},
       adsurl = {https://ui.adsabs.harvard.edu/abs/2020ApJ...902...46Y}
}

@ARTICLE{2018MNRAS.476.1299M,
       author = {{Mulligan}, Brian W. and {Wheeler}, J. Craig},
        title = "{A compact circumstellar shell as the source of high-velocity features in SN 2011fe}",
      journal = {\mnras},
         year = 2018,
        month = may,
       volume = {476},
       number = {1},
        pages = {1299-1309},
          doi = {10.1093/mnras/sty027},
       adsurl = {https://ui.adsabs.harvard.edu/abs/2018MNRAS.476.1299M}
}

@ARTICLE{2019MNRAS.484.4785M,
       author = {{Mulligan}, Brian W. and {Zhang}, Kaicheng and {Wheeler}, J. Craig},
        title = "{Exploring the shell model of high-velocity features of Type Ia supernovae using TARDIS}",
      journal = {\mnras},
         year = 2019,
        month = apr,
       volume = {484},
       number = {4},
        pages = {4785-4800},
          doi = {10.1093/mnras/stz235},
       adsurl = {https://ui.adsabs.harvard.edu/abs/2019MNRAS.484.4785M}
}

@ARTICLE{2022ApJ...934L...7R,
       author = {{Riess}, Adam G. and {Yuan}, Wenlong and {Macri}, Lucas M. and {Scolnic}, Dan and {Brout}, Dillon and {Casertano}, Stefano and {Jones}, David O. and {Murakami}, Yukei and {Anand}, Gagandeep S. and {Breuval}, Louise and {Brink}, Thomas G. and {Filippenko}, Alexei V. and {Hoffmann}, Samantha and {Jha}, Saurabh W. and {D'arcy Kenworthy}, W. and {Mackenty}, John and {Stahl}, Benjamin E. and {Zheng}, WeiKang},
        title = "{A Comprehensive Measurement of the Local Value of the Hubble Constant with 1 km s$^{-1}$ Mpc$^{-1}$ Uncertainty from the Hubble Space Telescope and the SH0ES Team}",
      journal = {\apjl},
         year = 2022,
        month = jul,
       volume = {934},
       number = {1},
          eid = {L7},
        pages = {L7},
          doi = {10.3847/2041-8213/ac5c5b},
archivePrefix = {arXiv},
       eprint = {2112.04510},
 primaryClass = {astro-ph.CO},
       adsurl = {https://ui.adsabs.harvard.edu/abs/2022ApJ...934L...7R}
}

@ARTICLE{2008AJ....135.1598M,
       author = {{Matheson}, T. and {Kirshner}, R.~P. and {Challis}, P. and {Jha}, S. and {Garnavich}, P.~M. and {Berlind}, P. and {Calkins}, M.~L. and {Blondin}, S. and {Balog}, Z. and {Bragg}, A.~E. and {Caldwell}, N. and {Dendy Concannon}, K. and {Falco}, E.~E. and {Graves}, G.~J.~M. and {Huchra}, J.~P. and {Kuraszkiewicz}, J. and {Mader}, J.~A. and {Mahdavi}, A. and {Phelps}, M. and {Rines}, K. and {Song}, I. and {Wilkes}, B.~J.},
        title = "{Optical Spectroscopy of Type ia Supernovae}",
      journal = {\aj},
         year = 2008,
        month = apr,
       volume = {135},
       number = {4},
        pages = {1598-1615},
          doi = {10.1088/0004-6256/135/4/1598},
archivePrefix = {arXiv},
       eprint = {0803.1705},
 primaryClass = {astro-ph},
       adsurl = {https://ui.adsabs.harvard.edu/abs/2008AJ....135.1598M}
}

@ARTICLE{2003ApJ...591.1110W,
       author = {{Wang}, Lifan and {Baade}, Dietrich and {H{\"o}flich}, Peter and {Khokhlov}, Alexei and {Wheeler}, J. Craig and {Kasen}, D. and {Nugent}, Peter E. and {Perlmutter}, Saul and {Fransson}, Claes and {Lundqvist}, Peter},
        title = "{Spectropolarimetry of SN 2001el in NGC 1448: Asphericity of a Normal Type Ia Supernova}",
      journal = {\apj},
         year = 2003,
        month = jul,
       volume = {591},
       number = {2},
        pages = {1110-1128},
          doi = {10.1086/375444},
archivePrefix = {arXiv},
       eprint = {astro-ph/0303397},
 primaryClass = {astro-ph},
       adsurl = {https://ui.adsabs.harvard.edu/abs/2003ApJ...591.1110W}
}

@ARTICLE{2010ApJ...722.1162M,
       author = {{Maund}, Justyn R. and {Wheeler}, J. Craig and {Wang}, Lifan and {Baade}, Dietrich and {Clocchiatti}, Alejandro and {Patat}, Ferdinando and {H{\"o}flich}, Peter and {Quinn}, Jason and {Zelaya}, Paula},
        title = "{A Spectropolarimetric View on the Nature of the Peculiar Type I SN 2005hk}",
      journal = {\apj},
         year = 2010,
        month = oct,
       volume = {722},
       number = {2},
        pages = {1162-1174},
          doi = {10.1088/0004-637X/722/2/1162},
archivePrefix = {arXiv},
       eprint = {1008.3985},
 primaryClass = {astro-ph.SR},
       adsurl = {https://ui.adsabs.harvard.edu/abs/2010ApJ...722.1162M}
}

@ARTICLE{1993A&A...268..570H,
       author = {{Hoeflich}, P. and {Mueller}, E. and {Khokhlov}, A.},
        title = "{Light curve models for type IA supernovae - Physical assumptions, their influence and validity}",
      journal = {\aap},
         year = 1993,
        month = feb,
       volume = {268},
       number = {2},
        pages = {570-590},
       adsurl = {https://ui.adsabs.harvard.edu/abs/1993A&A...268..570H}
}

@INCOLLECTION{2017hsn..book.1151H,
       author = {{Hoeflich}, Peter},
        title = "{Explosion Physics of Thermonuclear Supernovae and Their Signatures}",
    booktitle = {Handbook of Supernovae},
         year = 2017,
       editor = {{Alsabti}, Athem W. and {Murdin}, Paul},
        pages = {1151},
          doi = {10.1007/978-3-319-21846-5_56},
       adsurl = {https://ui.adsabs.harvard.edu/abs/2017hsn..book.1151H},
    publisher = {Springer Cham}
}

@ARTICLE{2013MNRAS.433L..20M,
       author = {{Maund}, J.~R. and {Spyromilio}, J. and {Hoflich}, P.~A. and {Wheeler}, J.~C. and {Baade}, D. and {Clocchiatti}, A. and {Patat}, F. and {Reilly}, E. and {Wang}, L. and {Zelaya}, P.},
        title = "{Spectropolarimetry of the type Ia supernova 2012fr.}",
      journal = {\mnras},
         year = 2013,
        month = jun,
       volume = {433},
        pages = {L20-L24},
          doi = {10.1093/mnrasl/slt050},
archivePrefix = {arXiv},
       eprint = {1302.0166},
 primaryClass = {astro-ph.SR},
       adsurl = {https://ui.adsabs.harvard.edu/abs/2013MNRAS.433L..20M}
}

@ARTICLE{2001ApJ...556..302H,
       author = {{Howell}, D. Andrew and {H{\"o}flich}, Peter and {Wang}, Lifan and {Wheeler}, J. Craig},
        title = "{Evidence for Asphericity in a Subluminous Type Ia Supernova: Spectropolarimetry of SN 1999by}",
      journal = {\apj},
         year = 2001,
        month = jul,
       volume = {556},
       number = {1},
        pages = {302-321},
          doi = {10.1086/321584},
archivePrefix = {arXiv},
       eprint = {astro-ph/0101520},
 primaryClass = {astro-ph},
       adsurl = {https://ui.adsabs.harvard.edu/abs/2001ApJ...556..302H}
}

@ARTICLE{2017MNRAS.469.1897S,
       author = {{Stevance}, H.~F. and {Maund}, J.~R. and {Baade}, D. and {H{\"o}flich}, P. and {Howerton}, S. and {Patat}, F. and {Rose}, M. and {Spyromilio}, J. and {Wheeler}, J.~C. and {Wang}, L.},
        title = "{The evolution of the 3D shape of the broad-lined Type Ic SN 2014ad}",
      journal = {\mnras},
         year = 2017,
        month = aug,
       volume = {469},
       number = {2},
        pages = {1897-1911},
          doi = {10.1093/mnras/stx970},
archivePrefix = {arXiv},
       eprint = {1704.06270},
 primaryClass = {astro-ph.HE},
       adsurl = {https://ui.adsabs.harvard.edu/abs/2017MNRAS.469.1897S}
}

@ARTICLE{2007Sci...315..212W,
   author = {{Wang}, L. and {Baade}, D. and {Patat}, F.},
    title = "{Spectropolarimetric Diagnostics of Thermonuclear Supernova Explosions}",
  journal = {Science},
   eprint = {astro-ph/0611902},
     year = 2007,
    month = jan,
   volume = 315,
    pages = {212},
      doi = {10.1126/science.1121656},
   adsurl = {http://adsabs.harvard.edu/abs/2007Sci...315..212W}
}

@ARTICLE{2010ApJ...725L.167M,
   author = {{Maund}, J.~R. and {H{\"o}flich}, P. and {Patat}, F. and {Wheeler}, J.~C. and 
	{Zelaya}, P. and {Baade}, D. and {Wang}, L. and {Clocchiatti}, A. and 
	{Quinn}, J.},
    title = "{The Unification of Asymmetry Signatures of Type Ia Supernovae}",
  journal = {ApJL},
archivePrefix = "arXiv",
   eprint = {1008.0651},
 primaryClass = "astro-ph.SR",
     year = 2010,
    month = dec,
   volume = 725,
    pages = {L167-L171},
      doi = {10.1088/2041-8205/725/2/L167},
   adsurl = {http://adsabs.harvard.edu/abs/2010ApJ...725L.167M}
}

@INCOLLECTION{2017hsn..book..317T,
       author = {{Taubenberger}, Stefan},
        title = "{The Extremes of Thermonuclear Supernovae}",
    booktitle = {Handbook of Supernovae},
         year = 2017,
       editor = {{Alsabti}, Athem W. and {Murdin}, Paul},
        pages = {317},
          doi = {10.1007/978-3-319-21846-5_37},
       adsurl = {https://ui.adsabs.harvard.edu/abs/2017hsn..book..317T},
publisher = {Springer Cham}
}

@ARTICLE{2023RAA....23h2001L,
       author = {{Liu}, Zheng-Wei and {R{\"o}pke}, Friedrich K. and {Han}, Zhanwen},
        title = "{Type Ia Supernova Explosions in Binary Systems: A Review}",
      journal = {Research in Astronomy and Astrophysics},
         year = 2023,
        month = aug,
       volume = {23},
       number = {8},
          eid = {082001},
        pages = {082001},
          doi = {10.1088/1674-4527/acd89e},
archivePrefix = {arXiv},
       eprint = {2305.13305},
 primaryClass = {astro-ph.HE},
       adsurl = {https://ui.adsabs.harvard.edu/abs/2023RAA....23h2001L}
}

@ARTICLE{2014MNRAS.440.1498S,
       author = {{Scalzo}, R. and {Aldering}, G. and {Antilogus}, P. and {Aragon}, C. and {Bailey}, S. and {Baltay}, C. and {Bongard}, S. and {Buton}, C. and {Cellier-Holzem}, F. and {Childress}, M. and {Chotard}, N. and {Copin}, Y. and {Fakhouri}, H.~K. and {Gangler}, E. and {Guy}, J. and {Kim}, A.~G. and {Kowalski}, M. and {Kromer}, M. and {Nordin}, J. and {Nugent}, P. and {Paech}, K. and {Pain}, R. and {Pecontal}, E. and {Pereira}, R. and {Perlmutter}, S. and {Rabinowitz}, D. and {Rigault}, M. and {Runge}, K. and {Saunders}, C. and {Sim}, S.~A. and {Smadja}, G. and {Tao}, C. and {Taubenberger}, S. and {Thomas}, R.~C. and {Weaver}, B.~A. and {Nearby Supernova Factory}},
        title = "{Type Ia supernova bolometric light curves and ejected mass estimates from the Nearby Supernova Factory}",
      journal = {\mnras},
         year = 2014,
        month = may,
       volume = {440},
       number = {2},
        pages = {1498-1518},
          doi = {10.1093/mnras/stu350},
archivePrefix = {arXiv},
       eprint = {1402.6842},
 primaryClass = {astro-ph.CO},
       adsurl = {https://ui.adsabs.harvard.edu/abs/2014MNRAS.440.1498S}
}

@ARTICLE{2014MNRAS.445..711S,
       author = {{Sasdelli}, Michele and {Mazzali}, P.~A. and {Pian}, E. and {Nomoto}, K. and {Hachinger}, S. and {Cappellaro}, E. and {Benetti}, S.},
        title = "{Abundance stratification in Type Ia supernovae - IV. The luminous, peculiar SN 1991T}",
      journal = {\mnras},
         year = 2014,
        month = nov,
       volume = {445},
       number = {1},
        pages = {711-725},
          doi = {10.1093/mnras/stu1777},
archivePrefix = {arXiv},
       eprint = {1409.0116},
 primaryClass = {astro-ph.SR},
       adsurl = {https://ui.adsabs.harvard.edu/abs/2014MNRAS.445..711S}
}

@ARTICLE{1992AJ....103.1632P,
       author = {{Phillips}, M.~M. and {Wells}, Lisa A. and {Suntzeff}, Nicholas B. and {Hamuy}, Mario and {Leibundgut}, Bruno and {Kirshner}, Robert P. and {Foltz}, Craig B.},
        title = "{SN 1991T: Further Evidence of the Heterogeneous Nature of Type IA Supernovae}",
      journal = {\aj},
         year = 1992,
        month = may,
       volume = {103},
        pages = {1632},
          doi = {10.1086/116177},
       adsurl = {https://ui.adsabs.harvard.edu/abs/1992AJ....103.1632P}
}

@ARTICLE{2019TNSTR2470....1S,
       author = {{Stanek}, K.~Z.},
        title = "{ASAS-SN Transient Discovery Report for 2019-11-28}",
      journal = {Transient Name Server Discovery Report},
         year = 2019,
        month = nov,
       volume = {2019-2470},
        pages = {1},
       adsurl = {https://ui.adsabs.harvard.edu/abs/2019TNSTR2470....1S}
}

@ARTICLE{2014ApJ...788...48S,
       author = {{Shappee}, B.~J. and {Prieto}, J.~L. and {Grupe}, D. and {Kochanek}, C.~S. and {Stanek}, K.~Z. and {De Rosa}, G. and {Mathur}, S. and {Zu}, Y. and {Peterson}, B.~M. and {Pogge}, R.~W. and {Komossa}, S. and {Im}, M. and {Jencson}, J. and {Holoien}, T.~W. -S. and {Basu}, U. and {Beacom}, J.~F. and {Szczygie{\l}}, D.~M. and {Brimacombe}, J. and {Adams}, S. and {Campillay}, A. and {Choi}, C. and {Contreras}, C. and {Dietrich}, M. and {Dubberley}, M. and {Elphick}, M. and {Foale}, S. and {Giustini}, M. and {Gonzalez}, C. and {Hawkins}, E. and {Howell}, D.~A. and {Hsiao}, E.~Y. and {Koss}, M. and {Leighly}, K.~M. and {Morrell}, N. and {Mudd}, D. and {Mullins}, D. and {Nugent}, J.~M. and {Parrent}, J. and {Phillips}, M.~M. and {Pojmanski}, G. and {Rosing}, W. and {Ross}, R. and {Sand}, D. and {Terndrup}, D.~M. and {Valenti}, S. and {Walker}, Z. and {Yoon}, Y.},
        title = "{The Man behind the Curtain: X-Rays Drive the UV through NIR Variability in the 2013 Active Galactic Nucleus Outburst in NGC 2617}",
      journal = {\apj},
         year = 2014,
        month = jun,
       volume = {788},
       number = {1},
          eid = {48},
        pages = {48},
          doi = {10.1088/0004-637X/788/1/48},
archivePrefix = {arXiv},
       eprint = {1310.2241},
 primaryClass = {astro-ph.HE},
       adsurl = {https://ui.adsabs.harvard.edu/abs/2014ApJ...788...48S}
}

@ARTICLE{2019TNSCR2483....1I,
       author = {{Ihanec}, N. and {Wevers}, T. and {Callis}, E. and {Gromadzki}, M. and {Yaron}, O.},
        title = "{ePESSTO+ Transient Classification Report for 2019-11-29}",
      journal = {Transient Name Server Classification Report},
         year = 2019,
        month = nov,
       volume = {2019-2483},
        pages = {1},
       adsurl = {https://ui.adsabs.harvard.edu/abs/2019TNSCR2483....1I}
}

@ARTICLE{2019TNSCR2898....1Z,
       author = {{Zhang}, J. and {Xu}, L. and {Wang}, X.},
        title = "{Transient Classification Report for 2019-11-30}",
      journal = {Transient Name Server Classification Report},
         year = 2019,
        month = nov,
       volume = {2019-2898},
        pages = {1},
       adsurl = {https://ui.adsabs.harvard.edu/abs/2019TNSCR2898....1Z}
}

@ARTICLE{2009MNRAS.399..683J,
       author = {{Jones}, D. Heath and {Read}, Mike A. and {Saunders}, Will and {Colless}, Matthew and {Jarrett}, Tom and {Parker}, Quentin A. and {Fairall}, Anthony P. and {Mauch}, Thomas and {Sadler}, Elaine M. and {Watson}, Fred G. and {Burton}, Donna and {Campbell}, Lachlan A. and {Cass}, Paul and {Croom}, Scott M. and {Dawe}, John and {Fiegert}, Kristin and {Frankcombe}, Leela and {Hartley}, Malcolm and {Huchra}, John and {James}, Dionne and {Kirby}, Emma and {Lahav}, Ofer and {Lucey}, John and {Mamon}, Gary A. and {Moore}, Lesa and {Peterson}, Bruce A. and {Prior}, Sayuri and {Proust}, Dominique and {Russell}, Ken and {Safouris}, Vicky and {Wakamatsu}, Ken-Ichi and {Westra}, Eduard and {Williams}, Mary},
        title = "{The 6dF Galaxy Survey: final redshift release (DR3) and southern large-scale structures}",
      journal = {\mnras},
         year = 2009,
        month = oct,
       volume = {399},
       number = {2},
        pages = {683-698},
          doi = {10.1111/j.1365-2966.2009.15338.x},
archivePrefix = {arXiv},
       eprint = {0903.5451},
 primaryClass = {astro-ph.CO},
       adsurl = {https://ui.adsabs.harvard.edu/abs/2009MNRAS.399..683J}
}

@ARTICLE{1992ApJ...384L..15F,
       author = {{Filippenko}, Alexei V. and {Richmond}, Michael W. and {Matheson}, Thomas and {Shields}, Joseph C. and {Burbidge}, E. Margaret and {Cohen}, Ross D. and {Dickinson}, Mark and {Malkan}, Matthew A. and {Nelson}, Brant and {Pietz}, Jochen and {Schlegel}, David and {Schmeer}, Patrick and {Spinrad}, Hyron and {Steidel}, Charles C. and {Tran}, Hien D. and {Wren}, William},
        title = "{The Peculiar Type IA SN 1991T: Detonation of a White Dwarf?}",
      journal = {\apjl},
         year = 1992,
        month = jan,
       volume = {384},
        pages = {L15},
          doi = {10.1086/186252},
       adsurl = {https://ui.adsabs.harvard.edu/abs/1992ApJ...384L..15F}
}

@ARTICLE{2011AJ....141...19B,
       author = {{Burns}, Christopher R. and {Stritzinger}, Maximilian and {Phillips}, M.~M. and {Kattner}, ShiAnne and {Persson}, S.~E. and {Madore}, Barry F. and {Freedman}, Wendy L. and {Boldt}, Luis and {Campillay}, Abdo and {Contreras}, Carlos and {Folatelli}, Gaston and {Gonzalez}, Sergio and {Krzeminski}, Wojtek and {Morrell}, Nidia and {Salgado}, Francisco and {Suntzeff}, Nicholas B.},
        title = "{The Carnegie Supernova Project: Light-curve Fitting with SNooPy}",
      journal = {\aj},
         year = 2011,
        month = jan,
       volume = {141},
       number = {1},
          eid = {19},
        pages = {19},
          doi = {10.1088/0004-6256/141/1/19},
archivePrefix = {arXiv},
       eprint = {1010.4040},
 primaryClass = {astro-ph.CO},
       adsurl = {https://ui.adsabs.harvard.edu/abs/2011AJ....141...19B}
}

@ARTICLE{2022ApJ...938...83Y,
       author = {{Yang}, Jiawen and {Wang}, Lifan and {Suntzeff}, Nicholas and {Hu}, Lei and {Aldoroty}, Lauren and {Brown}, Peter J. and {Krisciunas}, Kevin and {Arcavi}, Iair and {Burke}, Jamison and {Galbany}, Llu{\'\i}s and {Hiramatsu}, Daichi and {Hosseinzadeh}, Griffin and {Howell}, D. Andrew and {McCully}, Curtis and {Pellegrino}, Craig and {Valenti}, Stefano},
        title = "{Using 1991T/1999aa-like Type Ia Supernovae as Standardizable Candles}",
      journal = {\apj},
         year = 2022,
        month = oct,
       volume = {938},
       number = {1},
          eid = {83},
        pages = {83},
          doi = {10.3847/1538-4357/ac8c97},
archivePrefix = {arXiv},
       eprint = {2209.06301},
 primaryClass = {astro-ph.HE},
       adsurl = {https://ui.adsabs.harvard.edu/abs/2022ApJ...938...83Y}
}

@ARTICLE{2000ApJ...539..658K,
       author = {{Krisciunas}, Kevin and {Hastings}, N.~C. and {Loomis}, Karen and {McMillan}, Russet and {Rest}, Armin and {Riess}, Adam G. and {Stubbs}, Christopher},
        title = "{Uniformity of (V-Near-Infrared) Color Evolution of Type Ia Supernovae and Implications for Host Galaxy Extinction Determination}",
      journal = {\apj},
         year = 2000,
        month = aug,
       volume = {539},
       number = {2},
        pages = {658-674},
          doi = {10.1086/309263},
archivePrefix = {arXiv},
       eprint = {astro-ph/9912219},
 primaryClass = {astro-ph},
       adsurl = {https://ui.adsabs.harvard.edu/abs/2000ApJ...539..658K}
}

@ARTICLE{2004AJ....128..387G,
       author = {{Garavini}, G. and {Folatelli}, G. and {Goobar}, A. and {Nobili}, S. and {Aldering}, G. and {Amadon}, A. and {Amanullah}, R. and {Astier}, P. and {Balland}, C. and {Blanc}, G. and {Burns}, M.~S. and {Conley}, A. and {Dahl{\'e}n}, T. and {Deustua}, S.~E. and {Ellis}, R. and {Fabbro}, S. and {Fan}, X. and {Frye}, B. and {Gates}, E.~L. and {Gibbons}, R. and {Goldhaber}, G. and {Goldman}, B. and {Groom}, D.~E. and {Haissinski}, J. and {Hardin}, D. and {Hook}, I.~M. and {Howell}, D.~A. and {Kasen}, D. and {Kent}, S. and {Kim}, A.~G. and {Knop}, R.~A. and {Lee}, B.~C. and {Lidman}, C. and {Mendez}, J. and {Miller}, G.~J. and {Moniez}, M. and {Mour{\~a}o}, A. and {Newberg}, H. and {Nugent}, P.~E. and {Pain}, R. and {Perdereau}, O. and {Perlmutter}, S. and {Prasad}, V. and {Quimby}, R. and {Raux}, J. and {Regnault}, N. and {Rich}, J. and {Richards}, G.~T. and {Ruiz-Lapuente}, P. and {Sainton}, G. and {Schaefer}, B.~E. and {Schahmaneche}, K. and {Smith}, E. and {Spadafora}, A.~L. and {Stanishev}, V. and {Walton}, N.~A. and {Wang}, L. and {Wood-Vasey}, W.~M. and {Supernova Cosmology Project}},
        title = "{Spectroscopic Observations and Analysis of the Peculiar SN 1999aa}",
      journal = {\aj},
         year = 2004,
        month = jul,
       volume = {128},
       number = {1},
        pages = {387-404},
          doi = {10.1086/421747},
archivePrefix = {arXiv},
       eprint = {astro-ph/0404393},
 primaryClass = {astro-ph},
       adsurl = {https://ui.adsabs.harvard.edu/abs/2004AJ....128..387G}
}

@ARTICLE{2001PASP..113.1420V,
       author = {{van Dokkum}, Pieter G.},
        title = "{Cosmic-Ray Rejection by Laplacian Edge Detection}",
      journal = {\pasp},
         year = 2001,
        month = nov,
       volume = {113},
       number = {789},
        pages = {1420-1427},
          doi = {10.1086/323894},
archivePrefix = {arXiv},
       eprint = {astro-ph/0108003},
 primaryClass = {astro-ph},
       adsurl = {https://ui.adsabs.harvard.edu/abs/2001PASP..113.1420V}
}

@INPROCEEDINGS{2007ASPC..364..503F,
       author = {{Fossati}, L. and {Bagnulo}, S. and {Mason}, E. and {Landi Degl'Innocenti}, E.},
        title = "{Standard Stars for Linear Polarization Observed with FORS1}",
    booktitle = {The Future of Photometric, Spectrophotometric and Polarimetric Standardization},
         year = 2007,
       editor = {{Sterken}, C.},
       series = {Astronomical Society of the Pacific Conference Series},
       volume = {364},
        month = apr,
        pages = {503},
       adsurl = {https://ui.adsabs.harvard.edu/abs/2007ASPC..364..503F}
}

@ARTICLE{2014A&A...561A..82S,
       author = {{Siebenmorgen}, R. and {Voshchinnikov}, N.~V. and {Bagnulo}, S.},
        title = "{Dust in the diffuse interstellar medium. Extinction, emission, linear and circular polarisation}",
      journal = {\aap},
         year = 2014,
        month = jan,
       volume = {561},
          eid = {A82},
        pages = {A82},
          doi = {10.1051/0004-6361/201321716},
archivePrefix = {arXiv},
       eprint = {1308.3148},
 primaryClass = {astro-ph.GA},
       adsurl = {https://ui.adsabs.harvard.edu/abs/2014A&A...561A..82S}
}

@INPROCEEDINGS{2018SPIE10707E..0KM,
       author = {{McCully}, Curtis and {Volgenau}, Nikolaus H. and {Harbeck}, Daniel-Rolf and {Lister}, Tim A. and {Saunders}, Eric S. and {Turner}, Monica L. and {Siiverd}, Robert J. and {Bowman}, Mark},
        title = "{Real-time processing of the imaging data from the network of Las Cumbres Observatory Telescopes using BANZAI}",
    booktitle = {Software and Cyberinfrastructure for Astronomy V},
         year = 2018,
       editor = {{Guzman}, Juan C. and {Ibsen}, Jorge},
       series = {Society of Photo-Optical Instrumentation Engineers (SPIE) Conference Series},
       volume = {10707},
        month = jul,
          eid = {107070K},
        pages = {107070K},
          doi = {10.1117/12.2314340},
archivePrefix = {arXiv},
       eprint = {1811.04163},
 primaryClass = {astro-ph.IM},
       adsurl = {https://ui.adsabs.harvard.edu/abs/2018SPIE10707E..0KM}
}

@INPROCEEDINGS{2015AAS...22533616H,
       author = {{Henden}, Arne A. and {Levine}, Stephen and {Terrell}, Dirk and {Welch}, Douglas L.},
        title = "{APASS - The Latest Data Release}",
    booktitle = {American Astronomical Society Meeting Abstracts \#225},
         year = 2015,
       series = {American Astronomical Society Meeting Abstracts},
       volume = {225},
        month = jan,
          eid = {336.16},
        pages = {336.16},
       adsurl = {https://ui.adsabs.harvard.edu/abs/2015AAS...22533616H},
       editor = {American Astronomical Society},
       publisher = {American Astronomical Society}
}

@ARTICLE{1966CoLPL...4...99J,
       author = {{Johnson}, H.~L. and {Mitchell}, R.~I. and {Iriarte}, B. and {Wisniewski}, W.~Z.},
        title = "{UBVRIJKL Photometry of the Bright Stars}",
      journal = {Communications of the Lunar and Planetary Laboratory},
         year = 1966,
        month = jan,
       volume = {4},
        pages = {99-110},
       adsurl = {https://ui.adsabs.harvard.edu/abs/1966CoLPL...4...99J}
}

@ARTICLE{1996AJ....111.1748F,
       author = {{Fukugita}, M. and {Ichikawa}, T. and {Gunn}, J.~E. and {Doi}, M. and {Shimasaku}, K. and {Schneider}, D.~P.},
        title = "{The Sloan Digital Sky Survey Photometric System}",
      journal = {\aj},
         year = 1996,
        month = apr,
       volume = {111},
        pages = {1748},
          doi = {10.1086/117915},
       adsurl = {https://ui.adsabs.harvard.edu/abs/1996AJ....111.1748F}
}

@ARTICLE{1983ApJ...266..713O,
       author = {{Oke}, J.~B. and {Gunn}, J.~E.},
        title = "{Secondary standard stars for absolute spectrophotometry.}",
      journal = {\apj},
         year = 1983,
        month = mar,
       volume = {266},
        pages = {713-717},
          doi = {10.1086/160817},
       adsurl = {https://ui.adsabs.harvard.edu/abs/1983ApJ...266..713O}
}

@ARTICLE{2007ApJ...663.1187H,
       author = {{Hsiao}, E.~Y. and {Conley}, A. and {Howell}, D.~A. and {Sullivan}, M. and {Pritchet}, C.~J. and {Carlberg}, R.~G. and {Nugent}, P.~E. and {Phillips}, M.~M.},
        title = "{K-Corrections and Spectral Templates of Type Ia Supernovae}",
      journal = {\apj},
         year = 2007,
        month = jul,
       volume = {663},
       number = {2},
        pages = {1187-1200},
          doi = {10.1086/518232},
archivePrefix = {arXiv},
       eprint = {astro-ph/0703529},
 primaryClass = {astro-ph},
       adsurl = {https://ui.adsabs.harvard.edu/abs/2007ApJ...663.1187H}
}

@ARTICLE{2016AJ....152..102B,
       author = {{Brown}, Peter J. and {Breeveld}, Alice and {Roming}, Peter W.~A. and {Siegel}, Michael},
        title = "{Interpreting Flux from Broadband Photometry}",
      journal = {\aj},
         year = 2016,
        month = oct,
       volume = {152},
       number = {4},
          eid = {102},
        pages = {102},
          doi = {10.3847/0004-6256/152/4/102},
archivePrefix = {arXiv},
       eprint = {1608.02599},
 primaryClass = {astro-ph.IM},
       adsurl = {https://ui.adsabs.harvard.edu/abs/2016AJ....152..102B}
}

@ARTICLE{2022ApJ...927...78D,
       author = {{Dimitriadis}, Georgios and {Foley}, Ryan J. and {Arendse}, Nikki and {Coulter}, David A. and {Jacobson-Gal{\'a}n}, Wynn V. and {Siebert}, Matthew R. and {Izzo}, Luca and {Jones}, David O. and {Kilpatrick}, Charles D. and {Pan}, Yen-Chen and {Taggart}, Kirsty and {Auchettl}, Katie and {Gall}, Christa and {Hjorth}, Jens and {Kasen}, Daniel and {Piro}, Anthony L. and {Raimundo}, Sandra I. and {Ramirez-Ruiz}, Enrico and {Rest}, Armin and {Swift}, Jonathan J. and {Woosley}, Stan E.},
        title = "{A Carbon/Oxygen-dominated Atmosphere Days after Explosion for the ``Super-Chandrasekhar'' Type Ia SN 2020esm}",
      journal = {\apj},
         year = 2022,
        month = mar,
       volume = {927},
       number = {1},
          eid = {78},
        pages = {78},
          doi = {10.3847/1538-4357/ac4780},
archivePrefix = {arXiv},
       eprint = {2112.09930},
 primaryClass = {astro-ph.HE},
       adsurl = {https://ui.adsabs.harvard.edu/abs/2022ApJ...927...78D}
}

@ARTICLE{2011MNRAS.412.2735T,
       author = {{Taubenberger}, S. and {Benetti}, S. and {Childress}, M. and {Pakmor}, R. and {Hachinger}, S. and {Mazzali}, P.~A. and {Stanishev}, V. and {Elias-Rosa}, N. and {Agnoletto}, I. and {Bufano}, F. and {Ergon}, M. and {Harutyunyan}, A. and {Inserra}, C. and {Kankare}, E. and {Kromer}, M. and {Navasardyan}, H. and {Nicolas}, J. and {Pastorello}, A. and {Prosperi}, E. and {Salgado}, F. and {Sollerman}, J. and {Stritzinger}, M. and {Turatto}, M. and {Valenti}, S. and {Hillebrandt}, W.},
        title = "{High luminosity, slow ejecta and persistent carbon lines: SN 2009dc challenges thermonuclear explosion scenarios}",
      journal = {\mnras},
         year = 2011,
        month = apr,
       volume = {412},
       number = {4},
        pages = {2735-2762},
          doi = {10.1111/j.1365-2966.2010.18107.x},
archivePrefix = {arXiv},
       eprint = {1011.5665},
 primaryClass = {astro-ph.SR},
       adsurl = {https://ui.adsabs.harvard.edu/abs/2011MNRAS.412.2735T}
}

@ARTICLE{2014Ap&SS.354...89B,
       author = {{Brown}, Peter J. and {Breeveld}, Alice A. and {Holland}, Stephen and {Kuin}, Paul and {Pritchard}, Tyler},
        title = "{SOUSA: the Swift Optical/Ultraviolet Supernova Archive}",
      journal = {\apss},
         year = 2014,
        month = nov,
       volume = {354},
       number = {1},
        pages = {89-96},
          doi = {10.1007/s10509-014-2059-8},
archivePrefix = {arXiv},
       eprint = {1407.3808},
 primaryClass = {astro-ph.HE},
       adsurl = {https://ui.adsabs.harvard.edu/abs/2014Ap&SS.354...89B}
}

@ARTICLE{2011MNRAS.410..585S,
       author = {{Silverman}, Jeffrey M. and {Ganeshalingam}, Mohan and {Li}, Weidong and {Filippenko}, Alexei V. and {Miller}, Adam A. and {Poznanski}, Dovi},
        title = "{Fourteen months of observations of the possible super-Chandrasekhar mass Type Ia Supernova 2009dc}",
      journal = {\mnras},
         year = 2011,
        month = jan,
       volume = {410},
       number = {1},
        pages = {585-611},
          doi = {10.1111/j.1365-2966.2010.17474.x},
archivePrefix = {arXiv},
       eprint = {1003.2417},
 primaryClass = {astro-ph.HE},
       adsurl = {https://ui.adsabs.harvard.edu/abs/2011MNRAS.410..585S}
}

@ARTICLE{2015ApJS..220....9F,
       author = {{Friedman}, Andrew S. and {Wood-Vasey}, W.~M. and {Marion}, G.~H. and {Challis}, Peter and {Mandel}, Kaisey S. and {Bloom}, Joshua S. and {Modjaz}, Maryam and {Narayan}, Gautham and {Hicken}, Malcolm and {Foley}, Ryan J. and {Klein}, Christopher R. and {Starr}, Dan L. and {Morgan}, Adam and {Rest}, Armin and {Blake}, Cullen H. and {Miller}, Adam A. and {Falco}, Emilio E. and {Wyatt}, William F. and {Mink}, Jessica and {Skrutskie}, Michael F. and {Kirshner}, Robert P.},
        title = "{CfAIR2: Near-infrared Light Curves of 94 Type Ia Supernovae}",
      journal = {\apjs},
         year = 2015,
        month = sep,
       volume = {220},
       number = {1},
          eid = {9},
        pages = {9},
          doi = {10.1088/0067-0049/220/1/9},
archivePrefix = {arXiv},
       eprint = {1408.0465},
 primaryClass = {astro-ph.HE},
       adsurl = {https://ui.adsabs.harvard.edu/abs/2015ApJS..220....9F}
}

@ARTICLE{2012ApJ...757...12S,
       author = {{Scalzo}, R. and {Aldering}, G. and {Antilogus}, P. and {Aragon}, C. and {Bailey}, S. and {Baltay}, C. and {Bongard}, S. and {Buton}, C. and {Canto}, A. and {Cellier-Holzem}, F. and {Childress}, M. and {Chotard}, N. and {Copin}, Y. and {Fakhouri}, H.~K. and {Gangler}, E. and {Guy}, J. and {Hsiao}, E.~Y. and {Kerschhaggl}, M. and {Kowalski}, M. and {Nugent}, P. and {Paech}, K. and {Pain}, R. and {Pecontal}, E. and {Pereira}, R. and {Perlmutter}, S. and {Rabinowitz}, D. and {Rigault}, M. and {Runge}, K. and {Smadja}, G. and {Tao}, C. and {Thomas}, R.~C. and {Weaver}, B.~A. and {Wu}, C. and {Nearby Supernova Factory}, The},
        title = "{A Search for New Candidate Super-Chandrasekhar-mass Type Ia Supernovae in the Nearby Supernova Factory Data Set}",
      journal = {\apj},
         year = 2012,
        month = sep,
       volume = {757},
       number = {1},
          eid = {12},
        pages = {12},
          doi = {10.1088/0004-637X/757/1/12},
archivePrefix = {arXiv},
       eprint = {1207.2695},
 primaryClass = {astro-ph.CO},
       adsurl = {https://ui.adsabs.harvard.edu/abs/2012ApJ...757...12S}
}

@ARTICLE{2009MNRAS.400..531S,
       author = {{Seitenzahl}, I.~R. and {Taubenberger}, S. and {Sim}, S.~A.},
        title = "{Late-time supernova light curves: the effect of internal conversion and Auger electrons}",
      journal = {\mnras},
         year = 2009,
        month = nov,
       volume = {400},
       number = {1},
        pages = {531-535},
          doi = {10.1111/j.1365-2966.2009.15478.x},
archivePrefix = {arXiv},
       eprint = {0908.0247},
 primaryClass = {astro-ph.SR},
       adsurl = {https://ui.adsabs.harvard.edu/abs/2009MNRAS.400..531S}
}

@ARTICLE{2014ApJ...792...10S,
       author = {{Seitenzahl}, Ivo R. and {Timmes}, F.~X. and {Magkotsios}, Georgios},
        title = "{The Light Curve of SN 1987A Revisited: Constraining Production Masses of Radioactive Nuclides}",
      journal = {\apj},
         year = 2014,
        month = sep,
       volume = {792},
       number = {1},
          eid = {10},
        pages = {10},
          doi = {10.1088/0004-637X/792/1/10},
archivePrefix = {arXiv},
       eprint = {1408.5986},
 primaryClass = {astro-ph.SR},
       adsurl = {https://ui.adsabs.harvard.edu/abs/2014ApJ...792...10S}
}

@ARTICLE{1989ApJ...345..245C,
       author = {{Cardelli}, Jason A. and {Clayton}, Geoffrey C. and {Mathis}, John S.},
        title = "{The Relationship between Infrared, Optical, and Ultraviolet Extinction}",
      journal = {\apj},
         year = 1989,
        month = oct,
       volume = {345},
        pages = {245},
          doi = {10.1086/167900},
       adsurl = {https://ui.adsabs.harvard.edu/abs/1989ApJ...345..245C}
}

@ARTICLE{2011ApJ...737..103S,
       author = {{Schlafly}, Edward F. and {Finkbeiner}, Douglas P.},
        title = "{Measuring Reddening with Sloan Digital Sky Survey Stellar Spectra and Recalibrating SFD}",
      journal = {\apj},
         year = 2011,
        month = aug,
       volume = {737},
       number = {2},
          eid = {103},
        pages = {103},
          doi = {10.1088/0004-637X/737/2/103},
archivePrefix = {arXiv},
       eprint = {1012.4804},
 primaryClass = {astro-ph.GA},
       adsurl = {https://ui.adsabs.harvard.edu/abs/2011ApJ...737..103S}
}

@ARTICLE{2011AJ....142..156S,
       author = {{Stritzinger}, Maximilian D. and {Phillips}, M.~M. and {Boldt}, Luis N. and {Burns}, Chris and {Campillay}, Abdo and {Contreras}, Carlos and {Gonzalez}, Sergio and {Folatelli}, Gast{\'o}n and {Morrell}, Nidia and {Krzeminski}, Wojtek and {Roth}, Miguel and {Salgado}, Francisco and {DePoy}, D.~L. and {Hamuy}, Mario and {Freedman}, Wendy L. and {Madore}, Barry F. and {Marshall}, J.~L. and {Persson}, Sven E. and {Rheault}, Jean-Philippe and {Suntzeff}, Nicholas B. and {Villanueva}, Steven and {Li}, Weidong and {Filippenko}, Alexei V.},
        title = "{The Carnegie Supernova Project: Second Photometry Data Release of Low-redshift Type Ia Supernovae}",
      journal = {\aj},
         year = 2011,
        month = nov,
       volume = {142},
       number = {5},
          eid = {156},
        pages = {156},
          doi = {10.1088/0004-6256/142/5/156},
archivePrefix = {arXiv},
       eprint = {1108.3108},
 primaryClass = {astro-ph.CO},
       adsurl = {https://ui.adsabs.harvard.edu/abs/2011AJ....142..156S}
}

@ARTICLE{2017AJ....154..211K,
       author = {{Krisciunas}, Kevin and {Contreras}, Carlos and {Burns}, Christopher R. and {Phillips}, M.~M. and {Stritzinger}, Maximilian D. and {Morrell}, Nidia and {Hamuy}, Mario and {Anais}, Jorge and {Boldt}, Luis and {Busta}, Luis and {Campillay}, Abdo and {Castell{\'o}n}, Sergio and {Folatelli}, Gast{\'o}n and {Freedman}, Wendy L. and {Gonz{\'a}lez}, Consuelo and {Hsiao}, Eric Y. and {Krzeminski}, Wojtek and {Persson}, Sven Eric and {Roth}, Miguel and {Salgado}, Francisco and {Ser{\'o}n}, Jacqueline and {Suntzeff}, Nicholas B. and {Torres}, Sim{\'o}n and {Filippenko}, Alexei V. and {Li}, Weidong and {Madore}, Barry F. and {DePoy}, D.~L. and {Marshall}, Jennifer L. and {Rheault}, Jean-Philippe and {Villanueva}, Steven},
        title = "{The Carnegie Supernova Project. I. Third Photometry Data Release of Low-redshift Type Ia Supernovae and Other White Dwarf Explosions}",
      journal = {\aj},
         year = 2017,
        month = nov,
       volume = {154},
       number = {5},
          eid = {211},
        pages = {211},
          doi = {10.3847/1538-3881/aa8df0},
archivePrefix = {arXiv},
       eprint = {1709.05146},
 primaryClass = {astro-ph.IM},
       adsurl = {https://ui.adsabs.harvard.edu/abs/2017AJ....154..211K}
}

@ARTICLE{2011MNRAS.416.2607G,
       author = {{Ganeshalingam}, Mohan and {Li}, Weidong and {Filippenko}, Alexei V.},
        title = "{The rise-time distribution of nearby Type Ia supernovae}",
      journal = {\mnras},
         year = 2011,
        month = oct,
       volume = {416},
       number = {4},
        pages = {2607-2622},
          doi = {10.1111/j.1365-2966.2011.19213.x},
archivePrefix = {arXiv},
       eprint = {1107.2404},
 primaryClass = {astro-ph.CO},
       adsurl = {https://ui.adsabs.harvard.edu/abs/2011MNRAS.416.2607G}
}

@ARTICLE{1994ApJS...92..527N,
       author = {{Nadyozhin}, D.~K.},
        title = "{The Properties of NI CO Fe Decay}",
      journal = {\apjs},
         year = 1994,
        month = jun,
       volume = {92},
        pages = {527},
          doi = {10.1086/192008},
       adsurl = {https://ui.adsabs.harvard.edu/abs/1994ApJS...92..527N}
}

@ARTICLE{1995A&A...297..509M,
       author = {{Mazzali}, P.~A. and {Danziger}, I.~J. and {Turatto}, M.},
        title = "{A study of the properties of the peculiar SN IA 1991T through models of its evolving early-time spectrum.}",
      journal = {\aap},
         year = 1995,
        month = may,
       volume = {297},
        pages = {509},
       adsurl = {https://ui.adsabs.harvard.edu/abs/1995A&A...297..509M}
}

@ARTICLE{2024ApJS..273...16P,
       author = {{Phillips}, M.~M. and {Ashall}, C. and {Brown}, Peter J. and {Galbany}, L. and {Tucker}, M.~A. and {Burns}, Christopher R. and {Contreras}, Carlos and {Hoeflich}, P. and {Hsiao}, E.~Y. and {Kumar}, S. and {Morrell}, Nidia and {Uddin}, Syed A. and {Baron}, E. and {Freedman}, Wendy L. and {Krisciunas}, Kevin and {Persson}, S.~E. and {Piro}, Anthony L. and {Shappee}, B.~J. and {Stritzinger}, Maximilian and {Suntzeff}, Nicholas B. and {Chakraborty}, Sudeshna and {Kirshner}, R.~P. and {Lu}, J. and {Marion}, G.~H. and {Polin}, Abigail and {Shahbandeh}, M.},
        title = "{1991T-like Supernovae}",
      journal = {\apjs},
         year = 2024,
        month = jul,
       volume = {273},
       number = {1},
          eid = {16},
        pages = {16},
          doi = {10.3847/1538-4365/ad4f7e},
archivePrefix = {arXiv},
       eprint = {2405.15027},
 primaryClass = {astro-ph.HE},
       adsurl = {https://ui.adsabs.harvard.edu/abs/2024ApJS..273...16P}
}

@ARTICLE{2013MNRAS.430.1030S,
       author = {{Silverman}, Jeffrey M. and {Ganeshalingam}, Mohan and {Filippenko}, Alexei V.},
        title = "{Berkeley Supernova Ia Program - V. Late-time spectra of Type Ia Supernovae}",
      journal = {\mnras},
         year = 2013,
        month = apr,
       volume = {430},
       number = {2},
        pages = {1030-1041},
          doi = {10.1093/mnras/sts674},
archivePrefix = {arXiv},
       eprint = {1211.0279},
 primaryClass = {astro-ph.CO},
       adsurl = {https://ui.adsabs.harvard.edu/abs/2013MNRAS.430.1030S}
}

@ARTICLE{2012MNRAS.425.1789S,
       author = {{Silverman}, Jeffrey M. and {Foley}, Ryan J. and {Filippenko}, Alexei V. and {Ganeshalingam}, Mohan and {Barth}, Aaron J. and {Chornock}, Ryan and {Griffith}, Christopher V. and {Kong}, Jason J. and {Lee}, Nicholas and {Leonard}, Douglas C. and {Matheson}, Thomas and {Miller}, Emily G. and {Steele}, Thea N. and {Barris}, Brian J. and {Bloom}, Joshua S. and {Cobb}, Bethany E. and {Coil}, Alison L. and {Desroches}, Louis-Benoit and {Gates}, Elinor L. and {Ho}, Luis C. and {Jha}, Saurabh W. and {Kandrashoff}, Michael T. and {Li}, Weidong and {Mandel}, Kaisey S. and {Modjaz}, Maryam and {Moore}, Matthew R. and {Mostardi}, Robin E. and {Papenkova}, Marina S. and {Park}, Sung and {Perley}, Daniel A. and {Poznanski}, Dovi and {Reuter}, Cassie A. and {Scala}, James and {Serduke}, Franklin J.~D. and {Shields}, Joseph C. and {Swift}, Brandon J. and {Tonry}, John L. and {Van Dyk}, Schuyler D. and {Wang}, Xiaofeng and {Wong}, Diane S.},
        title = "{Berkeley Supernova Ia Program - I. Observations, data reduction and spectroscopic sample of 582 low-redshift Type Ia supernovae}",
      journal = {\mnras},
         year = 2012,
        month = sep,
       volume = {425},
       number = {3},
        pages = {1789-1818},
          doi = {10.1111/j.1365-2966.2012.21270.x},
archivePrefix = {arXiv},
       eprint = {1202.2128},
 primaryClass = {astro-ph.CO},
       adsurl = {https://ui.adsabs.harvard.edu/abs/2012MNRAS.425.1789S}
}

@ARTICLE{2013A&A...554A..27P,
       author = {{Pereira}, R. and {Thomas}, R.~C. and {Aldering}, G. and {Antilogus}, P. and {Baltay}, C. and {Benitez-Herrera}, S. and {Bongard}, S. and {Buton}, C. and {Canto}, A. and {Cellier-Holzem}, F. and {Chen}, J. and {Childress}, M. and {Chotard}, N. and {Copin}, Y. and {Fakhouri}, H.~K. and {Fink}, M. and {Fouchez}, D. and {Gangler}, E. and {Guy}, J. and {Hillebrandt}, W. and {Hsiao}, E.~Y. and {Kerschhaggl}, M. and {Kowalski}, M. and {Kromer}, M. and {Nordin}, J. and {Nugent}, P. and {Paech}, K. and {Pain}, R. and {P{\'e}contal}, E. and {Perlmutter}, S. and {Rabinowitz}, D. and {Rigault}, M. and {Runge}, K. and {Saunders}, C. and {Smadja}, G. and {Tao}, C. and {Taubenberger}, S. and {Tilquin}, A. and {Wu}, C.},
        title = "{Spectrophotometric time series of SN 2011fe from the Nearby Supernova Factory}",
      journal = {\aap},
         year = 2013,
        month = jun,
       volume = {554},
          eid = {A27},
        pages = {A27},
          doi = {10.1051/0004-6361/201221008},
archivePrefix = {arXiv},
       eprint = {1302.1292},
 primaryClass = {astro-ph.CO},
       adsurl = {https://ui.adsabs.harvard.edu/abs/2013A&A...554A..27P}
}

@ARTICLE{2007ApJ...656..661K,
       author = {{Kasen}, Daniel and {Woosley}, S.~E.},
        title = "{On the Origin of the Type Ia Supernova Width-Luminosity Relation}",
      journal = {\apj},
         year = 2007,
        month = feb,
       volume = {656},
       number = {2},
        pages = {661-665},
          doi = {10.1086/510375},
archivePrefix = {arXiv},
       eprint = {astro-ph/0609540},
 primaryClass = {astro-ph},
       adsurl = {https://ui.adsabs.harvard.edu/abs/2007ApJ...656..661K}
}

@ARTICLE{2006PASP..118..722C,
       author = {{Chornock}, Ryan and {Filippenko}, Alexei V. and {Branch}, David and {Foley}, Ryan J. and {Jha}, Saurabh and {Li}, Weidong},
        title = "{Spectropolarimetry of the Peculiar Type Ia Supernova 2005hk}",
      journal = {\pasp},
         year = 2006,
        month = may,
       volume = {118},
       number = {843},
        pages = {722-732},
          doi = {10.1086/504117},
archivePrefix = {arXiv},
       eprint = {astro-ph/0603083},
 primaryClass = {astro-ph},
       adsurl = {https://ui.adsabs.harvard.edu/abs/2006PASP..118..722C}
}

@ARTICLE{2009A&A...508..229P,
       author = {{Patat}, F. and {Baade}, D. and {H{\"o}flich}, P. and {Maund}, J.~R. and {Wang}, L. and {Wheeler}, J.~C.},
        title = "{VLT spectropolarimetry of the fast expanding type Ia SN 2006X}",
      journal = {\aap},
         year = 2009,
        month = dec,
       volume = {508},
       number = {1},
        pages = {229-246},
          doi = {10.1051/0004-6361/200810651},
archivePrefix = {arXiv},
       eprint = {0909.5564},
 primaryClass = {astro-ph.SR},
       adsurl = {https://ui.adsabs.harvard.edu/abs/2009A&A...508..229P}
}

@ARTICLE{1975ApJ...196..261S,
       author = {{Serkowski}, K. and {Mathewson}, D.~S. and {Ford}, V.~L.},
        title = "{Wavelength dependence of interstellar polarization and ratio of total to selective extinction.}",
      journal = {\apj},
         year = 1975,
        month = feb,
       volume = {196},
        pages = {261-290},
          doi = {10.1086/153410},
       adsurl = {https://ui.adsabs.harvard.edu/abs/1975ApJ...196..261S}
}

@ARTICLE{2023MNRAS.519.1618Y,
       author = {{Yang}, Yi and {Baade}, Dietrich and {Hoeflich}, Peter and {Wang}, Lifan and {Cikota}, Aleksandar and {Chen}, Ting-Wan and {Burke}, Jamison and {Hiramatsu}, Daichi and {Pellegrino}, Craig and {Howell}, D. Andrew and {McCully}, Curtis and {Valenti}, Stefano and {Schulze}, Steve and {Gal-Yam}, Avishay and {Wang}, Lingzhi and {Filippenko}, Alexei V. and {Maeda}, Keiichi and {Bulla}, Mattia and {Yao}, Yuhan and {Maund}, Justyn R. and {Patat}, Ferdinando and {Spyromilio}, Jason and {Wheeler}, J. Craig and {Rau}, Arne and {Hu}, Lei and {Li}, Wenxiong and {Andrews}, Jennifer E. and {Galbany}, Ll{\'u}is and {Sand}, David J. and {Shahbandeh}, Melissa and {Hsiao}, Eric Y. and {Wang}, Xiaofeng},
        title = "{The interaction of supernova 2018evt with a substantial amount of circumstellar matter - An SN 1997cy-like event}",
      journal = {\mnras},
         year = 2023,
        month = feb,
       volume = {519},
       number = {2},
        pages = {1618-1647},
          doi = {10.1093/mnras/stac3477},
archivePrefix = {arXiv},
       eprint = {2211.04423},
 primaryClass = {astro-ph.HE},
       adsurl = {https://ui.adsabs.harvard.edu/abs/2023MNRAS.519.1618Y}
}

@ARTICLE{1995ApJ...440..821H,
       author = {{H{\"o}flich}, Peter},
        title = "{Remarks on the Polarization Observed in SN 1993J}",
      journal = {\apj},
         year = 1995,
        month = feb,
       volume = {440},
        pages = {821},
          doi = {10.1086/175317},
       adsurl = {https://ui.adsabs.harvard.edu/abs/1995ApJ...440..821H}
}

@ARTICLE{1995ApJ...444..831H,
       author = {{Hoeflich}, P. and {Khokhlov}, A.~M. and {Wheeler}, J.~C.},
        title = "{Delayed Detonation Models for Normal and Subluminous Type IA Supernovae: Absolute Brightness, Light Curves, and Molecule Formation}",
      journal = {\apj},
         year = 1995,
        month = may,
       volume = {444},
        pages = {831},
          doi = {10.1086/175656},
       adsurl = {https://ui.adsabs.harvard.edu/abs/1995ApJ...444..831H}
}

@ARTICLE{2013MNRAS.433.2240G,
       author = {{Ganeshalingam}, Mohan and {Li}, Weidong and {Filippenko}, Alexei V.},
        title = "{Constraints on dark energy with the LOSS SN Ia sample}",
      journal = {\mnras},
         year = 2013,
        month = aug,
       volume = {433},
       number = {3},
        pages = {2240-2258},
          doi = {10.1093/mnras/stt893},
archivePrefix = {arXiv},
       eprint = {1307.0824},
 primaryClass = {astro-ph.CO},
       adsurl = {https://ui.adsabs.harvard.edu/abs/2013MNRAS.433.2240G}
}

@ARTICLE{2016A&A...594A..13P,
       author = {{Planck Collaboration} and {Ade}, P.~A.~R. and {Aghanim}, N. and {Arnaud}, M. and {Ashdown}, M. and {Aumont}, J. and {Baccigalupi}, C. and {Banday}, A.~J. and {Barreiro}, R.~B. and {Bartlett}, J.~G. and {Bartolo}, N. and {Battaner}, E. and {Battye}, R. and {Benabed}, K. and {Beno{\^\i}t}, A. and {Benoit-L{\'e}vy}, A. and {Bernard}, J. -P. and {Bersanelli}, M. and {Bielewicz}, P. and {Bock}, J.~J. and {Bonaldi}, A. and {Bonavera}, L. and {Bond}, J.~R. and {Borrill}, J. and {Bouchet}, F.~R. and {Boulanger}, F. and {Bucher}, M. and {Burigana}, C. and {Butler}, R.~C. and {Calabrese}, E. and {Cardoso}, J. -F. and {Catalano}, A. and {Challinor}, A. and {Chamballu}, A. and {Chary}, R. -R. and {Chiang}, H.~C. and {Chluba}, J. and {Christensen}, P.~R. and {Church}, S. and {Clements}, D.~L. and {Colombi}, S. and {Colombo}, L.~P.~L. and {Combet}, C. and {Coulais}, A. and {Crill}, B.~P. and {Curto}, A. and {Cuttaia}, F. and {Danese}, L. and {Davies}, R.~D. and {Davis}, R.~J. and {de Bernardis}, P. and {de Rosa}, A. and {de Zotti}, G. and {Delabrouille}, J. and {D{\'e}sert}, F. -X. and {Di Valentino}, E. and {Dickinson}, C. and {Diego}, J.~M. and {Dolag}, K. and {Dole}, H. and {Donzelli}, S. and {Dor{\'e}}, O. and {Douspis}, M. and {Ducout}, A. and {Dunkley}, J. and {Dupac}, X. and {Efstathiou}, G. and {Elsner}, F. and {En{\ss}lin}, T.~A. and {Eriksen}, H.~K. and {Farhang}, M. and {Fergusson}, J. and {Finelli}, F. and {Forni}, O. and {Frailis}, M. and {Fraisse}, A.~A. and {Franceschi}, E. and {Frejsel}, A. and {Galeotta}, S. and {Galli}, S. and {Ganga}, K. and {Gauthier}, C. and {Gerbino}, M. and {Ghosh}, T. and {Giard}, M. and {Giraud-H{\'e}raud}, Y. and {Giusarma}, E. and {Gjerl{\o}w}, E. and {Gonz{\'a}lez-Nuevo}, J. and {G{\'o}rski}, K.~M. and {Gratton}, S. and {Gregorio}, A. and {Gruppuso}, A. and {Gudmundsson}, J.~E. and {Hamann}, J. and {Hansen}, F.~K. and {Hanson}, D. and {Harrison}, D.~L. and {Helou}, G. and {Henrot-Versill{\'e}}, S. and {Hern{\'a}ndez-Monteagudo}, C. and {Herranz}, D. and {Hildebrandt}, S.~R. and {Hivon}, E. and {Hobson}, M. and {Holmes}, W.~A. and {Hornstrup}, A. and {Hovest}, W. and {Huang}, Z. and {Huffenberger}, K.~M. and {Hurier}, G. and {Jaffe}, A.~H. and {Jaffe}, T.~R. and {Jones}, W.~C. and {Juvela}, M. and {Keih{\"a}nen}, E. and {Keskitalo}, R. and {Kisner}, T.~S. and {Kneissl}, R. and {Knoche}, J. and {Knox}, L. and {Kunz}, M. and {Kurki-Suonio}, H. and {Lagache}, G. and {L{\"a}hteenm{\"a}ki}, A. and {Lamarre}, J. -M. and {Lasenby}, A. and {Lattanzi}, M. and {Lawrence}, C.~R. and {Leahy}, J.~P. and {Leonardi}, R. and {Lesgourgues}, J. and {Levrier}, F. and {Lewis}, A. and {Liguori}, M. and {Lilje}, P.~B. and {Linden-V{\o}rnle}, M. and {L{\'o}pez-Caniego}, M. and {Lubin}, P.~M. and {Mac{\'\i}as-P{\'e}rez}, J.~F. and {Maggio}, G. and {Maino}, D. and {Mandolesi}, N. and {Mangilli}, A. and {Marchini}, A. and {Maris}, M. and {Martin}, P.~G. and {Martinelli}, M. and {Mart{\'\i}nez-Gonz{\'a}lez}, E. and {Masi}, S. and {Matarrese}, S. and {McGehee}, P. and {Meinhold}, P.~R. and {Melchiorri}, A. and {Melin}, J. -B. and {Mendes}, L. and {Mennella}, A. and {Migliaccio}, M. and {Millea}, M. and {Mitra}, S. and {Miville-Desch{\^e}nes}, M. -A. and {Moneti}, A. and {Montier}, L. and {Morgante}, G. and {Mortlock}, D. and {Moss}, A. and {Munshi}, D. and {Murphy}, J.~A. and {Naselsky}, P. and {Nati}, F. and {Natoli}, P. and {Netterfield}, C.~B. and {N{\o}rgaard-Nielsen}, H.~U. and {Noviello}, F. and {Novikov}, D. and {Novikov}, I. and {Oxborrow}, C.~A. and {Paci}, F. and {Pagano}, L. and {Pajot}, F. and {Paladini}, R. and {Paoletti}, D. and {Partridge}, B. and {Pasian}, F. and {Patanchon}, G. and {Pearson}, T.~J. and {Perdereau}, O. and {Perotto}, L. and {Perrotta}, F. and {Pettorino}, V. and {Piacentini}, F. and {Piat}, M. and {Pierpaoli}, E. and {Pietrobon}, D. and {Plaszczynski}, S. and {Pointecouteau}, E. and {Polenta}, G. and {Popa}, L. and {Pratt}, G.~W. and {Pr{\'e}zeau}, G.},
        title = "{Planck 2015 results. XIII. Cosmological parameters}",
      journal = {\aap},
         year = 2016,
        month = sep,
       volume = {594},
          eid = {A13},
        pages = {A13},
          doi = {10.1051/0004-6361/201525830},
archivePrefix = {arXiv},
       eprint = {1502.01589},
 primaryClass = {astro-ph.CO},
       adsurl = {https://ui.adsabs.harvard.edu/abs/2016A&A...594A..13P}
}

@ARTICLE{2010ApJS..190..418G,
       author = {{Ganeshalingam}, Mohan and {Li}, Weidong and {Filippenko}, Alexei V. and {Anderson}, Carmen and {Foster}, Griffin and {Gates}, Elinor L. and {Griffith}, Christopher V. and {Grigsby}, Bryant J. and {Joubert}, Niels and {Leja}, Joel and {Lowe}, Thomas B. and {Macomber}, Brent and {Pritchard}, Tyler and {Thrasher}, Patrick and {Winslow}, Dustin},
        title = "{Results of the Lick Observatory Supernova Search Follow-up Photometry Program: BVRI Light Curves of 165 Type Ia Supernovae}",
      journal = {\apjs},
         year = 2010,
        month = oct,
       volume = {190},
       number = {2},
        pages = {418-448},
          doi = {10.1088/0067-0049/190/2/418},
       adsurl = {https://ui.adsabs.harvard.edu/abs/2010ApJS..190..418G}
}

@ARTICLE{2018ApJ...859...24C,
       author = {{Contreras}, Carlos and {Phillips}, M.~M. and {Burns}, Christopher R. and {Piro}, Anthony L. and {Shappee}, B.~J. and {Stritzinger}, Maximilian D. and {Baltay}, C. and {Brown}, Peter J. and {Conseil}, Emmanuel and {Klotz}, Alain and {Nugent}, Peter E. and {Turpin}, Damien and {Parker}, Stu and {Rabinowitz}, D. and {Hsiao}, Eric Y. and {Morrell}, Nidia and {Campillay}, Abdo and {Castell{\'o}n}, Sergio and {Corco}, Carlos and {Gonz{\'a}lez}, Consuelo and {Krisciunas}, Kevin and {Ser{\'o}n}, Jacqueline and {Tucker}, Brad E. and {Walker}, E.~S. and {Baron}, E. and {Cain}, C. and {Childress}, Michael J. and {Folatelli}, Gast{\'o}n and {Freedman}, Wendy L. and {Hamuy}, Mario and {Hoeflich}, P. and {Persson}, S.~E. and {Scalzo}, Richard and {Schmidt}, Brian and {Suntzeff}, Nicholas B.},
        title = "{SN 2012fr: Ultraviolet, Optical, and Near-infrared Light Curves of a Type Ia Supernova Observed within a Day of Explosion}",
      journal = {\apj},
         year = 2018,
        month = may,
       volume = {859},
       number = {1},
          eid = {24},
        pages = {24},
          doi = {10.3847/1538-4357/aabaf8},
archivePrefix = {arXiv},
       eprint = {1803.10095},
 primaryClass = {astro-ph.HE},
       adsurl = {https://ui.adsabs.harvard.edu/abs/2018ApJ...859...24C}
}

@ARTICLE{1993ApJ...413L.105P,
       author = {{Phillips}, M.~M.},
        title = "{The Absolute Magnitudes of Type IA Supernovae}",
      journal = {\apjl},
         year = 1993,
        month = aug,
       volume = {413},
        pages = {L105},
          doi = {10.1086/186970},
       adsurl = {https://ui.adsabs.harvard.edu/abs/1993ApJ...413L.105P}
}

@ARTICLE{1992ApJ...387L..33R,
       author = {{Ruiz-Lapuente}, P. and {Cappellaro}, E. and {Turatto}, M. and {Gouiffes}, C. and {Danziger}, I.~J. and {della Valle}, M. and {Lucy}, L.~B.},
        title = "{Modeling the Iron-dominated Spectra of the Type IA Supernova SN 1991T at Premaximum}",
      journal = {\apjl},
         year = 1992,
        month = mar,
       volume = {387},
        pages = {L33},
          doi = {10.1086/186299},
       adsurl = {https://ui.adsabs.harvard.edu/abs/1992ApJ...387L..33R}
}

@ARTICLE{2024ApJ...969...80C,
       author = {{Chakraborty}, Sudeshna and {Sadler}, Benjamin and {Hoeflich}, Peter and {Hsiao}, Eric Y. and {Phillips}, M.~M. and {Burns}, C.~R. and {Diamond}, T. and {Dominguez}, I. and {Galbany}, L. and {Uddin}, S.~A. and {Ashall}, C. and {Krisciunas}, K. and {Kumar}, S. and {Mera}, T.~B. and {Morrell}, N. and {Baron}, E. and {Contreras}, C. and {Stritzinger}, M.~D. and {Suntzeff}, N.~B.},
        title = "{Type Ia Supernova Progenitor Properties and their Host Galaxies}",
      journal = {\apj},
         year = 2024,
        month = jul,
       volume = {969},
       number = {2},
          eid = {80},
        pages = {80},
          doi = {10.3847/1538-4357/ad4702},
archivePrefix = {arXiv},
       eprint = {2311.03473},
 primaryClass = {astro-ph.HE},
       adsurl = {https://ui.adsabs.harvard.edu/abs/2024ApJ...969...80C}
}

@ARTICLE{2023MNRAS.520..560H,
       author = {{Hoeflich}, Peter and {Yang}, Yi and {Baade}, Dietrich and {Cikota}, Aleksandar and {Maund}, Justyn R. and {Mishra}, Divya and {Patat}, Ferdinando and {Patra}, Kishore C. and {Wang}, Lifan and {Wheeler}, J. Craig and {Filippenko}, Alexei V. and {Gal-Yam}, Avishay and {Schulze}, Steven},
        title = "{The core normal Type Ia supernova 2019np - an overall spherical explosion with an aspherical surface layer and an aspherical $^{56}$Ni core}",
      journal = {\mnras},
         year = 2023,
        month = mar,
       volume = {520},
       number = {1},
        pages = {560-582},
          doi = {10.1093/mnras/stad172},
archivePrefix = {arXiv},
       eprint = {2301.04721},
 primaryClass = {astro-ph.SR},
       adsurl = {https://ui.adsabs.harvard.edu/abs/2023MNRAS.520..560H}
}

@ARTICLE{2001ApJ...546..734L,
       author = {{Li}, Weidong and {Filippenko}, Alexei V. and {Treffers}, Richard R. and {Riess}, Adam G. and {Hu}, Jingyao and {Qiu}, Yulei},
        title = "{A High Intrinsic Peculiarity Rate among Type IA Supernovae}",
      journal = {\apj},
         year = 2001,
        month = jan,
       volume = {546},
       number = {2},
        pages = {734-743},
          doi = {10.1086/318299},
archivePrefix = {arXiv},
       eprint = {astro-ph/0006292},
 primaryClass = {astro-ph},
       adsurl = {https://ui.adsabs.harvard.edu/abs/2001ApJ...546..734L}
}

@ARTICLE{2005SSRv..120...95R,
       author = {{Roming}, Peter W.~A. and {Kennedy}, Thomas E. and {Mason}, Keith O. and {Nousek}, John A. and {Ahr}, Lindy and {Bingham}, Richard E. and {Broos}, Patrick S. and {Carter}, Mary J. and {Hancock}, Barry K. and {Huckle}, Howard E. and {Hunsberger}, S.~D. and {Kawakami}, Hajime and {Killough}, Ronnie and {Koch}, T. Scott and {McLelland}, Michael K. and {Smith}, Kelly and {Smith}, Philip J. and {Soto}, Juan Carlos and {Boyd}, Patricia T. and {Breeveld}, Alice A. and {Holland}, Stephen T. and {Ivanushkina}, Mariya and {Pryzby}, Michael S. and {Still}, Martin D. and {Stock}, Joseph},
        title = "{The Swift Ultra-Violet/Optical Telescope}",
      journal = {\ssr},
         year = 2005,
        month = oct,
       volume = {120},
       number = {3-4},
        pages = {95-142},
          doi = {10.1007/s11214-005-5095-4},
archivePrefix = {arXiv},
       eprint = {astro-ph/0507413},
 primaryClass = {astro-ph},
       adsurl = {https://ui.adsabs.harvard.edu/abs/2005SSRv..120...95R}
}

@ARTICLE{2004ApJ...611.1005G,
       author = {{Gehrels}, N. and {Chincarini}, G. and {Giommi}, P. and {Mason}, K.~O. and {Nousek}, J.~A. and {Wells}, A.~A. and {White}, N.~E. and {Barthelmy}, S.~D. and {Burrows}, D.~N. and {Cominsky}, L.~R. and {Hurley}, K.~C. and {Marshall}, F.~E. and {M{\'e}sz{\'a}ros}, P. and {Roming}, P.~W.~A. and {Angelini}, L. and {Barbier}, L.~M. and {Belloni}, T. and {Campana}, S. and {Caraveo}, P.~A. and {Chester}, M.~M. and {Citterio}, O. and {Cline}, T.~L. and {Cropper}, M.~S. and {Cummings}, J.~R. and {Dean}, A.~J. and {Feigelson}, E.~D. and {Fenimore}, E.~E. and {Frail}, D.~A. and {Fruchter}, A.~S. and {Garmire}, G.~P. and {Gendreau}, K. and {Ghisellini}, G. and {Greiner}, J. and {Hill}, J.~E. and {Hunsberger}, S.~D. and {Krimm}, H.~A. and {Kulkarni}, S.~R. and {Kumar}, P. and {Lebrun}, F. and {Lloyd-Ronning}, N.~M. and {Markwardt}, C.~B. and {Mattson}, B.~J. and {Mushotzky}, R.~F. and {Norris}, J.~P. and {Osborne}, J. and {Paczynski}, B. and {Palmer}, D.~M. and {Park}, H. -S. and {Parsons}, A.~M. and {Paul}, J. and {Rees}, M.~J. and {Reynolds}, C.~S. and {Rhoads}, J.~E. and {Sasseen}, T.~P. and {Schaefer}, B.~E. and {Short}, A.~T. and {Smale}, A.~P. and {Smith}, I.~A. and {Stella}, L. and {Tagliaferri}, G. and {Takahashi}, T. and {Tashiro}, M. and {Townsley}, L.~K. and {Tueller}, J. and {Turner}, M.~J.~L. and {Vietri}, M. and {Voges}, W. and {Ward}, M.~J. and {Willingale}, R. and {Zerbi}, F.~M. and {Zhang}, W.~W.},
        title = "{The Swift Gamma-Ray Burst Mission}",
      journal = {\apj},
         year = 2004,
        month = aug,
       volume = {611},
       number = {2},
        pages = {1005-1020},
          doi = {10.1086/422091},
archivePrefix = {arXiv},
       eprint = {astro-ph/0405233},
 primaryClass = {astro-ph},
       adsurl = {https://ui.adsabs.harvard.edu/abs/2004ApJ...611.1005G}
}

@ARTICLE{1996PASP..108..190K,
       author = {{Kim}, Alex and {Goobar}, Ariel and {Perlmutter}, Saul},
        title = "{A Generalized K Correction for Type IA Supernovae: Comparing R-band Photometry beyond z=0.2 with B, V, and R-band Nearby Photometry}",
      journal = {\pasp},
         year = 1996,
        month = feb,
       volume = {108},
        pages = {190},
          doi = {10.1086/133709},
archivePrefix = {arXiv},
       eprint = {astro-ph/9505024},
 primaryClass = {astro-ph},
       adsurl = {https://ui.adsabs.harvard.edu/abs/1996PASP..108..190K}
}

@ARTICLE{1992ApJ...397..304J,
       author = {{Jeffery}, David J. and {Leibundgut}, Bruno and {Kirshner}, Robert P. and {Benetti}, Stefano and {Branch}, David and {Sonneborn}, George},
        title = "{Analysis of the Photospheric Epoch Spectra of Type IA Supernovae SN 1990N and SN 1991T}",
      journal = {\apj},
         year = 1992,
        month = sep,
       volume = {397},
        pages = {304},
          doi = {10.1086/171787},
       adsurl = {https://ui.adsabs.harvard.edu/abs/1992ApJ...397..304J}
}

@ARTICLE{1995ApJ...443...89H,
       author = {{H{\"o}flich}, P.},
        title = "{Analysis of the Type IA Supernova SN 1994D}",
      journal = {\apj},
         year = 1995,
        month = apr,
       volume = {443},
        pages = {89},
          doi = {10.1086/175505},
       adsurl = {https://ui.adsabs.harvard.edu/abs/1995ApJ...443...89H}
}

@ARTICLE{2007ApJ...666.1083Q,
       author = {{Quimby}, Robert and {H{\"o}flich}, Peter and {Wheeler}, J. Craig},
        title = "{SN 2005hj: Evidence for Two Classes of Normal-Bright SNe Ia and Implications for Cosmology}",
      journal = {\apj},
         year = 2007,
        month = sep,
       volume = {666},
       number = {2},
        pages = {1083-1092},
          doi = {10.1086/520527},
archivePrefix = {arXiv},
       eprint = {0705.4467},
 primaryClass = {astro-ph},
       adsurl = {https://ui.adsabs.harvard.edu/abs/2007ApJ...666.1083Q}
}

@ARTICLE{1992A&A...253L...9K,
       author = {{Khokhlov}, A. and {Mueller}, E. and {Hoeflich}, P.},
        title = "{Type IA supernovae : theoretical light curves with a slow pre-maximum rise.}",
      journal = {\aap},
         year = 1992,
        month = jan,
       volume = {253},
        pages = {L9-L12},
       adsurl = {https://ui.adsabs.harvard.edu/abs/1992A&A...253L...9K}
}

@ARTICLE{2024MNRAS.528.3875M,
       author = {{Maund}, Justyn R.},
        title = "{Exploring the polarization of axially symmetric supernovae with unsupervised deep learning}",
      journal = {\mnras},
         year = 2024,
        month = mar,
       volume = {528},
       number = {3},
        pages = {3875-3890},
          doi = {10.1093/mnras/stad2572},
archivePrefix = {arXiv},
       eprint = {2308.16686},
 primaryClass = {astro-ph.SR},
       adsurl = {https://ui.adsabs.harvard.edu/abs/2024MNRAS.528.3875M}
}

@ARTICLE{1998ApJ...496..454B,
       author = {{Blinnikov}, S.~I. and {Eastman}, R. and {Bartunov}, O.~S. and {Popolitov}, V.~A. and {Woosley}, S.~E.},
        title = "{A Comparative Modeling of Supernova 1993J}",
      journal = {\apj},
         year = 1998,
        month = mar,
       volume = {496},
       number = {1},
        pages = {454-472},
          doi = {10.1086/305375},
archivePrefix = {arXiv},
       eprint = {astro-ph/9711055},
 primaryClass = {astro-ph},
       adsurl = {https://ui.adsabs.harvard.edu/abs/1998ApJ...496..454B}
}

@ARTICLE{2013PASP..125.1031B,
       author = {{Brown}, T.~M. and {Baliber}, N. and {Bianco}, F.~B. and {Bowman}, M. and {Burleson}, B. and {Conway}, P. and {Crellin}, M. and {Depagne}, {\'E}. and {De Vera}, J. and {Dilday}, B. and {Dragomir}, D. and {Dubberley}, M. and {Eastman}, J.~D. and {Elphick}, M. and {Falarski}, M. and {Foale}, S. and {Ford}, M. and {Fulton}, B.~J. and {Garza}, J. and {Gomez}, E.~L. and {Graham}, M. and {Greene}, R. and {Haldeman}, B. and {Hawkins}, E. and {Haworth}, B. and {Haynes}, R. and {Hidas}, M. and {Hjelstrom}, A.~E. and {Howell}, D.~A. and {Hygelund}, J. and {Lister}, T.~A. and {Lobdill}, R. and {Martinez}, J. and {Mullins}, D.~S. and {Norbury}, M. and {Parrent}, J. and {Paulson}, R. and {Petry}, D.~L. and {Pickles}, A. and {Posner}, V. and {Rosing}, W.~E. and {Ross}, R. and {Sand}, D.~J. and {Saunders}, E.~S. and {Shobbrook}, J. and {Shporer}, A. and {Street}, R.~A. and {Thomas}, D. and {Tsapras}, Y. and {Tufts}, J.~R. and {Valenti}, S. and {Vander Horst}, K. and {Walker}, Z. and {White}, G. and {Willis}, M.},
        title = "{Las Cumbres Observatory Global Telescope Network}",
      journal = {\pasp},
         year = 2013,
        month = sep,
       volume = {125},
       number = {931},
        pages = {1031},
          doi = {10.1086/673168},
archivePrefix = {arXiv},
       eprint = {1305.2437},
 primaryClass = {astro-ph.IM},
       adsurl = {https://ui.adsabs.harvard.edu/abs/2013PASP..125.1031B}
}

@ARTICLE{2014MNRAS.438L.101V,
       author = {{Valenti}, S. and {Sand}, D. and {Pastorello}, A. and {Graham}, M.~L. and {Howell}, D.~A. and {Parrent}, J.~T. and {Tomasella}, L. and {Ochner}, P. and {Fraser}, M. and {Benetti}, S. and {Yuan}, F. and {Smartt}, S.~J. and {Maund}, J.~R. and {Arcavi}, I. and {Gal-Yam}, A. and {Inserra}, C. and {Young}, D.},
        title = "{The first month of evolution of the slow-rising Type IIP SN 2013ej in M74$^{★}$}",
      journal = {\mnras},
         year = 2014,
        month = feb,
       volume = {438},
       number = {1},
        pages = {L101-L105},
          doi = {10.1093/mnrasl/slt171},
archivePrefix = {arXiv},
       eprint = {1309.4269},
 primaryClass = {astro-ph.CO},
       adsurl = {https://ui.adsabs.harvard.edu/abs/2014MNRAS.438L.101V}
}

@ARTICLE{1982PASP...94..715F,
       author = {{Filippenko}, A.~V.},
        title = "{The importance of atmospheric differential refraction in spectrophotometry.}",
      journal = {\pasp},
         year = 1982,
        month = aug,
       volume = {94},
        pages = {715-721},
          doi = {10.1086/131052},
       adsurl = {https://ui.adsabs.harvard.edu/abs/1982PASP...94..715F}
}

@ARTICLE{2009ApJ...695.1244B,
       author = {{Bravo}, Eduardo and {Garc{\'\i}a-Senz}, Domingo},
        title = "{Pulsating Reverse Detonation Models of Type Ia Supernovae. I. Detonation Ignition}",
      journal = {\apj},
         year = 2009,
        month = apr,
       volume = {695},
       number = {2},
        pages = {1244-1256},
          doi = {10.1088/0004-637X/695/2/1244},
archivePrefix = {arXiv},
       eprint = {0901.3008},
 primaryClass = {astro-ph.SR},
       adsurl = {https://ui.adsabs.harvard.edu/abs/2009ApJ...695.1244B}
}

@ARTICLE{2009ApJ...695.1257B,
       author = {{Bravo}, Eduardo and {Garc{\'\i}a-Senz}, Domingo and {Cabez{\'o}n}, Rub{\'e}n M. and {Dom{\'\i}nguez}, Inmaculada},
        title = "{Pulsating Reverse Detonation Models of Type Ia Supernovae. II. Explosion}",
      journal = {\apj},
         year = 2009,
        month = apr,
       volume = {695},
       number = {2},
        pages = {1257-1272},
          doi = {10.1088/0004-637X/695/2/1257},
archivePrefix = {arXiv},
       eprint = {0901.3013},
 primaryClass = {astro-ph.SR},
       adsurl = {https://ui.adsabs.harvard.edu/abs/2009ApJ...695.1257B}
}

@ARTICLE{2000AJ....120.1479H,
       author = {{Hamuy}, Mario and {Trager}, S.~C. and {Pinto}, Philip A. and {Phillips}, M.~M. and {Schommer}, R.~A. and {Ivanov}, Valentin and {Suntzeff}, Nicholas B.},
        title = "{A Search for Environmental Effects on Type IA Supernovae}",
      journal = {\aj},
         year = 2000,
        month = sep,
       volume = {120},
       number = {3},
        pages = {1479-1486},
          doi = {10.1086/301527},
archivePrefix = {arXiv},
       eprint = {astro-ph/0005213},
 primaryClass = {astro-ph},
       adsurl = {https://ui.adsabs.harvard.edu/abs/2000AJ....120.1479H}
}

@ARTICLE{2022ApJ...938...47P,
       author = {{Phillips}, M.~M. and {Ashall}, C. and {Burns}, Christopher R. and {Contreras}, Carlos and {Galbany}, L. and {Hoeflich}, P. and {Hsiao}, E.~Y. and {Morrell}, Nidia and {Nugent}, Peter and {Uddin}, Syed A. and {Baron}, E. and {Freedman}, Wendy L. and {Harris}, Chelsea E. and {Krisciunas}, Kevin and {Kumar}, S. and {Lu}, J. and {Persson}, S.~E. and {Piro}, Anthony L. and {Polin}, Abigail and {Shahbandeh}, M. and {Stritzinger}, Maximilian and {Suntzeff}, Nicholas B.},
        title = "{The Absolute Magnitudes of 1991T-like Supernovae}",
      journal = {\apj},
         year = 2022,
        month = oct,
       volume = {938},
       number = {1},
          eid = {47},
        pages = {47},
          doi = {10.3847/1538-4357/ac9305},
archivePrefix = {arXiv},
       eprint = {2209.08031},
 primaryClass = {astro-ph.HE},
       adsurl = {https://ui.adsabs.harvard.edu/abs/2022ApJ...938...47P}
}

@ARTICLE{1994ApJ...427..315A,
       author = {{Arnett}, David and {Livne}, Eli},
        title = "{The Delayed-Detonation Model of Type IA Supernovae I. The Deflagration Phase}",
      journal = {\apj},
         year = 1994,
        month = may,
       volume = {427},
        pages = {314},
          doi = {10.1086/174142},
       adsurl = {https://ui.adsabs.harvard.edu/abs/1994ApJ...427..315A}
}

@ARTICLE{2022MNRAS.509.6028C,
       author = {{Chu}, Matthew R. and {Cikota}, Aleksandar and {Baade}, Dietrich and {Patat}, Ferdinando and {Filippenko}, Alexei V. and {Wheeler}, J. Craig and {Maund}, Justyn and {Bulla}, Mattia and {Yang}, Yi and {H{\"o}flich}, Peter and {Wang}, Lifan},
        title = "{An imaging polarimetry survey of Type Ia supernovae: are peculiar extinction and polarization properties produced by circumstellar or interstellar matter?}",
      journal = {\mnras},
         year = 2022,
        month = feb,
       volume = {509},
       number = {4},
        pages = {6028-6046},
          doi = {10.1093/mnras/stab3392},
archivePrefix = {arXiv},
       eprint = {2111.09980},
 primaryClass = {astro-ph.GA},
       adsurl = {https://ui.adsabs.harvard.edu/abs/2022MNRAS.509.6028C}
}

@ARTICLE{2001ApJ...550.1030W,
       author = {{Wang}, Lifan and {Howell}, D. Andrew and {H{\"o}flich}, Peter and {Wheeler}, J. Craig},
        title = "{Bipolar Supernova Explosions}",
      journal = {\apj},
         year = 2001,
        month = apr,
       volume = {550},
       number = {2},
        pages = {1030-1035},
          doi = {10.1086/319822},
       adsurl = {https://ui.adsabs.harvard.edu/abs/2001ApJ...550.1030W}
}

@ARTICLE{2020MNRAS.494..885S,
       author = {{Stevance}, H.~F. and {Baade}, D. and {Bruten}, J.~R. and {Cikota}, A. and {Clocchiatti}, A. and {Hines}, D.~C. and {H{\"o}flich}, P. and {Maund}, J.~R. and {Patat}, F. and {Vallely}, P.~J. and {Wheeler}, J.~C.},
        title = "{The shape of SN 1993J re-analysed}",
      journal = {\mnras},
         year = 2020,
        month = may,
       volume = {494},
       number = {1},
        pages = {885-901},
          doi = {10.1093/mnras/staa721},
archivePrefix = {arXiv},
       eprint = {2003.06032},
 primaryClass = {astro-ph.SR},
       adsurl = {https://ui.adsabs.harvard.edu/abs/2020MNRAS.494..885S}
}

@ARTICLE{1992A&A...259..549H,
       author = {{Hoeflich}, P. and {Khokhlov}, A. and {Mueller}, E.},
        title = "{Gamma-ray light curves and spectra for Type IA supernovae}",
      journal = {\aap},
         year = 1992,
        month = jun,
       volume = {259},
       number = {2},
        pages = {549-566},
       adsurl = {https://ui.adsabs.harvard.edu/abs/1992A&A...259..549H}
}

@ARTICLE{1997ARA&A..35..309F,
       author = {{Filippenko}, Alexei V.},
        title = "{Optical Spectra of Supernovae}",
      journal = {\araa},
         year = 1997,
        month = jan,
       volume = {35},
        pages = {309-355},
          doi = {10.1146/annurev.astro.35.1.309},
       adsurl = {https://ui.adsabs.harvard.edu/abs/1997ARA&A..35..309F}
}

@ARTICLE{1991ApJ...371L..23L,
       author = {{Leibundgut}, Bruno and {Kirshner}, Robert P. and {Filippenko}, Alexei V. and {Shields}, Joseph C. and {Foltz}, Craig B. and {Phillips}, Mark M. and {Sonneborn}, George},
        title = "{Premaximum Observations of the Type IA SN 1990N}",
      journal = {\apjl},
         year = 1991,
        month = apr,
       volume = {371},
        pages = {L23},
          doi = {10.1086/185993},
       adsurl = {https://ui.adsabs.harvard.edu/abs/1991ApJ...371L..23L}
}

@ARTICLE{2013ApJ...779...23M,
       author = {{Milne}, Peter A. and {Brown}, Peter J. and {Roming}, Peter W.~A. and {Bufano}, Filomena and {Gehrels}, Neil},
        title = "{Grouping Normal Type Ia Supernovae by UV to Optical Color Differences}",
      journal = {\apj},
         year = 2013,
        month = dec,
       volume = {779},
       number = {1},
          eid = {23},
        pages = {23},
          doi = {10.1088/0004-637X/779/1/23},
archivePrefix = {arXiv},
       eprint = {1308.2703},
 primaryClass = {astro-ph.CO},
       adsurl = {https://ui.adsabs.harvard.edu/abs/2013ApJ...779...23M}
}

@ARTICLE{2009A&A...507...85B,
       author = {{Balland}, C. and {Baumont}, S. and {Basa}, S. and {Mouchet}, M. and {Howell}, D.~A. and {Astier}, P. and {Carlberg}, R.~G. and {Conley}, A. and {Fouchez}, D. and {Guy}, J. and {Hardin}, D. and {Hook}, I.~M. and {Pain}, R. and {Perrett}, K. and {Pritchet}, C.~J. and {Regnault}, N. and {Rich}, J. and {Sullivan}, M. and {Antilogus}, P. and {Arsenijevic}, V. and {Le Du}, J. and {Fabbro}, S. and {Lidman}, C. and {Mour{\~a}o}, A. and {Palanque-Delabrouille}, N. and {P{\'e}contal}, E. and {Ruhlmann-Kleider}, V.},
        title = "{The ESO/VLT 3rd year Type Ia supernova data set from the supernova legacy survey}",
      journal = {\aap},
         year = 2009,
        month = nov,
       volume = {507},
       number = {1},
        pages = {85-103},
          doi = {10.1051/0004-6361/200912246},
archivePrefix = {arXiv},
       eprint = {0909.3316},
 primaryClass = {astro-ph.CO},
       adsurl = {https://ui.adsabs.harvard.edu/abs/2009A&A...507...85B}
}

@ARTICLE{2018ApJ...867...23H,
       author = {{Hounsell}, R. and {Scolnic}, D. and {Foley}, R.~J. and {Kessler}, R. and {Miranda}, V. and {Avelino}, A. and {Bohlin}, R.~C. and {Filippenko}, A.~V. and {Frieman}, J. and {Jha}, S.~W. and {Kelly}, P.~L. and {Kirshner}, R.~P. and {Mandel}, K. and {Rest}, A. and {Riess}, A.~G. and {Rodney}, S.~A. and {Strolger}, L.},
        title = "{Simulations of the WFIRST Supernova Survey and Forecasts of Cosmological Constraints}",
      journal = {\apj},
         year = 2018,
        month = nov,
       volume = {867},
       number = {1},
          eid = {23},
        pages = {23},
          doi = {10.3847/1538-4357/aac08b},
archivePrefix = {arXiv},
       eprint = {1702.01747},
 primaryClass = {astro-ph.IM},
       adsurl = {https://ui.adsabs.harvard.edu/abs/2018ApJ...867...23H}
}

@ARTICLE{2021arXiv211103081R,
       author = {{Rose}, B.~M. and {Baltay}, C. and {Hounsell}, R. and {Macias}, P. and {Rubin}, D. and {Scolnic}, D. and {Aldering}, G. and {Bohlin}, R. and {Dai}, M. and {Deustua}, S.~E. and {Foley}, R.~J. and {Fruchter}, A. and {Galbany}, L. and {Jha}, S.~W. and {Jones}, D.~O. and {Joshi}, B.~A. and {Kelly}, P.~L. and {Kessler}, R. and {Kirshner}, R.~P. and {Mandel}, K.~S. and {Perlmutter}, S. and {Pierel}, J. and {Qu}, H. and {Rabinowitz}, D. and {Rest}, A. and {Riess}, A.~G. and {Rodney}, S. and {Sako}, M. and {Siebert}, M.~R. and {Strolger}, L. and {Suzuki}, N. and {Thorp}, S. and {Van Dyk}, S.~D. and {Wang}, K. and {Ward}, S.~M. and {Wood-Vasey}, W.~M.},
        title = "{A Reference Survey for Supernova Cosmology with the Nancy Grace Roman Space Telescope}",
      journal = {arXiv e-prints},
         year = 2021,
        month = nov,
          eid = {arXiv:2111.03081},
        pages = {arXiv:2111.03081},
          doi = {10.48550/arXiv.2111.03081},
archivePrefix = {arXiv},
       eprint = {2111.03081},
 primaryClass = {astro-ph.CO},
       adsurl = {https://ui.adsabs.harvard.edu/abs/2021arXiv211103081R}
}

@ARTICLE{2023SCPMA..6629511L,
       author = {{Li}, Shi-Yu and {Li}, Yun-Long and {Zhang}, Tianmeng and {Vink{\'o}}, J{\'o}zsef and {Reg{\H{o}}s}, Enik{\H{o}} and {Wang}, Xiaofeng and {Xi}, Gaobo and {Zhan}, Hu},
        title = "{Forecast of cosmological constraints with type Ia supernovae from the Chinese Space Station Telescope}",
      journal = {Science China Physics, Mechanics, and Astronomy},
         year = 2023,
        month = feb,
       volume = {66},
       number = {2},
          eid = {229511},
        pages = {229511},
          doi = {10.1007/s11433-022-2018-0},
archivePrefix = {arXiv},
       eprint = {2210.05450},
 primaryClass = {astro-ph.CO},
       adsurl = {https://ui.adsabs.harvard.edu/abs/2023SCPMA..6629511L}
}

@ARTICLE{2026SCPMA..6939501C,
       author = {{CSST Collaboration} and {Gong}, Yan and {Miao}, Haitao and {Zhan}, Hu and {Li}, Zhao-Yu and {Shangguan}, Jinyi and {Li}, Haining and {Liu}, Chao and {Chen}, Xuefei and {Yuan}, Haibo and {Zhou}, Jilin and {Liu}, Hui-Gen and {Yu}, Cong and {Ji}, Jianghui and {Qi}, Zhaoxiang and {Liu}, Jiacheng and {Dai}, Zigao and {Wang}, Xiaofeng and {Zheng}, Zhenya and {Hao}, Lei and {Dou}, Jiangpei and {Ao}, Yiping and {Lin}, Zhenhui and {Zhang}, Kun and {Wang}, Wei and {Sun}, Guotong and {Li}, Ran and {Li}, Guoliang and {Xu}, Youhua and {Li}, Xinfeng and {Li}, Shengyang and {Wu}, Peng and {Zhang}, Jiuxing and {Wang}, Bo and {Bai}, Jinming and {Cai}, Yi-Fu and {Cai}, Zheng and {Cao}, Jie and {Chan}, Kwan Chuen and {Chang}, Jin and {Chen}, Xiaodian and {Chen}, Xuelei and {Chen}, Yuqin and {Chen}, Yun and {Cui}, Wei and {Dong}, Subo and {Du}, Pu and {Duan}, Wenying and {Fan}, Junhui and {Fan}, LuLu and {Fan}, Zhou and {Fan}, Zuhui and {Fang}, Taotao and {Fu}, Jianning and {Fu}, Liping and {Fu}, Zhensen and {Gao}, Jian and {Gu}, Shenghong and {Gu}, Yidong and {Guo}, Qi and {Han}, Zhanwen and {Hu}, Bin and {Huang}, Zhiqi and {Ho}, Luis C. and {Jiang}, Linhua and {Jiang}, Ning and {Jing}, Yipeng and {Kang}, Xi and {Kong}, Xu and {Li}, Cheng and {Li}, Chengyuan and {Li}, Di and {Li}, Jing and {Li}, Nan and {Li}, Yang A. and {Liao}, Shilong and {Lin}, Weipeng and {Liu}, Fengshan and {Liu}, Jifeng and {Liu}, Xiangkun and {Liu}, Zhuokai and {Mao}, Ruiqing and {Mao}, Shude and {Meng}, Xianmin and {Pang}, Xiaoying and {Peng}, Xiyan and {Peng}, Yingjie and {Shan}, Huanyuan and {Shen}, Juntai and {Shen}, Shiyin and {Shen}, Zhiqiang and {Shi}, Sheng-Cai and {Shi}, Yong and {Tan}, Siyuan and {Tian}, Hao and {Wang}, Jianmin and {Wang}, Jun-Xian and {Wang}, Xin and {Wang}, Yuting and {Wu}, Hong and {Wu}, Jingwen and {Wu}, Xuebing and {Xu}, Chun and {Xue}, Xiang-Xiang and {Xue}, Yongquan and {Yang}, Ji and {Yang}, Xiaohu and {Yao}, Qijun and {Yuan}, Fangting and {Yuan}, Zhen and {Zhang}, Jun and {Zhang}, Pengjie and {Zhang}, Tianmeng and {Zhang}, Wei and {Zhang}, Xin and {Zhao}, Gang and {Zhao}, Gongbo and {Zhong}, Hongen and {Zhong}, Jing and {Zhou}, Liyong and {Zhu}, Wei and {Zu}, Ying},
        title = "{Introduction to the Chinese Space Station Survey Telescope (CSST)}",
      journal = {Science China Physics, Mechanics, and Astronomy},
         year = 2026,
        month = jan,
       volume = {69},
       number = {3},
          eid = {239501},
        pages = {239501},
          doi = {10.1007/s11433-025-2809-0},
archivePrefix = {arXiv},
       eprint = {2507.04618},
 primaryClass = {astro-ph.IM},
       adsurl = {https://ui.adsabs.harvard.edu/abs/2026SCPMA..6939501C}
}

@ARTICLE{2026RAA....26e5020Z,
       author = {{Zheng}, Zhen-Ya and {Xu}, Chun and {Liu}, Xiaohua and {Chen}, Yong-He and {Xu}, Fang and {Zhan}, Hu and {Li}, Xinfeng and {Zheng}, Lixin and {Shan}, Huanyuan and {Zhong}, Jing and {Yan}, Zhaojun and {Yuan}, Fang-Ting and {Jiang}, Chunyan and {Peng}, Xiyan and {Chen}, Wei and {Cheng}, Xue and {Chen}, Zhen-Lei and {Zhu}, Shuairu and {Long}, Lin and {Zhang}, Xin and {Gong}, Yan and {Shao}, Li and {Wang}, Wei and {Zhang}, Tianyi and {Ju}, Guohao and {Li}, Chenghao and {Li}, Zhiyuan and {Wang}, Tao and {Wang}, Junfeng and {Li}, Chengyuan and {Ma}, Bin and {Wang}, Jianguo and {Wang}, Lei and {Liu}, Dezi and {Lin}, Nie and {Li}, Kexin and {Wen}, Xinrong and {Wu}, Maochun and {Lin}, Ruqiu and {Ji}, Xiang},
        title = "{MCI: Multi-channel Imager on the Chinese Space Station Survey Telescope}",
      journal = {Research in Astronomy and Astrophysics},
         year = 2026,
        month = may,
       volume = {26},
       number = {5},
          eid = {055020},
        pages = {055020},
          doi = {10.1088/1674-4527/ae4a05},
archivePrefix = {arXiv},
       eprint = {2509.14691},
 primaryClass = {astro-ph.IM},
       adsurl = {https://ui.adsabs.harvard.edu/abs/2026RAA....26e5020Z}
}

@ARTICLE{2025arXiv250510574O,
       author = {{Observations Time Allocation Committee}, Roman and {Community Survey Definition Committees}, Core},
        title = "{Roman Observations Time Allocation Committee: Final Report and Recommendations}",
      journal = {arXiv e-prints},
         year = 2025,
        month = may,
          eid = {arXiv:2505.10574},
        pages = {arXiv:2505.10574},
          doi = {10.48550/arXiv.2505.10574},
archivePrefix = {arXiv},
       eprint = {2505.10574},
 primaryClass = {astro-ph.IM},
       adsurl = {https://ui.adsabs.harvard.edu/abs/2025arXiv250510574O}
}

@ARTICLE{2002PASP..114..803N,
       author = {{Nugent}, Peter and {Kim}, Alex and {Perlmutter}, Saul},
        title = "{K-Corrections and Extinction Corrections for Type Ia Supernovae}",
      journal = {\pasp},
         year = 2002,
        month = aug,
       volume = {114},
       number = {798},
        pages = {803-819},
          doi = {10.1086/341707},
archivePrefix = {arXiv},
       eprint = {astro-ph/0205351},
 primaryClass = {astro-ph},
       adsurl = {https://ui.adsabs.harvard.edu/abs/2002PASP..114..803N}
}

@ARTICLE{2006ApJ...641...50W,
       author = {{Wang}, Lifan and {Strovink}, Mark and {Conley}, Alexander and {Goldhaber}, Gerson and {Kowalski}, Marek and {Perlmutter}, Saul and {Siegrist}, James},
        title = "{Nonlinear Decline-Rate Dependence and Intrinsic Variation of Type Ia Supernova Luminosities}",
      journal = {\apj},
         year = 2006,
        month = apr,
       volume = {641},
       number = {1},
        pages = {50-69},
          doi = {10.1086/500422},
archivePrefix = {arXiv},
       eprint = {astro-ph/0512370},
 primaryClass = {astro-ph},
       adsurl = {https://ui.adsabs.harvard.edu/abs/2006ApJ...641...50W}
}

@INPROCEEDINGS{2000AIPC..540..227F,
       author = {{Filippenko}, Alexei V. and {Riess}, Adam G. and {High-Z Supernova Search Team}},
        title = "{Evidence from Type Ia supernovae for an accelerating universe}",
    booktitle = {Particle Physics and Cosmology},
         year = 2000,
       editor = {{Nieves}, Jose F.},
       series = {American Institute of Physics Conference Series},
       volume = {540},
        month = oct,
    publisher = {AIP},
        pages = {227-246},
          doi = {10.1063/1.1328887},
archivePrefix = {arXiv},
       eprint = {astro-ph/0008057},
 primaryClass = {astro-ph},
       adsurl = {https://ui.adsabs.harvard.edu/abs/2000AIPC..540..227F}
}

@ARTICLE{2000ApJ...544L.111C,
       author = {{Coil}, Alison L. and {Matheson}, Thomas and {Filippenko}, Alexei V. and {Leonard}, Douglas C. and {Tonry}, John and {Riess}, Adam G. and {Challis}, Peter and {Clocchiatti}, Alejandro and {Garnavich}, Peter M. and {Hogan}, Craig J. and {Jha}, Saurabh and {Kirshner}, Robert P. and {Leibundgut}, B. and {Phillips}, M.~M. and {Schmidt}, Brian P. and {Schommer}, Robert A. and {Smith}, R. Chris and {Soderberg}, Alicia M. and {Spyromilio}, J. and {Stubbs}, Christopher and {Suntzeff}, Nicholas B. and {Woudt}, Patrick},
        title = "{Optical Spectra of Type IA Supernovae at Z=0.46 and Z=1.2}",
      journal = {\apjl},
         year = 2000,
        month = dec,
       volume = {544},
       number = {2},
        pages = {L111-L114},
          doi = {10.1086/317311},
archivePrefix = {arXiv},
       eprint = {astro-ph/0009102},
 primaryClass = {astro-ph},
       adsurl = {https://ui.adsabs.harvard.edu/abs/2000ApJ...544L.111C}
}

@ARTICLE{2001ApJ...560...49R,
       author = {{Riess}, Adam G. and {Nugent}, Peter E. and {Gilliland}, Ronald L. and {Schmidt}, Brian P. and {Tonry}, John and {Dickinson}, Mark and {Thompson}, Rodger I. and {Budav{\'a}ri}, Tam{\'a}s and {Casertano}, Stefano and {Evans}, Aaron S. and {Filippenko}, Alexei V. and {Livio}, Mario and {Sanders}, David B. and {Shapley}, Alice E. and {Spinrad}, Hyron and {Steidel}, Charles C. and {Stern}, Daniel and {Surace}, Jason and {Veilleux}, Sylvain},
        title = "{The Farthest Known Supernova: Support for an Accelerating Universe and a Glimpse of the Epoch of Deceleration}",
      journal = {\apj},
         year = 2001,
        month = oct,
       volume = {560},
       number = {1},
        pages = {49-71},
          doi = {10.1086/322348},
archivePrefix = {arXiv},
       eprint = {astro-ph/0104455},
 primaryClass = {astro-ph},
       adsurl = {https://ui.adsabs.harvard.edu/abs/2001ApJ...560...49R}
}

@ARTICLE{2004ApJ...607..665R,
       author = {{Riess}, Adam G. and {Strolger}, Louis-Gregory and {Tonry}, John and {Casertano}, Stefano and {Ferguson}, Henry C. and {Mobasher}, Bahram and {Challis}, Peter and {Filippenko}, Alexei V. and {Jha}, Saurabh and {Li}, Weidong and {Chornock}, Ryan and {Kirshner}, Robert P. and {Leibundgut}, Bruno and {Dickinson}, Mark and {Livio}, Mario and {Giavalisco}, Mauro and {Steidel}, Charles C. and {Ben{\'\i}tez}, Txitxo and {Tsvetanov}, Zlatan},
        title = "{Type Ia Supernova Discoveries at z > 1 from the Hubble Space Telescope: Evidence for Past Deceleration and Constraints on Dark Energy Evolution}",
      journal = {\apj},
         year = 2004,
        month = jun,
       volume = {607},
       number = {2},
        pages = {665-687},
          doi = {10.1086/383612},
archivePrefix = {arXiv},
       eprint = {astro-ph/0402512},
 primaryClass = {astro-ph},
       adsurl = {https://ui.adsabs.harvard.edu/abs/2004ApJ...607..665R}
}


\begin{table}[ht]
\normalsize	
\centering
\caption{Bolometric Light Curve of SN~2019vrq After Correcting for the UV and NIR Contributions}\label{Table_bolo_lc}
\begin{tabular}{c|c|c}
\hline
\hline
$t-t_{B{\rm max}}$ [day]  &  log\,$L^{\rm bol}$   &  Uncertainty \\
\hline
$-$10.656  &  43.0178  &   0.0226  \\
 $-$9.285  &  43.1312  &   0.0227  \\
 $-$7.773  &  43.2273  &   0.0232  \\
 $-$6.856  &  43.2647  &   0.0250  \\
 $-$5.813  &  43.3003  &   0.0228  \\
 $-$4.699  &  43.3096  &   0.0349  \\
 $-$3.125  &  43.3412  &   0.0272  \\
 $-$1.891  &  43.3458  &   0.0236  \\
 $-$1.856  &  43.3430  &   0.0227  \\
 $-$0.848  &  43.3450  &   0.0275  \\
    1.750  &  43.3001  &   0.0240  \\
    3.137  &  43.2899  &   0.0255  \\
    4.785  &  43.2560  &   0.0267  \\
    6.227  &  43.2323  &   0.0277  \\
    7.738  &  43.1958  &   0.0277  \\
    9.152  &  43.1664  &   0.0271  \\
   11.168  &  43.1236  &   0.0269  \\
   16.199  &  42.9525  &   0.0264  \\
   19.059  &  42.8746  &   0.0255  \\
   21.758  &  42.8104  &   0.0257  \\
   25.066  &  42.7495  &   0.0248  \\
   28.363  &  42.7102  &   0.0252  \\
   29.176  &  42.6994  &   0.0248  \\
   33.094  &  42.6484  &   0.0250  \\
   33.137  &  42.6446  &   0.0243  \\
   38.387  &  42.5575  &   0.0355  \\
   40.340  &  42.5130  &   0.0248  \\
   45.098  &  42.4400  &   0.0263  \\
   50.332  &  42.3422  &   0.0314  \\
   54.070  &  42.2929  &   0.0299  \\
   59.324  &  42.2438  &   0.0298  \\
   65.731  &  42.1753  &   0.0321  \\
   67.301  &  42.1505  &   0.0365  \\
   72.293  &  42.1071  &   0.0630  \\
   78.332  &  42.0353  &   0.0359  \\
   84.309  &  42.0138  &   0.0523  \\
   87.277  &  41.9888  &   0.0630  \\
   93.662  &  41.8922  &   0.0385  \\
   95.031  &  41.8747  &   0.0330  \\
  101.012  &  41.8148  &   0.0401  \\
 \hline
\end{tabular}
\label{tab:bolo_lc}
\end{table}

\end{document}